\documentclass{article}
\usepackage[margin=1in]{geometry}
\usepackage{amsmath,amssymb}
\usepackage{graphicx}

\DeclareMathOperator{\Bernoulli}{Bernoulli}
\DeclareMathOperator{\E}{Expectation}

\title{Plots of the cumulative differences between observed\\
       and expected values of ordered Bernoulli variates}
\author{Mark Tygert}

\begin{document}

\maketitle

\begin{abstract}
Many predictions are probabilistic in nature; for example,
a prediction could be for precipitation tomorrow, but with only a 30\% chance.
Given both the predictions and the actual outcomes,
``reliability diagrams'' (also known as ``calibration plots'') help detect
and diagnose statistically significant discrepancies
between the predictions and the outcomes.
The canonical reliability diagrams are based on histogramming
the observed and expected values of the predictions;
several variants of the standard reliability diagrams propose
to replace the hard histogram binning with soft kernel density estimation
using smooth convolutional kernels of widths similar to the widths of the bins.
In all cases, an important question naturally arises: which widths are best
(or are multiple plots with different widths better)?
Rather than answering this question, plots of the cumulative differences
between the observed and expected values largely avoid the question,
by displaying miscalibration directly as the slopes of secant lines
for the graphs. Slope is easy to perceive with quantitative precision
even when the constant offsets of the secant lines are irrelevant.
There is no need to bin or perform kernel density estimation
with a somewhat arbitrary kernel.
\end{abstract}

\begin{center}
\end{center}

\section{Introduction}

Given 100 independent observations of outcomes (``success'' or ``failure'')
of Bernoulli trials that are forecast to have an 80\% chance of success,
the forecasts are perfectly {\it calibrated} when 80 of the observations
report success. More generally, given some number, say $n$,
of independent observations of outcomes of Bernoulli trials
that are forecast to have a probability $p$ of success,
the predictions are perfectly calibrated when $np$ of the observations
report success. Needless to say, the actual number of observations of success
is likely to vary around $np$ randomly, so in practice we test not whether
$np$ is exactly equal to the actual number of observations, but rather
whether the difference between $np$ and the actual number of observations
is statistically significant. Such significance tests
can be found in any standard textbook on statistics in the case for which
all $n$ observations have to do with the same predicted probability $p$.
The present paper considers the following more general setting.

Suppose we have $n$ observations $C_1$,~$C_2$, \dots, $C_n$
of the outcomes of Bernoulli trials with corresponding predicted probabilities
of success, say $P_1$,~$P_2$, \dots, $P_n$.
For instance, each $P_k$ could be a classifier's probabilistic score 
and the corresponding $C_k$ could be the indicator of correct classification,
with $C_k = 1$ when the classification is correct
and $C_k = 0$ when the classification is incorrect
(so $C_k$ could also be regarded as a class label,
where class 1 corresponds to ``the classifier succeeded''
and class 0 corresponds to ``the classifier erred'').
We would then want to test the hypothesis
\begin{equation}
\label{null}
C_k \sim \Bernoulli(P_k)
\end{equation}
for all $k = 1$,~$2$, \dots, $n$; the null hypothesis~(\ref{null})
is the same as that considered in the previous paragraph
when $P_1 = P_2 = \dots = P_n = p$.
Let us reorder the samples
(preserving the pairing of $C_k$ with $P_k$ for every $k$)
such that $P_1 \le P_2 \le \dots \le P_n$, with any ties ordered randomly,
perturbing so that $P_1 < P_2 < \dots < P_n$.

The canonical graphical method for assessing~(\ref{null})
is to bin $P_1$,~$P_2$, \dots, $P_n$ into some number --- say $m$ ---
of disjoint, abutting intervals indexed by $I_1$,~$I_2$, \dots, $I_m$,
and calculate both the average of $P_k$ and the average of $C_k$
for each bin:
\begin{equation}
\label{averageP}
A_j = \frac{1}{\# I_j} \sum_{k \in I_j} P_k
\end{equation}
and
\begin{equation}
\label{averageC}
B_j = \frac{1}{\# I_j} \sum_{k \in I_j} C_k
\end{equation}
for $j = 1$,~$2$, \dots, $m$,
where $\# I_j$ is the number of integer indices in $I_j$.
If the probabilistic predictions are well-calibrated,
then $A_j$ and $B_j$ will be close for $j = 1$,~$2$, \dots, $m$.
The conventional visual means of displaying whether they are close
is known as a ``reliability diagram'' or ``calibration plot,''
which plot the pairs $(A_j, B_j)$ for $j = 1$,~$2$, \dots, $m$,
along with the line connecting the origin $(0, 0)$ to the point $(1, 1)$;
a pair $(A_j, B_j)$ falls on that line precisely when $A_j = B_j$.
Copious examples of such reliability diagrams are available
in the figures below, as well as in the works
of~\cite{brocker}, \cite{brocker-smith},
\cite{corbett-davies-pierson-feller-goel-huq},
\cite{crowson-atkinson-therneau}, \cite{gneiting-balabdaoui-raftery},
\cite{guo-pleiss-sun-weinberger}, \cite{murphy-winkler},
\cite{vaicenavicius-widmann-andersson-lindsten-roll-schoen}, \cite{wilks},
and many others; those works consider applications ranging
from weather forecasting to medical prognosis to fairness in criminal justice
to quantifying the uncertainty in predictions of artificial neural networks.
An approach closely related to reliability diagrams is to smooth over
the binning using kernel density estimation, as discussed by~\cite{brocker},
\cite{wilks}, and others.

A common concern in diagnostics for calibration and reliability
is the selection of widths for bins
or for convolutional kernels in kernel density estimation:
which width is best? Conveniently, this question never arises with the plots
of cumulative differences suggested in the present paper,
as they avoid any binning, kernel density estimation, or other procedures
for smoothing or regularization.
The present paper highlights the utility of cumulative plots,
at least when data is reasonably scarce for the assessment of calibration.

Other works, such as Section~3.2 of~\cite{gneiting-balabdaoui-raftery},
Chapter 8 of~\cite{wilks}, and Figure~1
of~\cite{gupta-rahimi-ajanthan-mensink-sminchisescu-hartley},
also point to the utility of cumulative reliability diagrams and plots
somewhat similar to those in the present paper.
The particular plots proposed below focus on calibration specifically,
encoding miscalibration directly as the slopes of secant lines for the graphs.
Such plots lucidly depict miscalibration
with significant quantitative precision.
Popular graphical methods for assessing calibration appear not to leverage
the key to the approach advocated below, namely that slope is easy to assess
visually even when the constant offset of the graph (or portion of the graph
under consideration) is arbitrary and meaningless.

The following, Section~\ref{methods}, details the construction of the plots
of cumulative differences. Then, Section~\ref{results} presents
several examples of such plots alongside classical reliability diagrams.
Finally, Section~\ref{conclusion} concludes the paper with a brief discussion
of the results and their consequences.

\section{Methods}
\label{methods}

We adopt the notation introduced above,
with $n$ observations $C_1$,~$C_2$, \dots, $C_n$
of the outcomes of Bernoulli trials with corresponding predicted probabilities
$P_1$,~$P_2$, \dots, $P_n$; we want to test the hypothesis~(\ref{null}),
via a graphical display of cumulative differences.
We order the samples (preserving the pairing of $C_k$ with $P_k$ for every $k$)
such that $P_1 \le P_2 \le \dots \le P_n$, ordering any ties at random,
perturbed so that $P_1 < P_2 < \dots < P_n$.

The cumulative function is
\begin{equation}
\label{F}
F(p) = \frac{1}{n} \sum_{P_k \le p} P_k.
\end{equation}

An empirical estimate is
\begin{equation}
\label{E}
E(p) = \frac{1}{n} \sum_{P_k \le p} C_k
     = \frac{\#\{k : P_k \le p \hbox{ and } C_k = 1\}}{n}.
\end{equation}

We will plot the difference between the following sequences:
\begin{equation}
\label{expected}
F_k = F(P_k) = \frac{1}{n} \sum_{j=1}^k P_j
\end{equation}
and
\begin{equation}
\label{empirical}
E_k = E(P_k) = \frac{1}{n} \sum_{j=1}^k C_j
    = \frac{\#\{j : 1 \le j \le k \hbox{ and } C_j = 1\}}{n}
\end{equation}
for $k = 1$,~$2$, \dots, $n$.

Although the accumulation from lower values for $p$ in~(\ref{F}) and~(\ref{E})
might appear to overwhelm the contributions from higher values for $p$,
a plot of $E(p)-F(p)$ as a function of $k$ with $p = P_k$ will reflect
calibration problems for any value of $p$ solely in slopes
that deviate significantly from 0;
problems accumulated from earlier, lower values of $p$ pertain only
to the constant offset from 0, not to the slope deviating from 0.
Indeed, the increment in the expected difference $E_j-F_j$
from $j = k-1$ to $j = k$ is
\begin{equation}
\E[ (E_k-F_k) - (E_{k-1}-F_{k-1}) ] = \frac{\tilde{P_k} - P_k}{n},
\end{equation}
where $\tilde{P_k}$ is the probability that the outcome is a success,
that is, the probability that $C_k = 1$;
thus, on a plot with the values for $k$ spaced $1/n$ apart,
the slope from $j = k-1$ to $j = k$ is
\begin{equation}
\Delta_k = \tilde{P_k} - P_k.
\end{equation}
Miscalibration for the probabilities near $P_k$ occurs
when $\Delta_k$ is significantly nonzero, that is, when the slope
of the plot of $E_k-F_k$ deviates significantly from horizontal
over a significantly long range.

To reiterate: {\it miscalibration over a contiguous range of $P_k$
is the slope of the secant line for the plot of $E_k-F_k$
as a function of $\frac{k}{n}$ over that range,
aside from the expected random fluctuations discussed next}.

The plot of $E_k-F_k$ as a function of $k/n$ automatically
includes some ``error bars'' courtesy of the discrepancy
$E_k-F_k$ fluctuating randomly as the index $k$ increments.
Of course, the standard deviation of a Bernoulli variate
whose expected value is $P_k$ is $\sqrt{P_k (1-P_k)}$ ---
smaller both for $P_k$ near 0 and for $P_k$ near 1.
To indicate the size of the fluctuations, the plots should include
a triangle centered at the origin whose height above the origin is $1/n$
times the standard deviation of the sum of independent Bernoulli variates
with success probabilities $P_1$, $P_2$, \dots, $P_n$;
thus, the height of the triangle above the origin 
(where the triangle itself is centered at the origin) is
$\sqrt{\sum_{k=1}^n P_k (1-P_k)} / n$.
The expected deviation from 0 of $|E_k-F_k|$ (at any specified value for $k$)
is no greater than this height, under the assumption that the samples
$C_1$,~$C_2$, \dots, $C_n$ are draws
from independent Bernoulli distributions with the correct success probabilities
$P_1$, $P_2$, \dots, $P_n$, that is, under the null hypothesis~(\ref{null}).
The triangle is similar to the classic confidence bands
around an empirical cumulative distribution function
given by Kolmogorov and Smirnov, as reviewed by~\cite{doksum}.

In addition to noting the size of the triangle at the origin,
interpreting such plots of the cumulative difference ($E_k-F_k$)
between observed and expected values of ordered Bernoulli variates
does require careful attention to one caveat:
avoid hallucination of minor miscalibrations
where in fact the calibration is good! The sample paths of random walks
and Brownian motion can look surprisingly non-random (drifting?)\ quite often
for short stints. The most trustworthy detections of miscalibration
are long ranges (as a function of $k/n$) of steep slopes for $E(P_k)-F(P_k)$.
The triangles centered at the origins of the plots give a sense
of the length scale for variations that are statistically significant.

For all plots, whether cumulative or classical,
bear in mind that even at 95\% confidence, one in twenty detections
is likely to be false. So, if there are a hundred bins,
each with a 95\% confidence interval, the reality is likely to violate
around 5 of those confidence intervals.
Beware when conducting multiple tests of significance
(or be sure to adjust the confidence level accordingly)!

\section{Results}
\label{results}

Via several numerical examples, we illustrate the methods
of the previous section together with the conventional diagrams
discussed in the introduction.
The figures display the classical calibration plots
as well as both the plots of cumulative differences and the exact expectations
in the absense of noise from random sampling.
To generate the figures, we specify values for $P_1$,~$P_2$, \dots, $P_n$
and for $\tilde{P}_1$,~$\tilde{P}_2$, \dots, $\tilde{P}_n$
differing from $P_1$,~$P_2$, \dots, $P_n$,
then independently draw $C_1$,~$C_2$, \dots, $C_n$
from the Bernoulli distributions with parameters
$\tilde{P}_1$,~$\tilde{P}_2$, \dots, $\tilde{P}_n$, respectively.
Ideally the plots would show how and where 
$\tilde{P}_1$,~$\tilde{P}_2$, \dots, $\tilde{P}_n$ differs
from $P_1$,~$P_2$, \dots, $P_n$.
The appendix considers the case in which $\tilde{P}_k = P_k$
for all $k = 1$,~$2$, \dots, $n$.

The top rows of the figures plot $E_k-F_k$
from~(\ref{expected}) and~(\ref{empirical}) as a function of $k/n$,
with the rightmost plot displaying its noiseless expected value
rather than using the samples $C_1$,~$C_2$, \dots, $C_n$.
In each of these plots,
the upper axis specifies $k/n$, while the lower axis specifies $P_k$
for the corresponding value of $k$.
The middle two rows of the figures plot the pairs
$(A_1, B_1)$,~$(A_2, B_2)$, \dots, $(A_m, B_m)$
from~(\ref{averageP}) and~(\ref{averageC}),
with the rightmost plots using an equal number of samples per bin.
The left and right plots in the middle rows of Figures~\ref{10000}--\ref{100}
are in fact identical, since $P_1$,~$P_2$, \dots, $P_n$ are equispaced
for those examples (so equally wide bins contain equal numbers
of samples). The bottom rows of the figures again plot pairs
$(A_1, B_1)$,~$(A_2, B_2)$, \dots, $(A_n, B_n)$
from~(\ref{averageP}) and~(\ref{averageC}),
but this time using their noiseless expected values
instead of the samples $C_1$,~$C_2$, \dots, $C_n$.

Perhaps the simplest, most straightforward method to gauge uncertainty
in the binned plots is to vary the number of bins
and observe how the plotted values vary.
All figures displayed employ this method, with the number of bins increased in
the second rows of plots beyond the number of bins in the third rows of plots.
The figures also include the ``error bars'' resulting
from one of the bootstrap resampling schemes proposed by~\cite{brocker-smith},
obtained by drawing $n$ samples independently and uniformly at random
with replacement from $(P_1, C_1)$, $(P_2, C_2)$, \dots, $(P_n, C_n)$
and then plotting (in light gray) the corresponding reliability diagram,
and repeating for a total of 20 times (thus displaying 20 gray lines per plot).
The chance that all 20 lines are unrepresentative
of the expected statistical variations would be roughly $1/20 = 5$\%,
so plotting these 20 lines corresponds to approximately 95\% confidence.
An alternative is to display the bin frequencies as suggested,
for example, by~\cite{murphy-winkler}.
Other possibilities often involve kernel density estimation,
as suggested, for example, by~\cite{brocker} and~\cite{wilks}.
All such methods require selecting widths for the bins or kernel smoothing;
avoiding having to make what is a necessarily somewhat arbitrary choice
is possible by varying the widths, as done in the plots of the present paper.
Chapter~8 of~\cite{wilks} comprehensively reviews the extant literature.

We may set the widths of the bins such that either
(1) the average of $P_k$ for $k$ in each bin is approximately equidistant
from the average of $P_k$ for $k$ in each neighboring bin or
(2) the range of $k$ for every bin has the same width.
Both options are natural; the first is the canonical choice,
whereas the second ensures that error bars would be roughly the same size
for every bin. The figures display both possibilities,
with the first on the left and the second on the right.
Setting the number of bins together with either of these choices
fully specifies the bins. As discussed earlier, we vary the number of bins
since there is no perfect setting --- using fewer bins offers estimates
with higher confidence yet limits the resolution for detecting miscalibration
and for assessing the dependence of calibration as a function of $P_k$.

Figures~\ref{10000}--\ref{100} all draw from the same underlying distribution
that deviates linearly as a function of $k$ from the distribution of $P_k$,
and $P_1$, $P_2$, \dots, $P_n$ are equispaced;
Figure~\ref{10000} sets $n =$ 10,000, Figure~\ref{1000} sets $n =$ 1,000,
and Figure~\ref{100} sets $n =$ 100.
All plots, whether cumulative or conventional, appear to work well
in Figures~\ref{10000} and~\ref{1000}. However, the conventional plots
become increasingly problematic as $n$ becomes 100 in Figure~\ref{100},
whereas the cumulative plot still detects roughly the right level
of miscalibration for $0 \lesssim P_k \lesssim 0.2$
and $0.8 \lesssim P_k \lesssim 1$; the cumulative plot
indicates that too little data is available for $0.2 \lesssim P_k \lesssim 0.8$
to detect any statistically significant miscalibration in that range of $P_k$.
Overall, the cumulative plots seem more informative
(or at least easier to interpret) in Figures~\ref{10000}--\ref{100},
but only mildly.

Figures~\ref{10000_0}--\ref{100_0} all draw
from the same underlying distribution that is overconfident
(lying above the perfectly calibrated ideal),
with the overconfidence peaking for $P_k$ around $0.25$
(aside from a perfectly calibrated notch right around $0.25$),
where $P_k$ is proportional to $(k-0.5)^2$;
Figure~\ref{10000_0} sets $n =$ 10,000, Figure~\ref{1000_0} sets $n =$ 1,000,
and Figure~\ref{100_0} sets $n =$ 100.
All plots, whether cumulative or conventional, work well enough
in Figures~\ref{10000_0} and~\ref{1000_0}, though the reliability diagrams
might be mistakingly misleading relative to the exact expectations,
at least without diligent attention to the significant variation
with the number of bins.
The plots, whether cumulative or conventional, reveal similar information
in Figure~\ref{100_0}, too, though the reliability diagram
with an equal number of samples per bin provides more reliable estimates
than the other reliability diagram.
The cumulative plot is perhaps the easiest to interpret:
the miscalibration is significant for $0.1 \lesssim P_k \lesssim 0.23$
and $0.27 \lesssim P_k \lesssim 0.6$,
with about the correct amount of miscalibration
(the amount is correct since the secant lines have the expected slopes).

Figures~\ref{10000_1}--\ref{100_1} all draw
from the same, relatively complicated underlying distribution,
with $P_k$ being proportional to $\sqrt{k-0.5}$;
Figure~\ref{10000_1} sets $n =$ 10,000, Figure~\ref{1000_1} sets $n =$ 1,000,
and Figure~\ref{100_1} sets $n =$ 100.
The cumulative plots capture more of the oscillations in the miscalibration,
as do to some extent the reliability diagrams with an equal number
of samples per bin; however, the variations in the reliability diagrams
could be difficult to interpret without access
to the ground-truth exact expectations.

The following section concludes the discussion of these results
and their implications.

\newlength{\vertsep}
\setlength{\vertsep}{.25in}
\newlength{\imsize}
\setlength{\imsize}{.41\textwidth}

\begin{figure}
\begin{centering}

\parbox{\imsize}{\includegraphics[width=\imsize]
                 {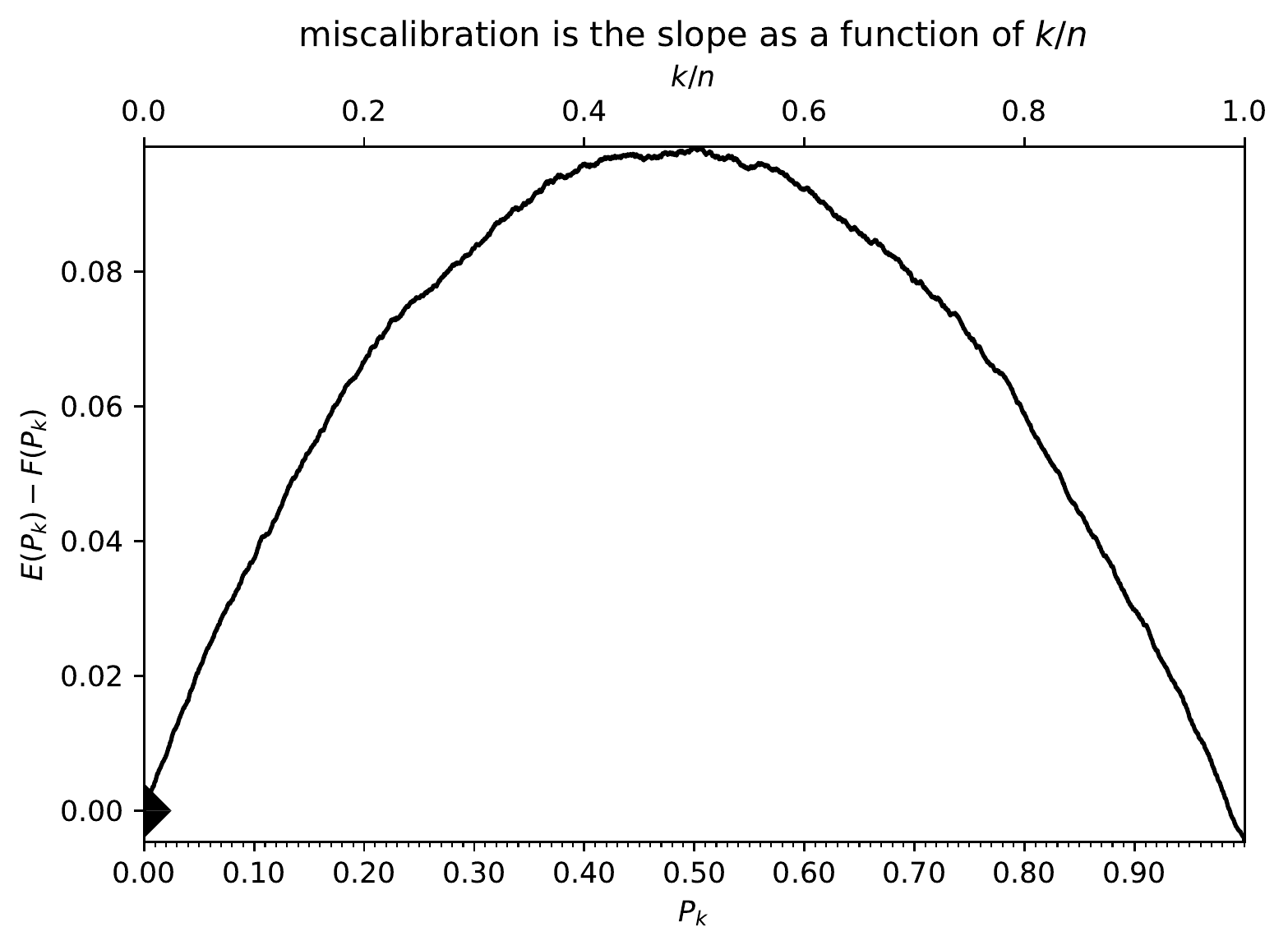}}
\quad\quad
\parbox{\imsize}{\includegraphics[width=\imsize]
                 {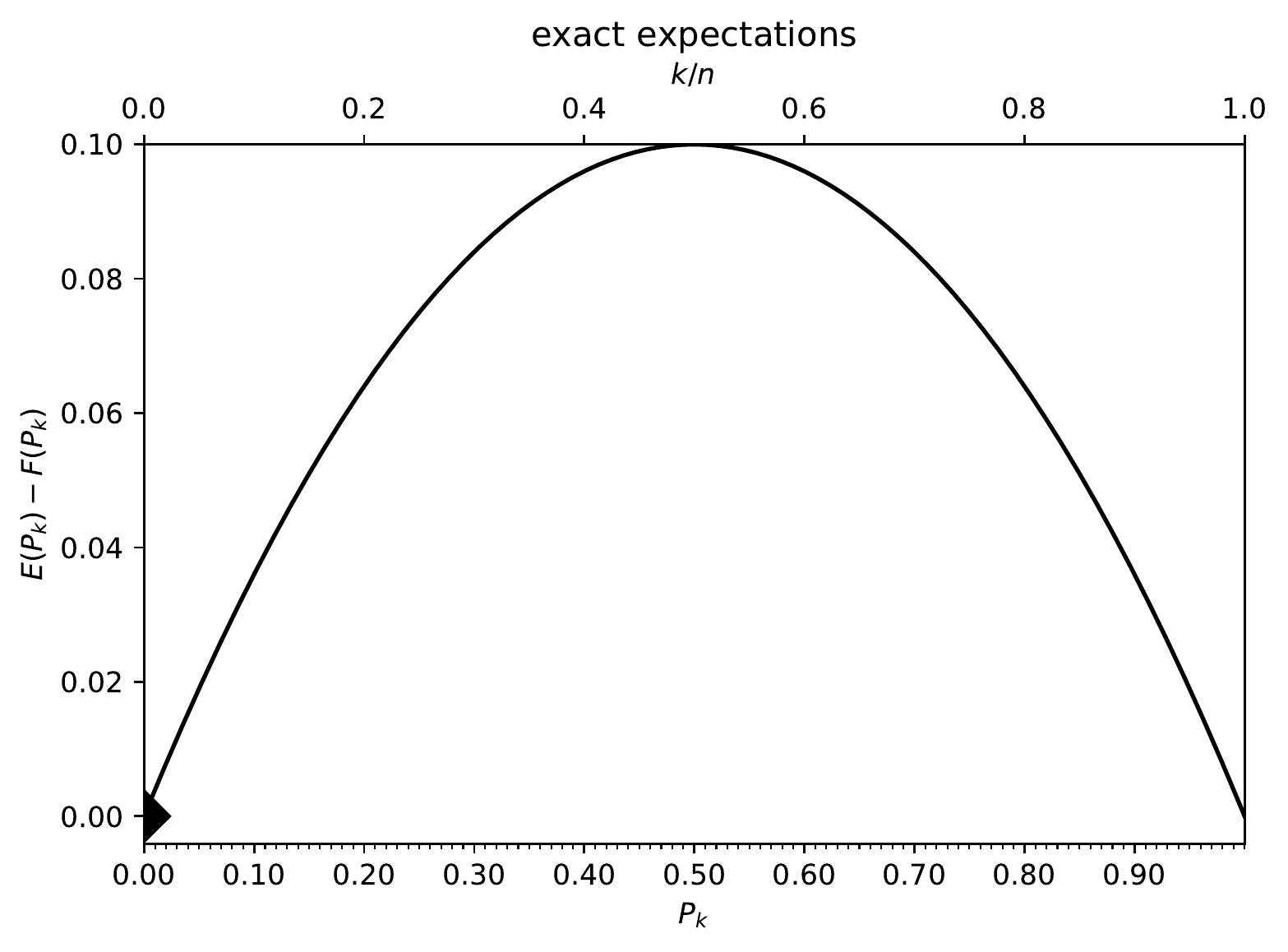}}

\vspace{\vertsep}

\parbox{\imsize}{\includegraphics[width=\imsize]
                 {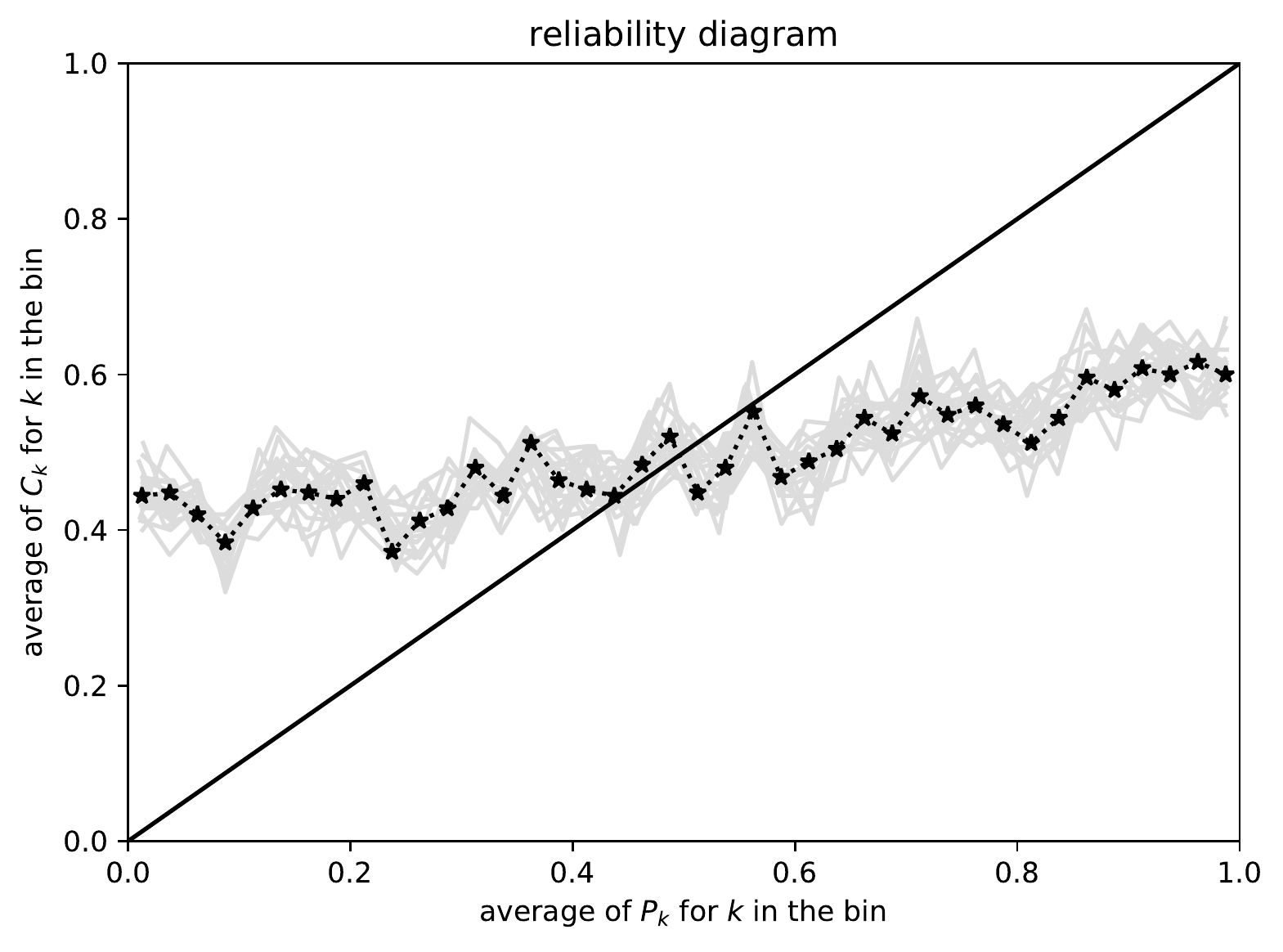}}
\quad\quad
\parbox{\imsize}{\includegraphics[width=\imsize]
                 {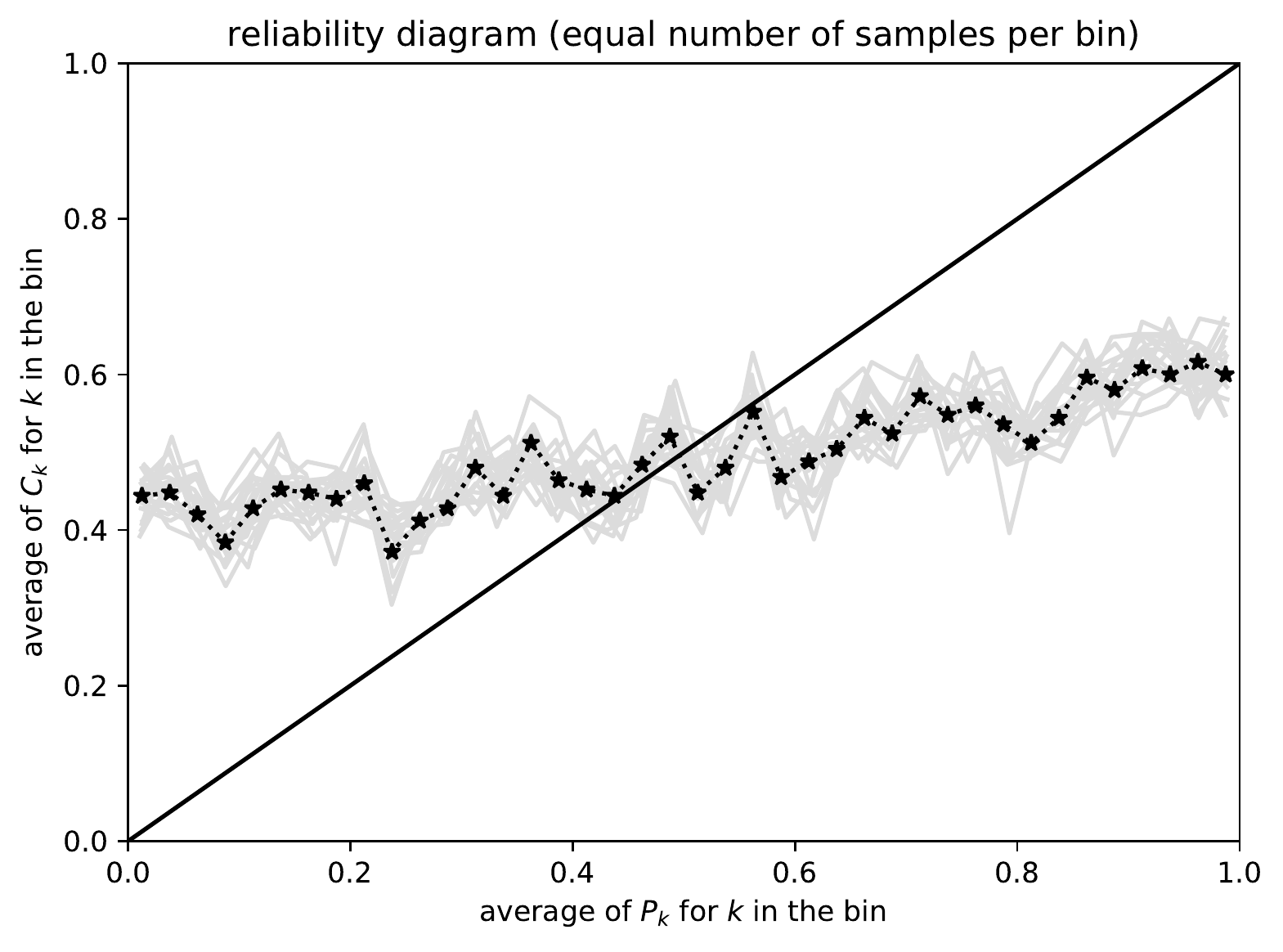}}

\vspace{\vertsep}

\parbox{\imsize}{\includegraphics[width=\imsize]
                 {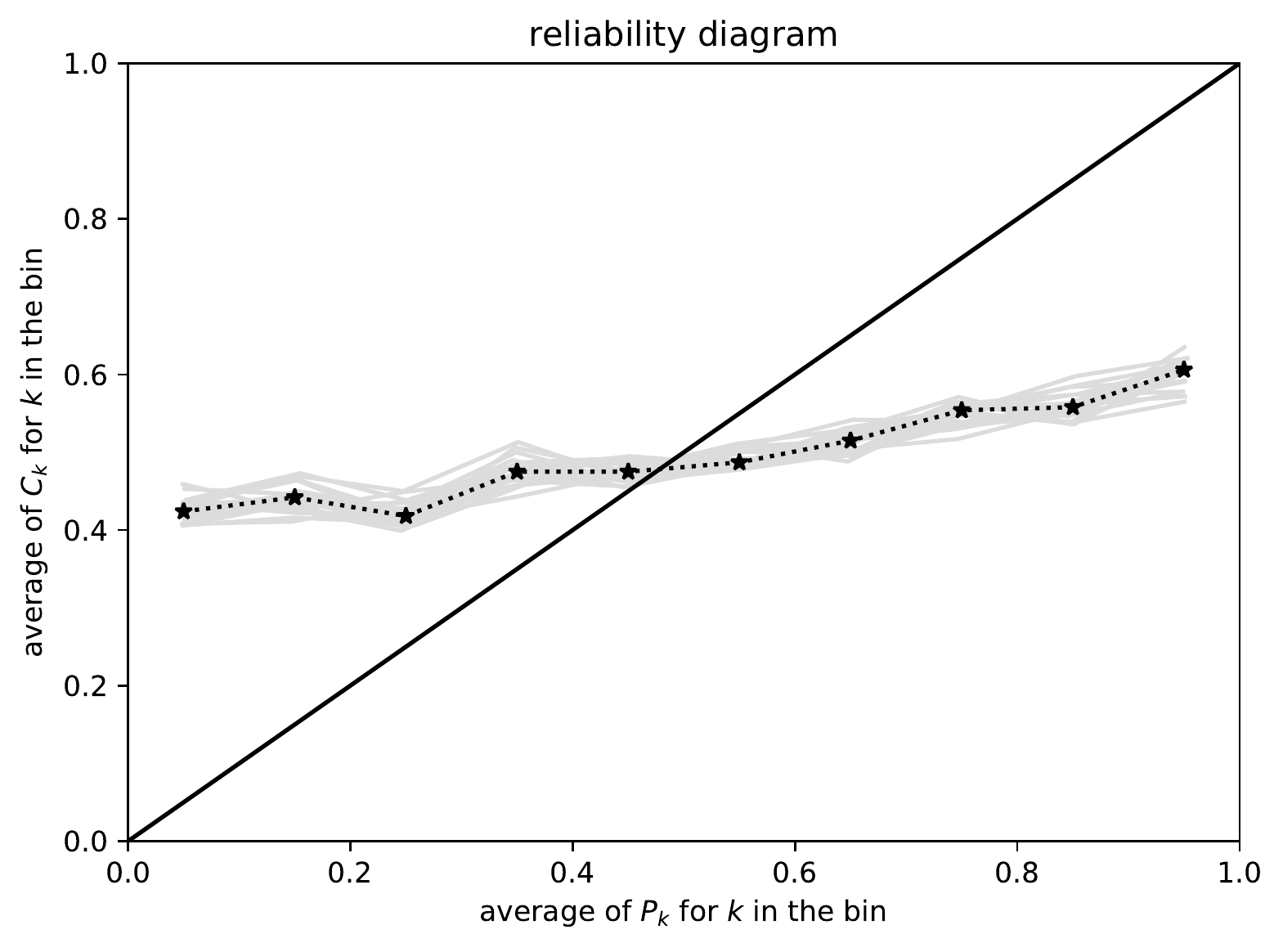}}
\quad\quad
\parbox{\imsize}{\includegraphics[width=\imsize]
                 {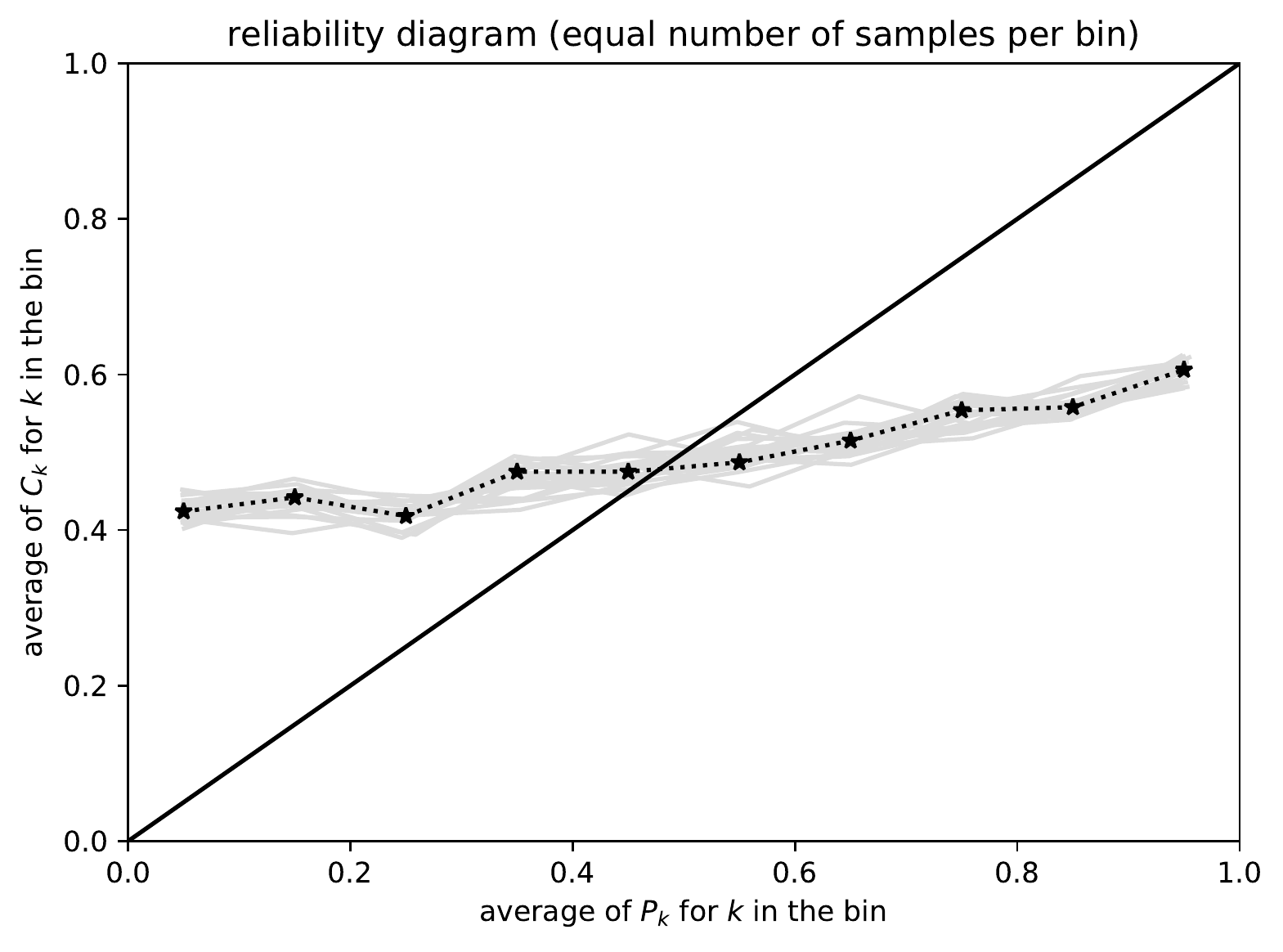}}

\vspace{\vertsep}

\parbox{\imsize}{\includegraphics[width=\imsize]
                 {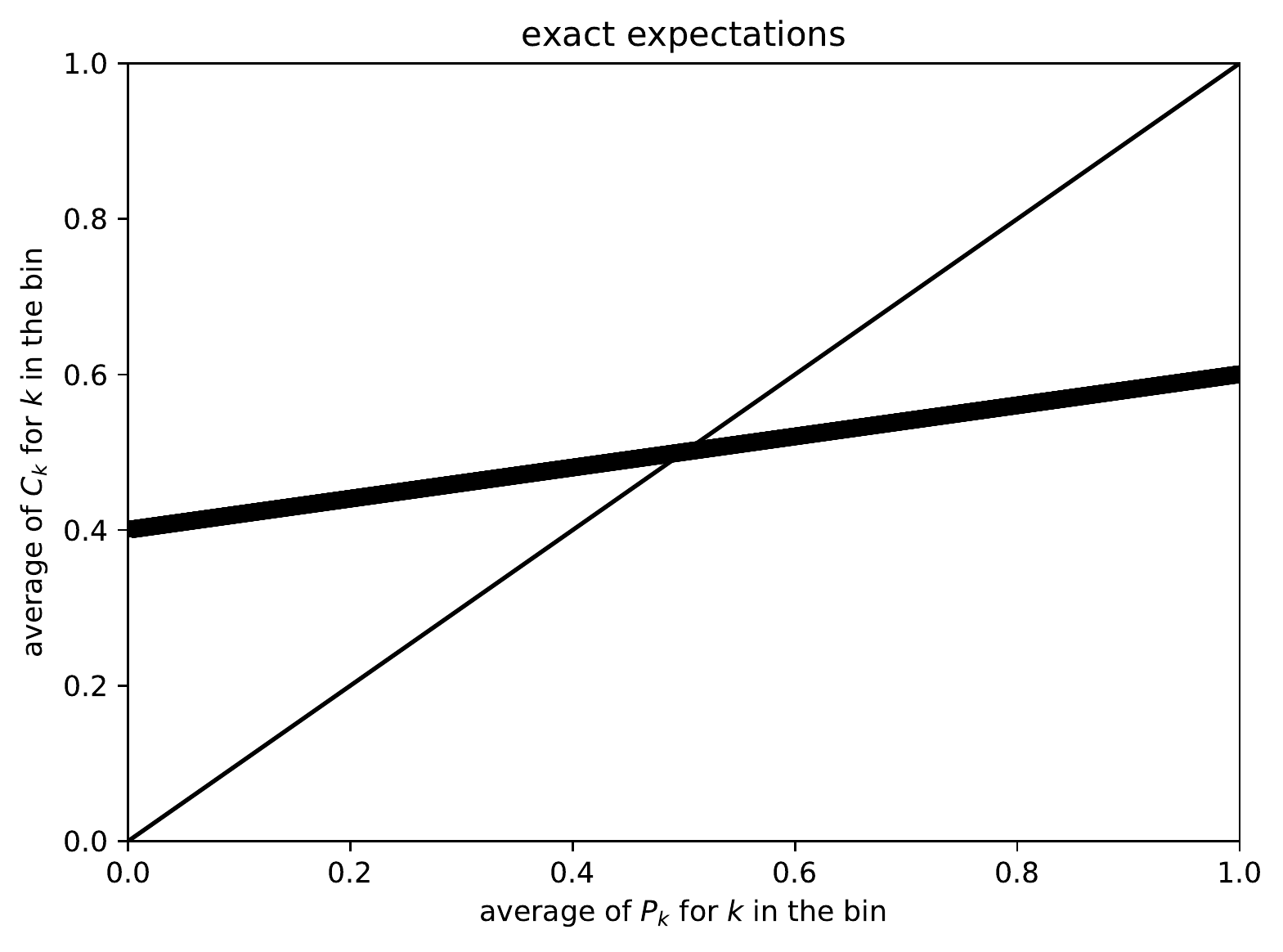}}

\end{centering}
\caption{$n =$ 10,000; $P_1$, $P_2$, \dots, $P_n$ are equispaced}
\label{10000}
\end{figure}

\begin{figure}
\begin{centering}

\parbox{\imsize}{\includegraphics[width=\imsize]
                 {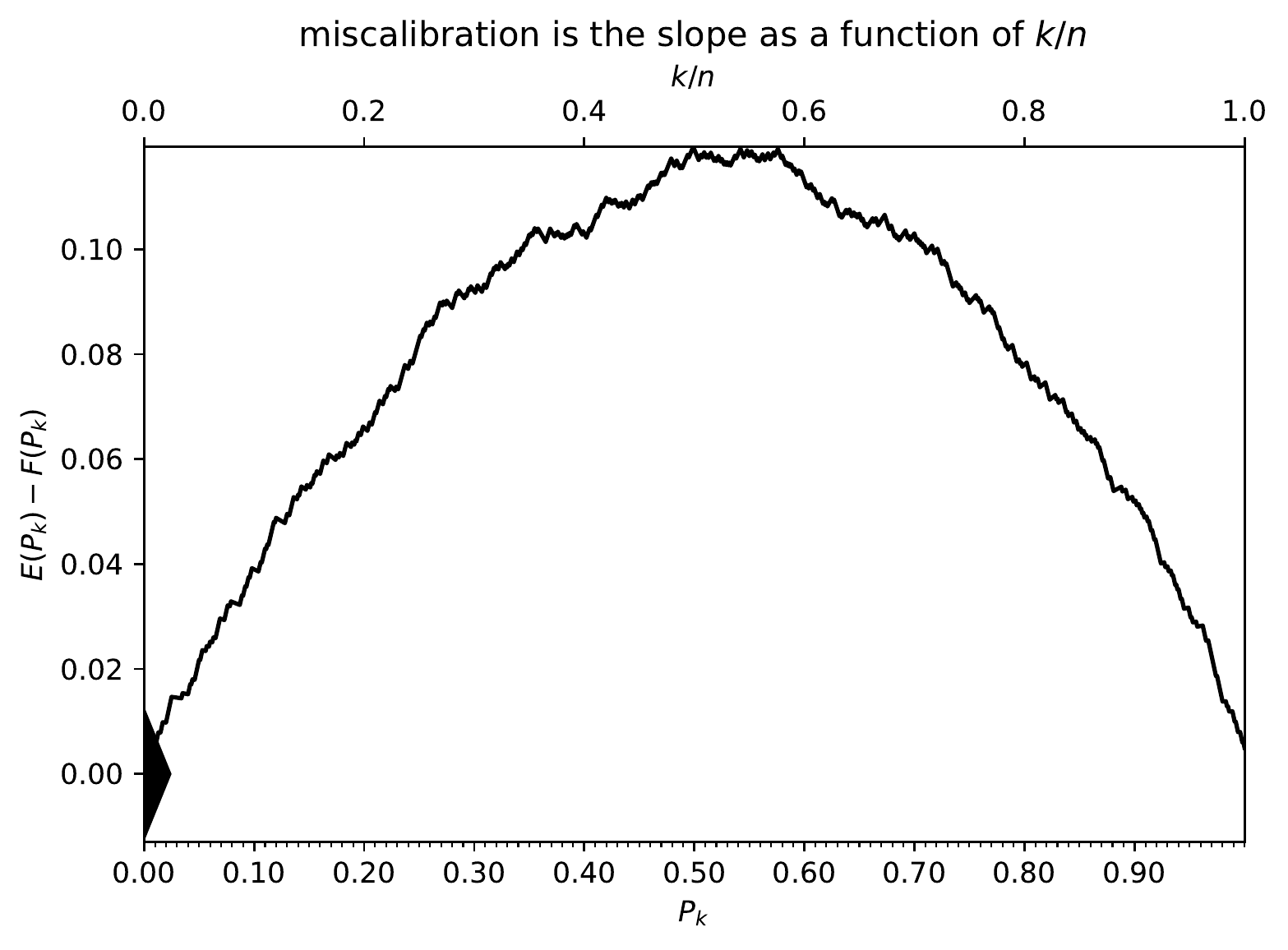}}
\quad\quad
\parbox{\imsize}{\includegraphics[width=\imsize]
                 {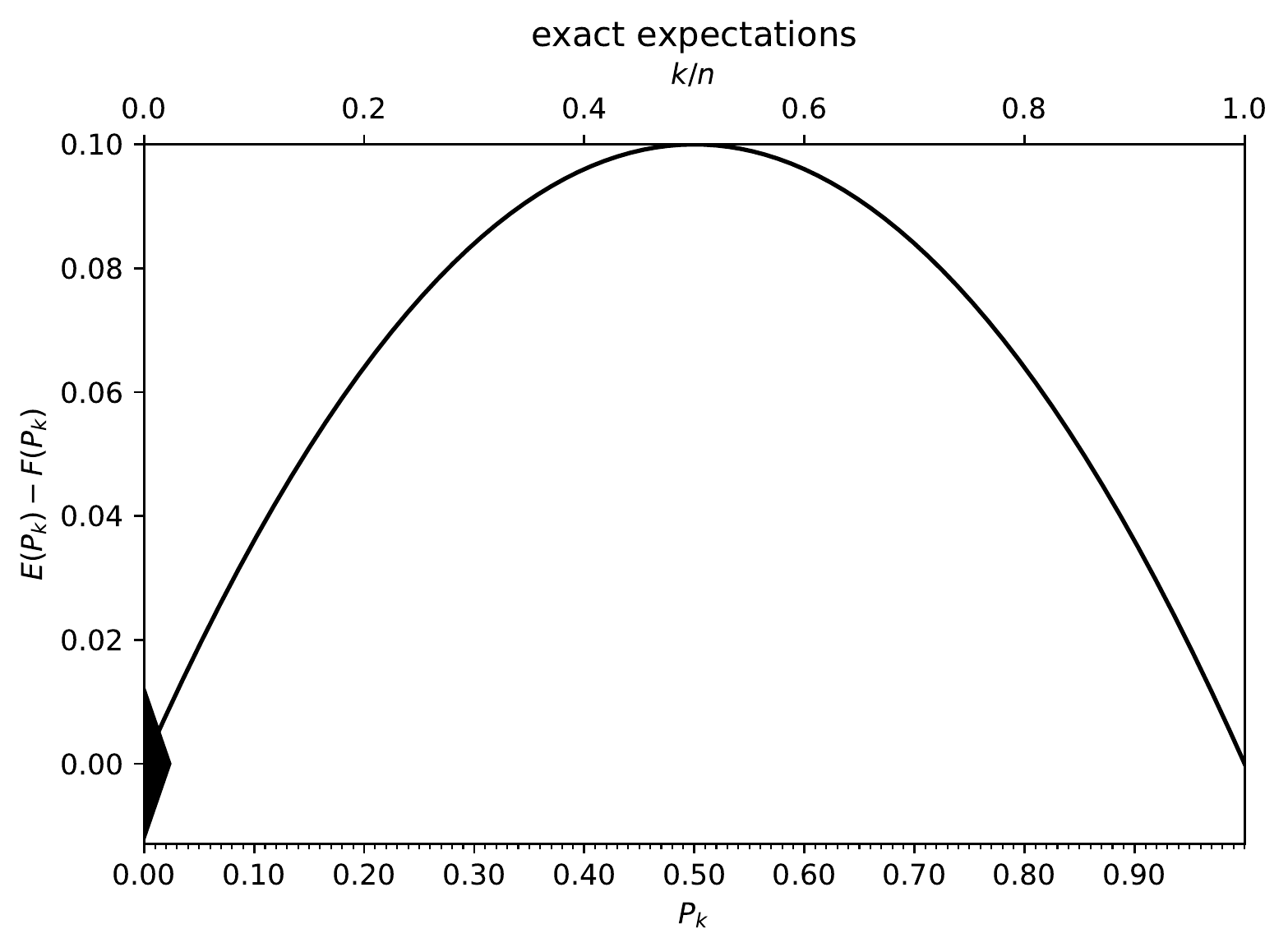}}

\vspace{\vertsep}

\parbox{\imsize}{\includegraphics[width=\imsize]
                 {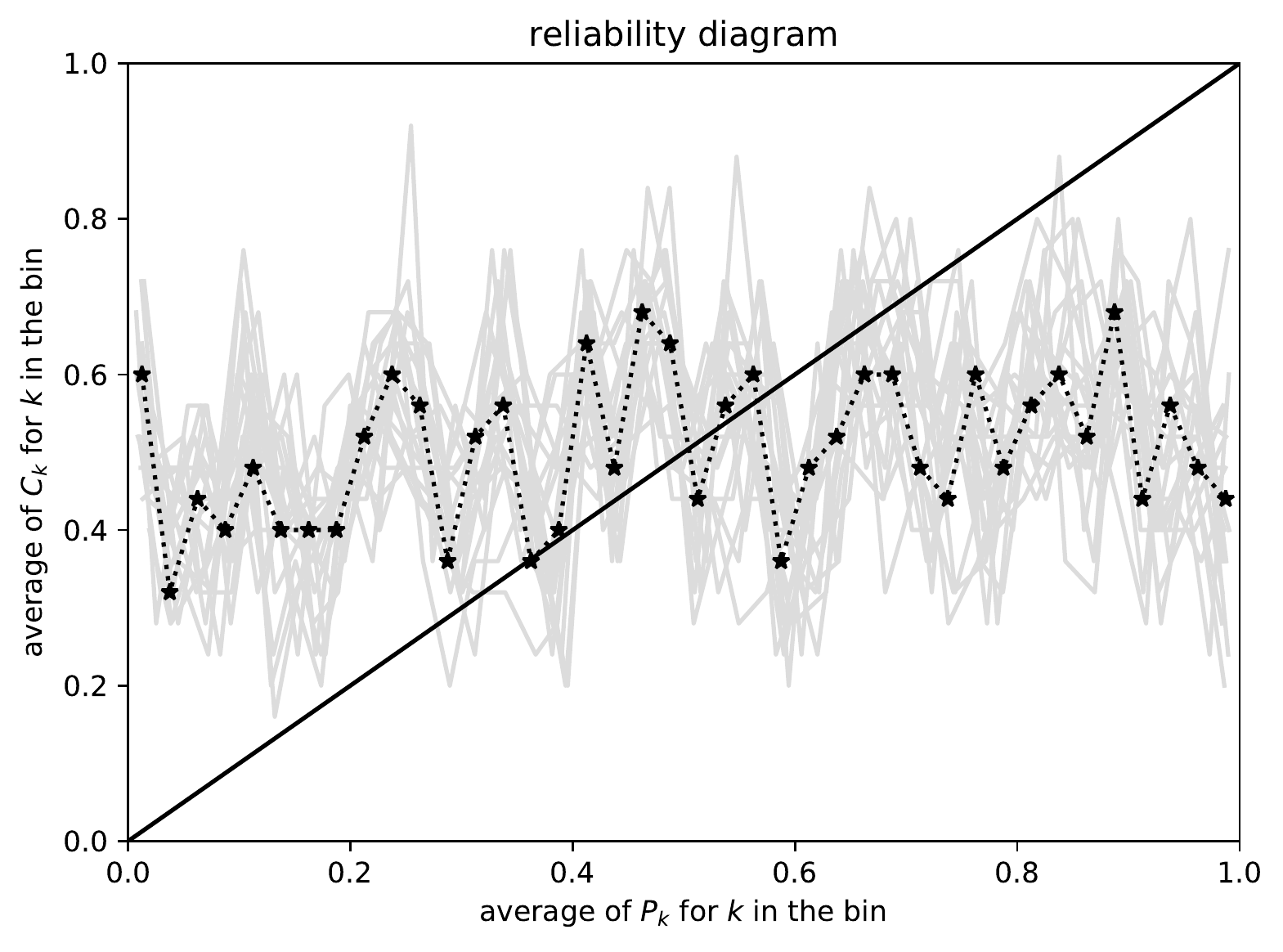}}
\quad\quad
\parbox{\imsize}{\includegraphics[width=\imsize]
                 {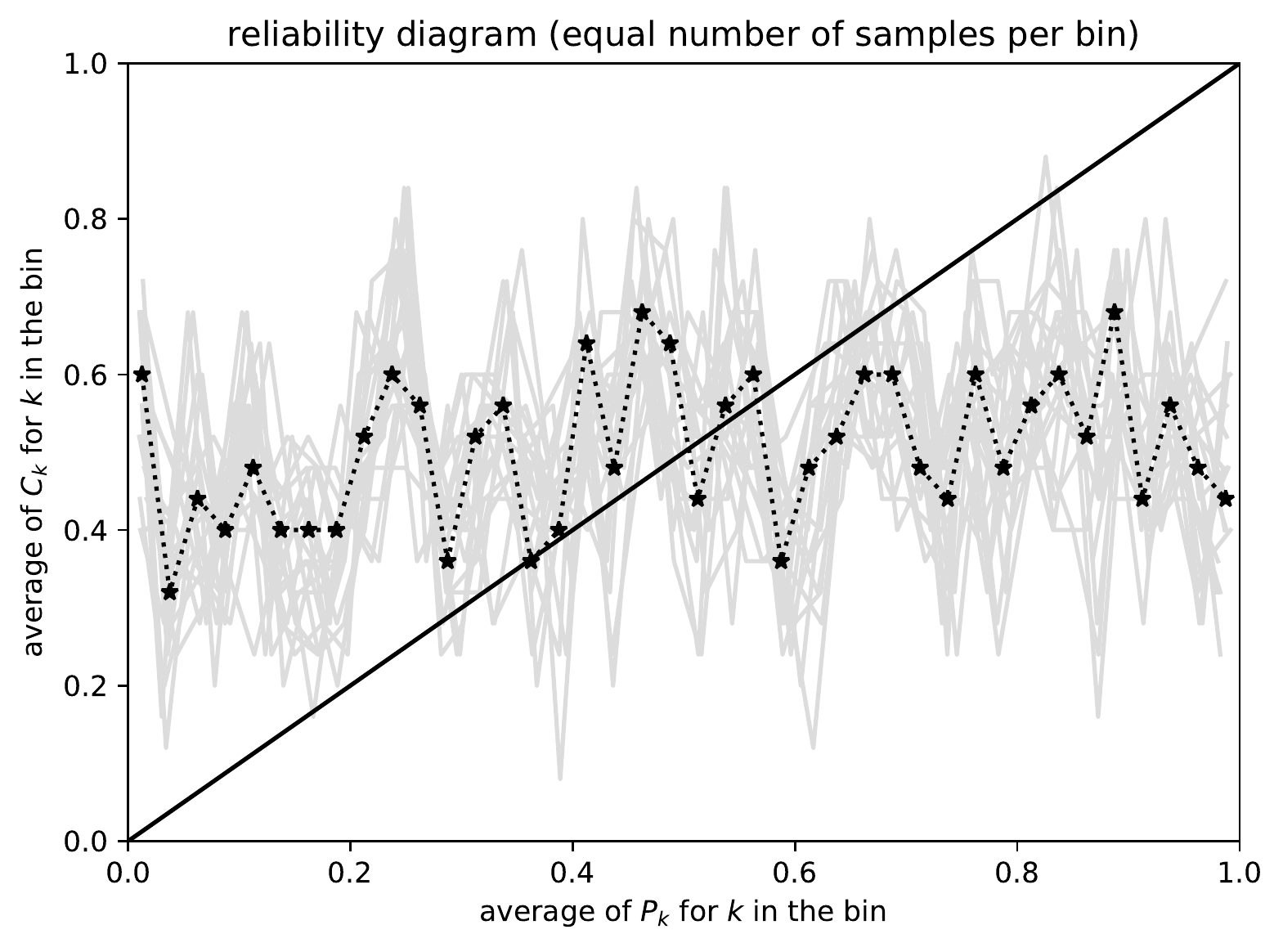}}

\vspace{\vertsep}

\parbox{\imsize}{\includegraphics[width=\imsize]
                 {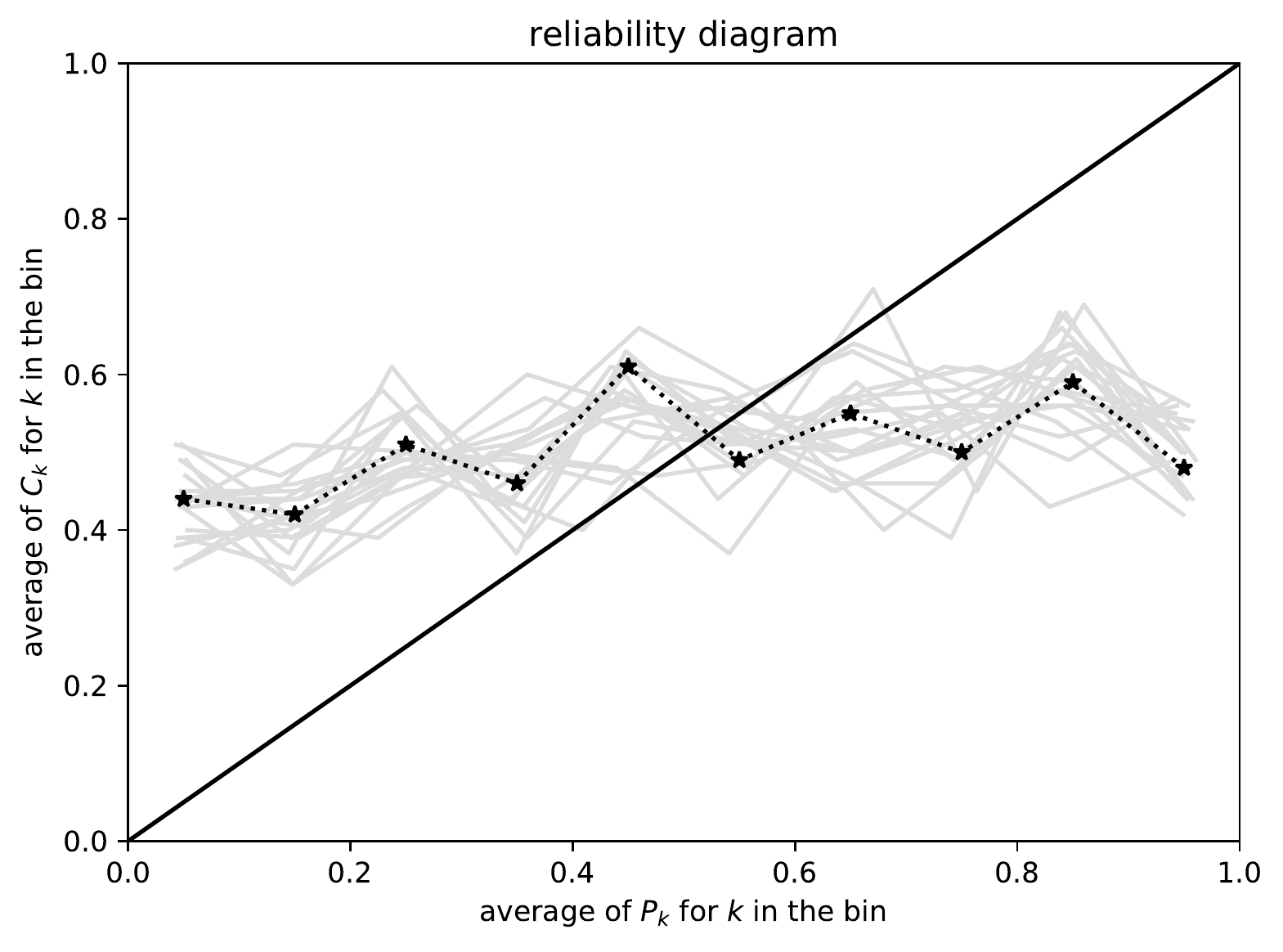}}
\quad\quad
\parbox{\imsize}{\includegraphics[width=\imsize]
                 {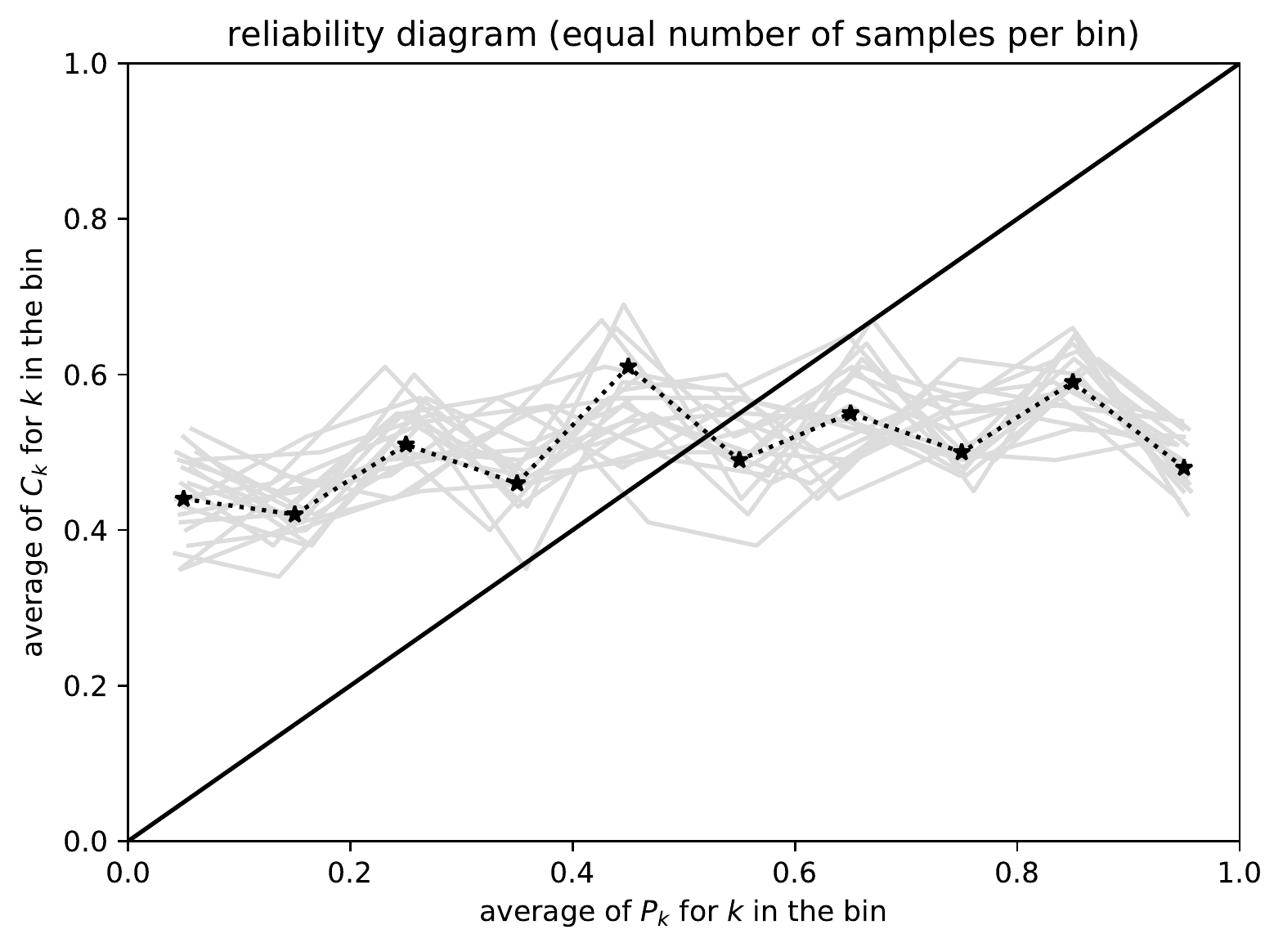}}

\vspace{\vertsep}

\parbox{\imsize}{\includegraphics[width=\imsize]
                 {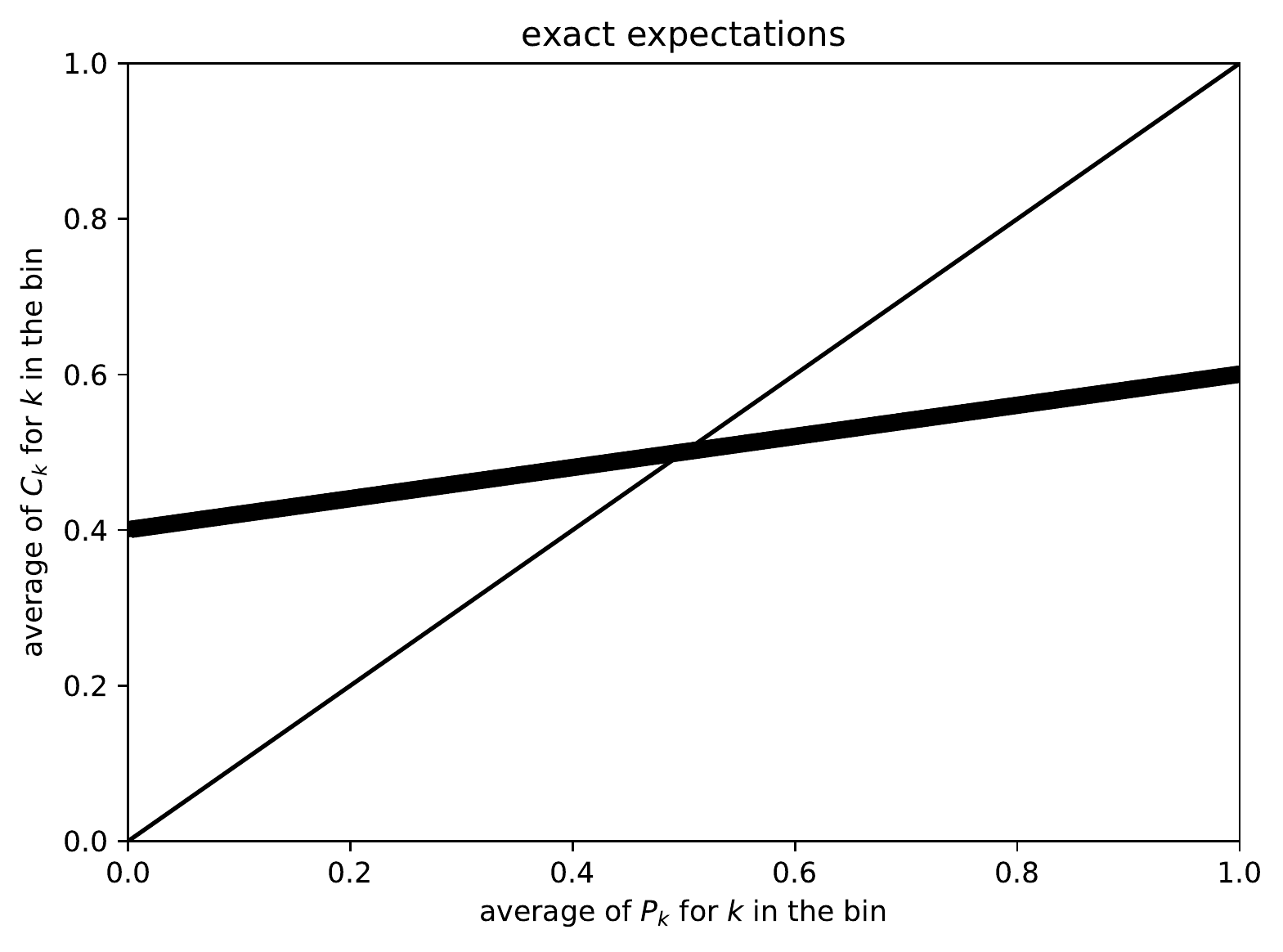}}

\end{centering}
\caption{$n =$ 1,000; $P_1$, $P_2$, \dots, $P_n$ are equispaced}
\label{1000}
\end{figure}

\begin{figure}
\begin{centering}

\parbox{\imsize}{\includegraphics[width=\imsize]
                 {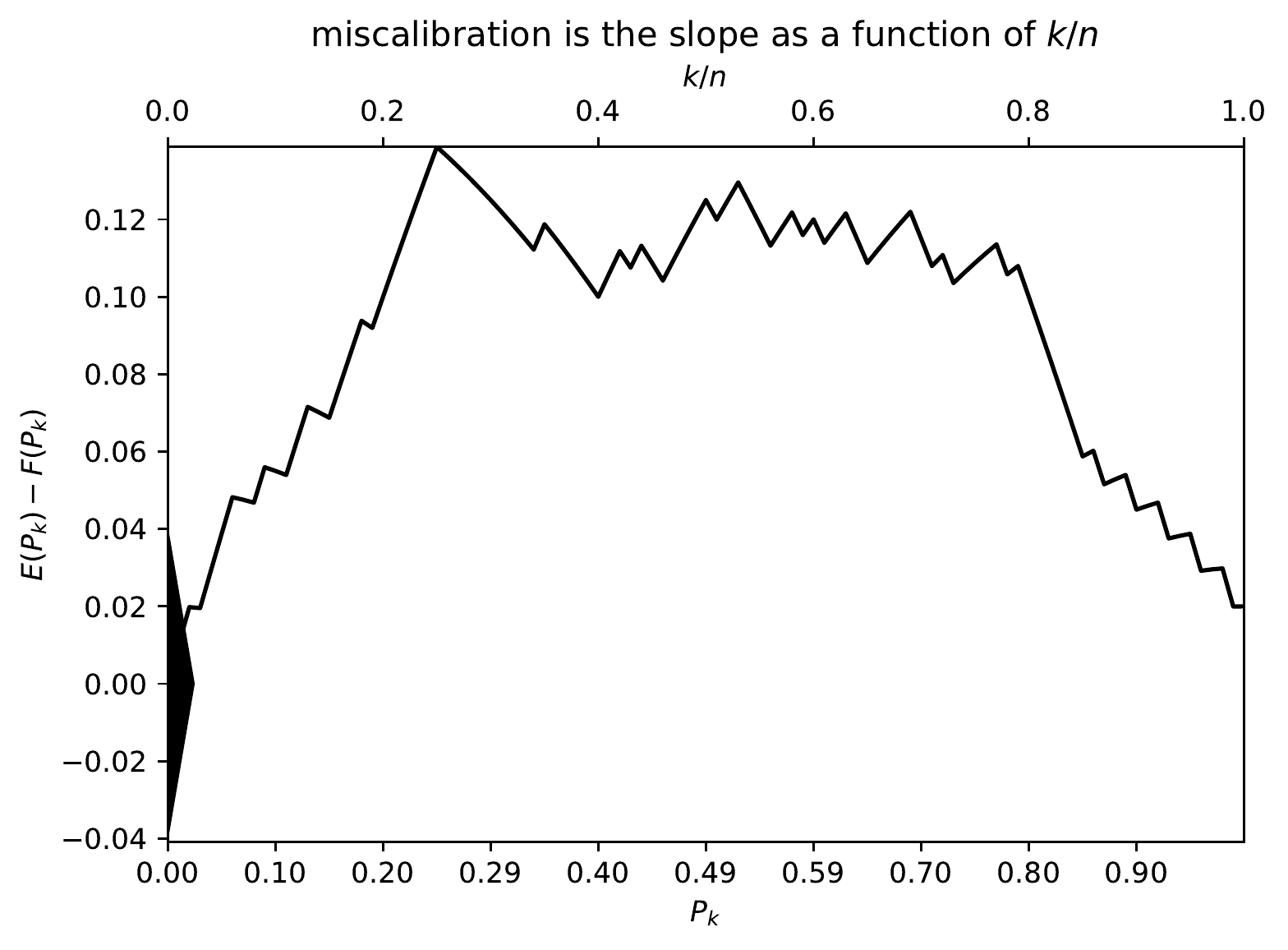}}
\quad\quad
\parbox{\imsize}{\includegraphics[width=\imsize]
                 {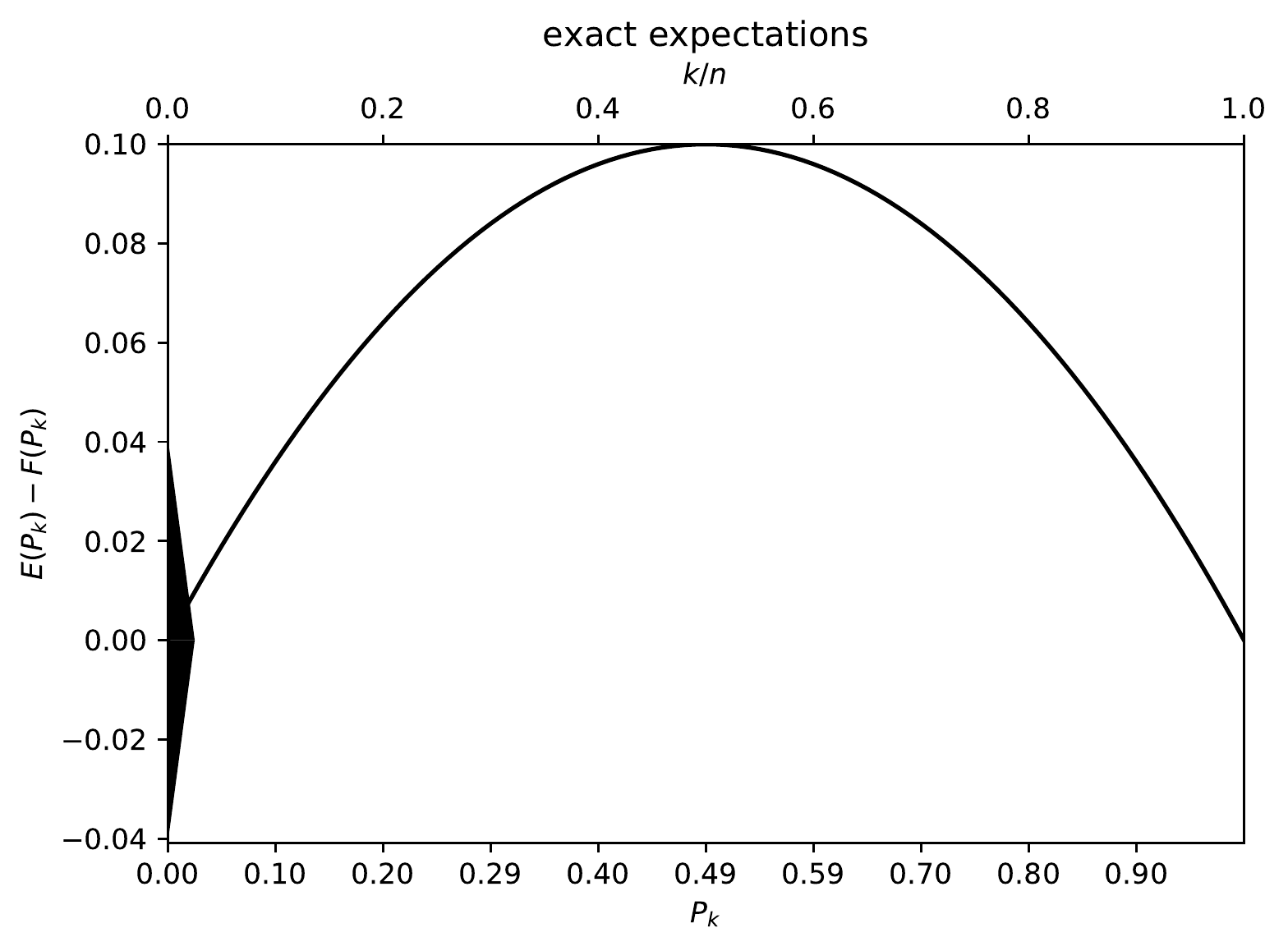}}

\vspace{\vertsep}

\parbox{\imsize}{\includegraphics[width=\imsize]
                 {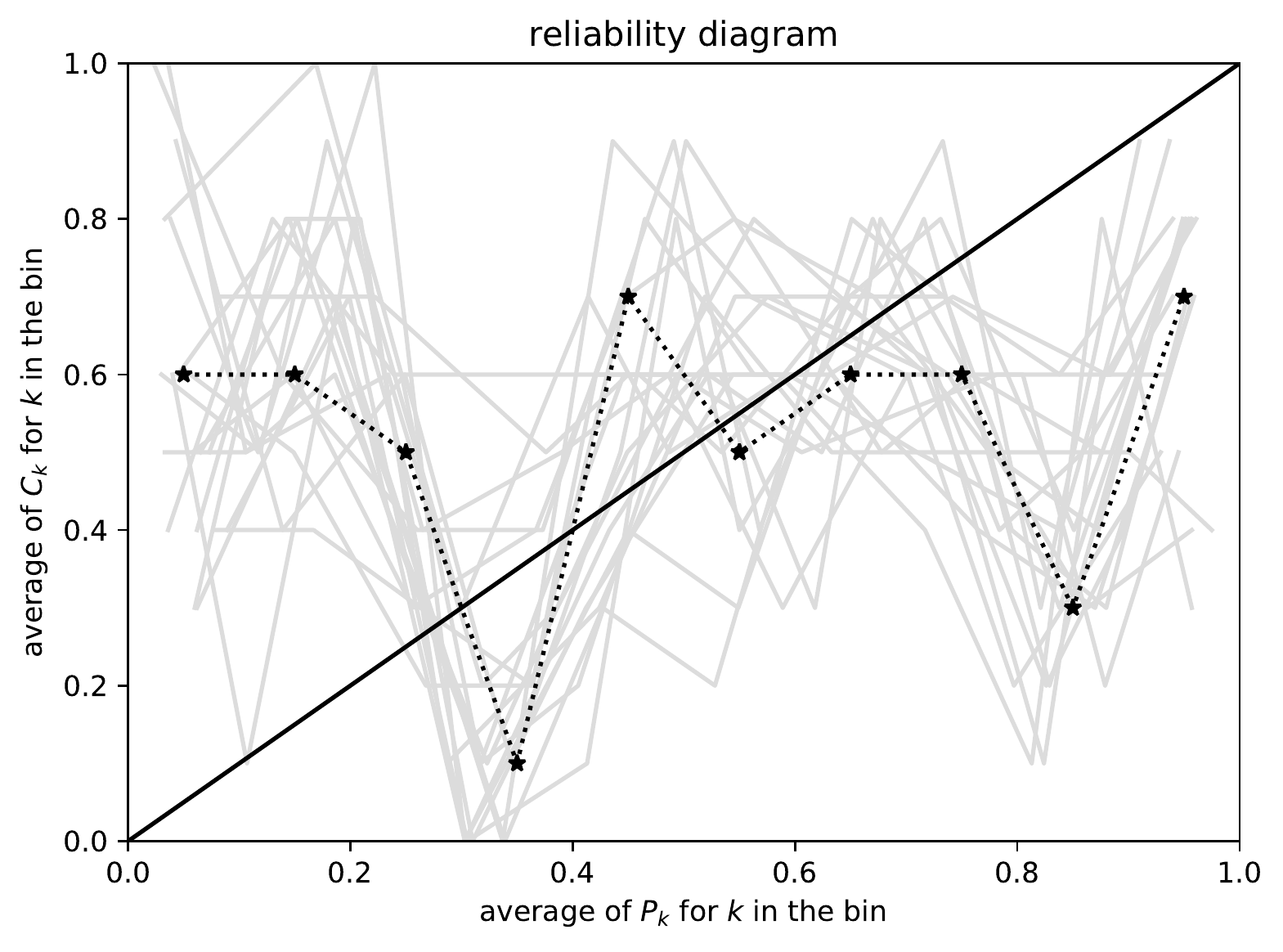}}
\quad\quad
\parbox{\imsize}{\includegraphics[width=\imsize]
                 {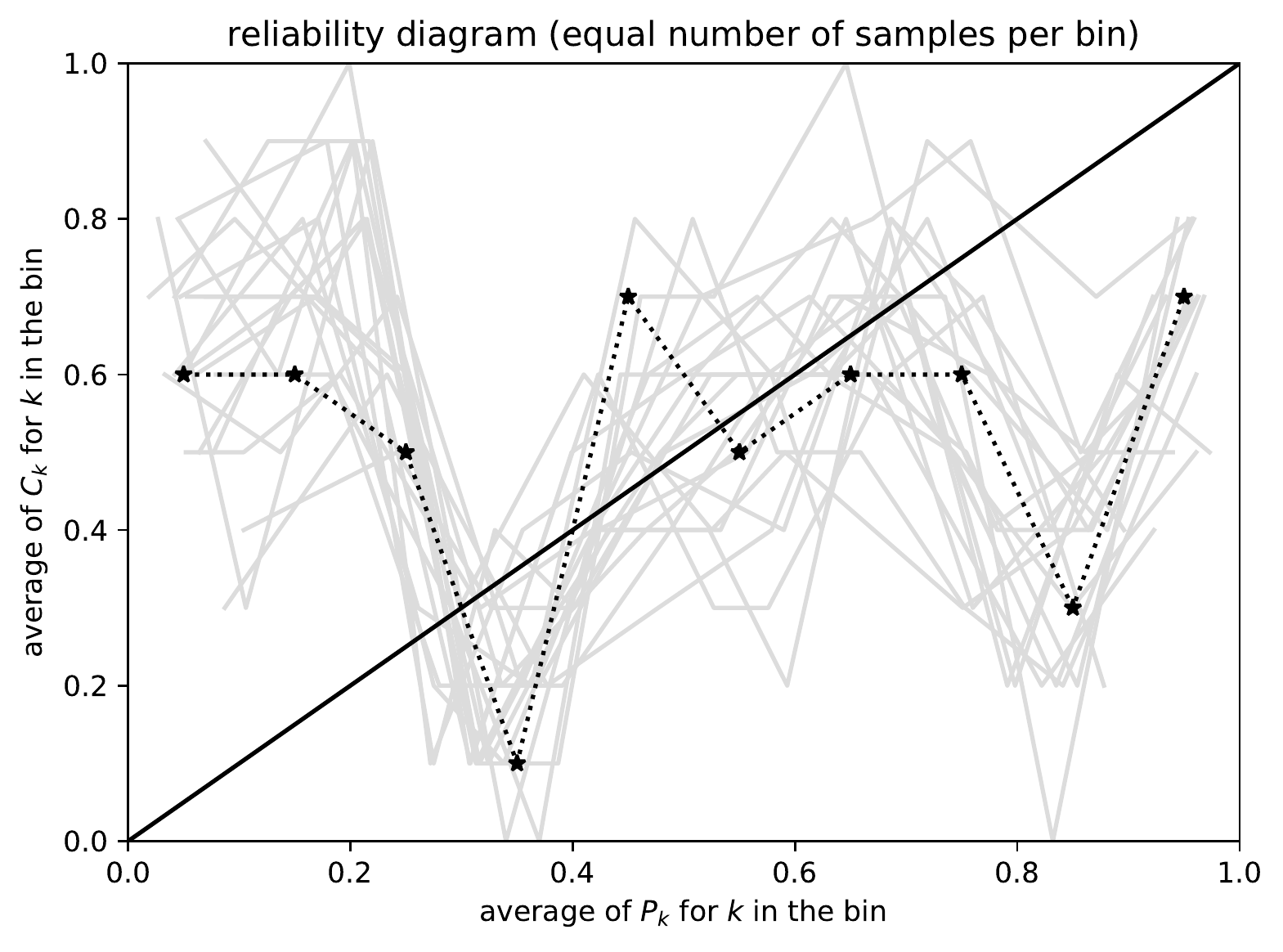}}

\vspace{\vertsep}

\parbox{\imsize}{\includegraphics[width=\imsize]
                 {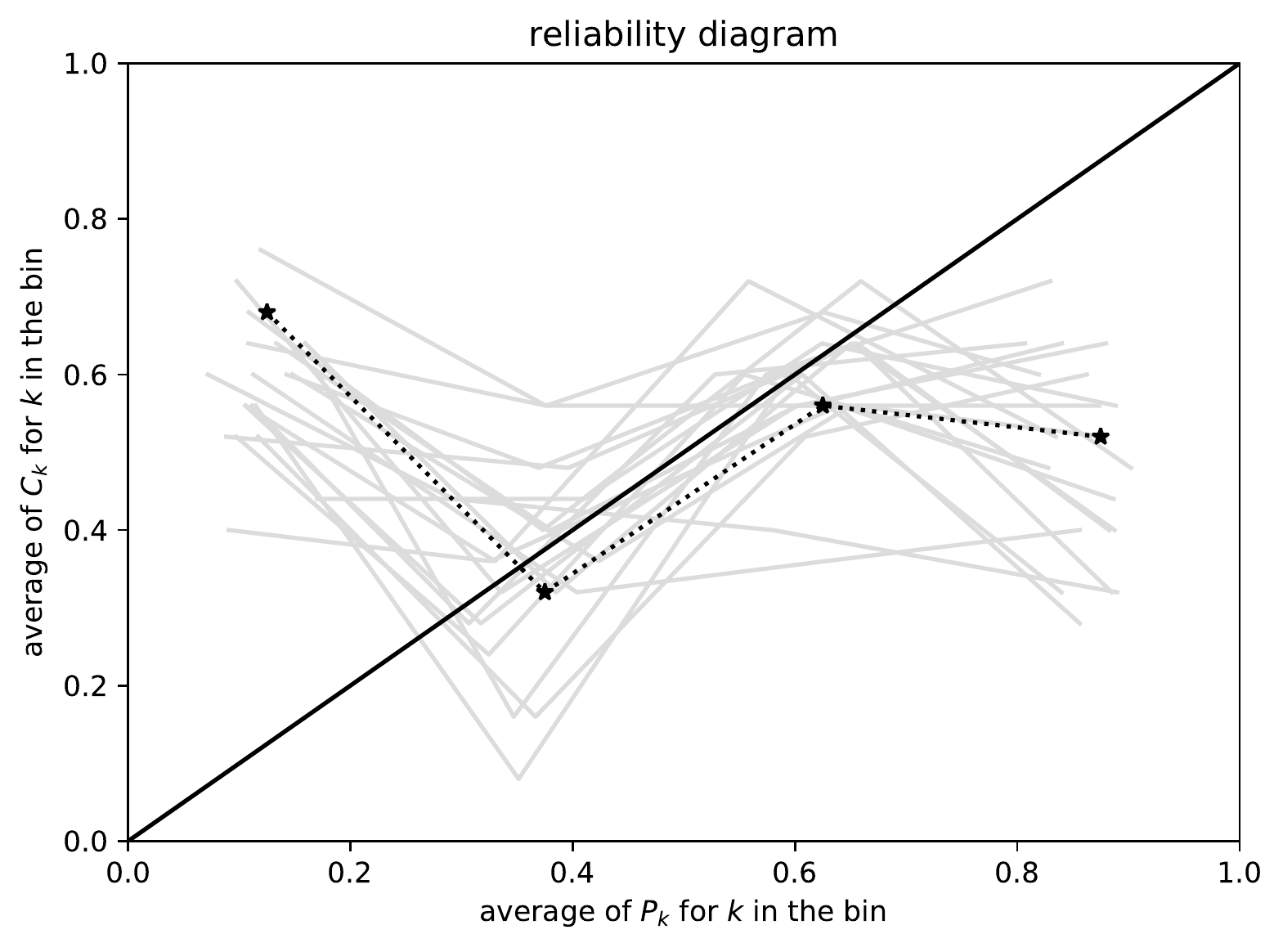}}
\quad\quad
\parbox{\imsize}{\includegraphics[width=\imsize]
                 {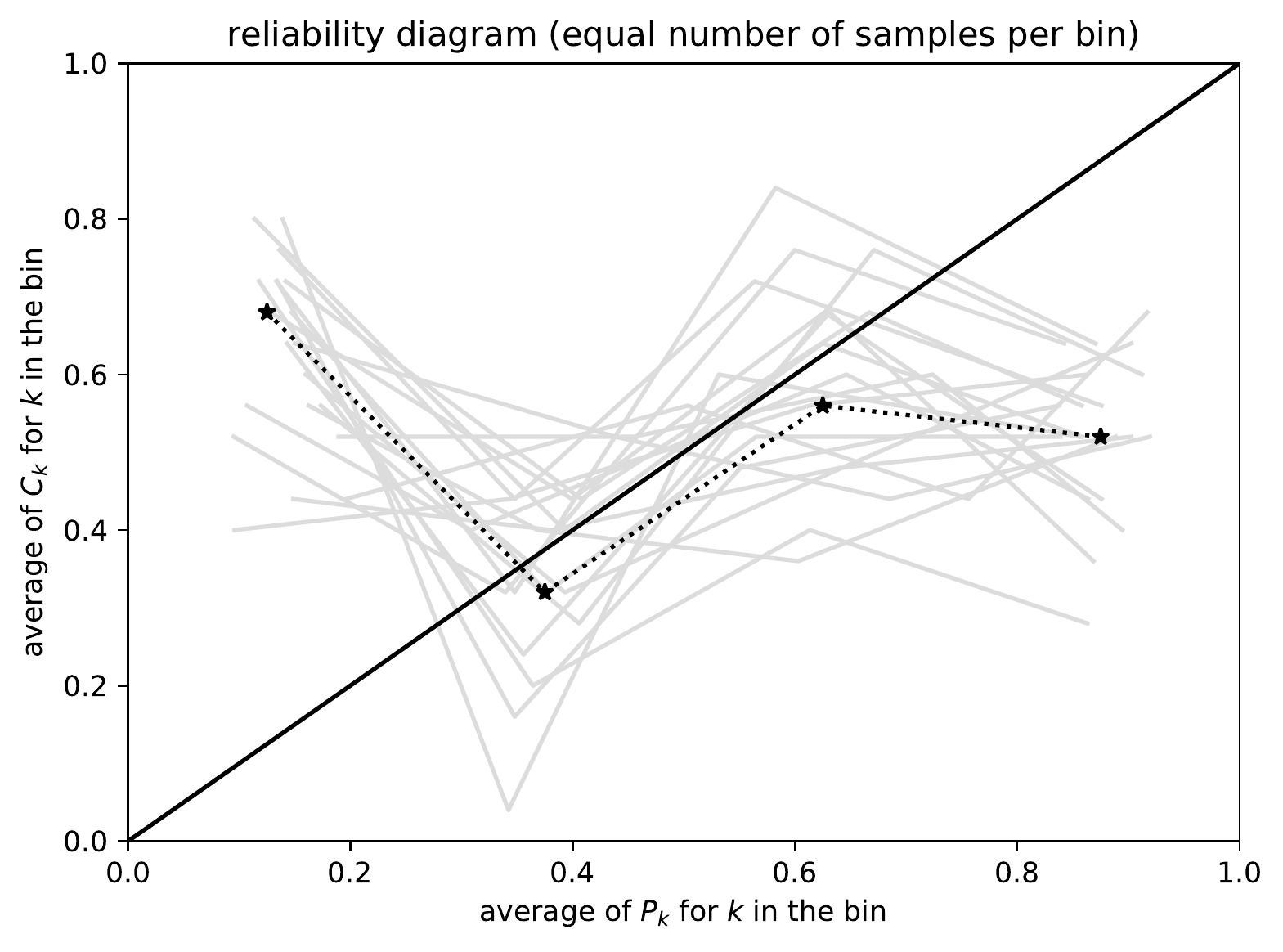}}

\vspace{\vertsep}

\parbox{\imsize}{\includegraphics[width=\imsize]
                 {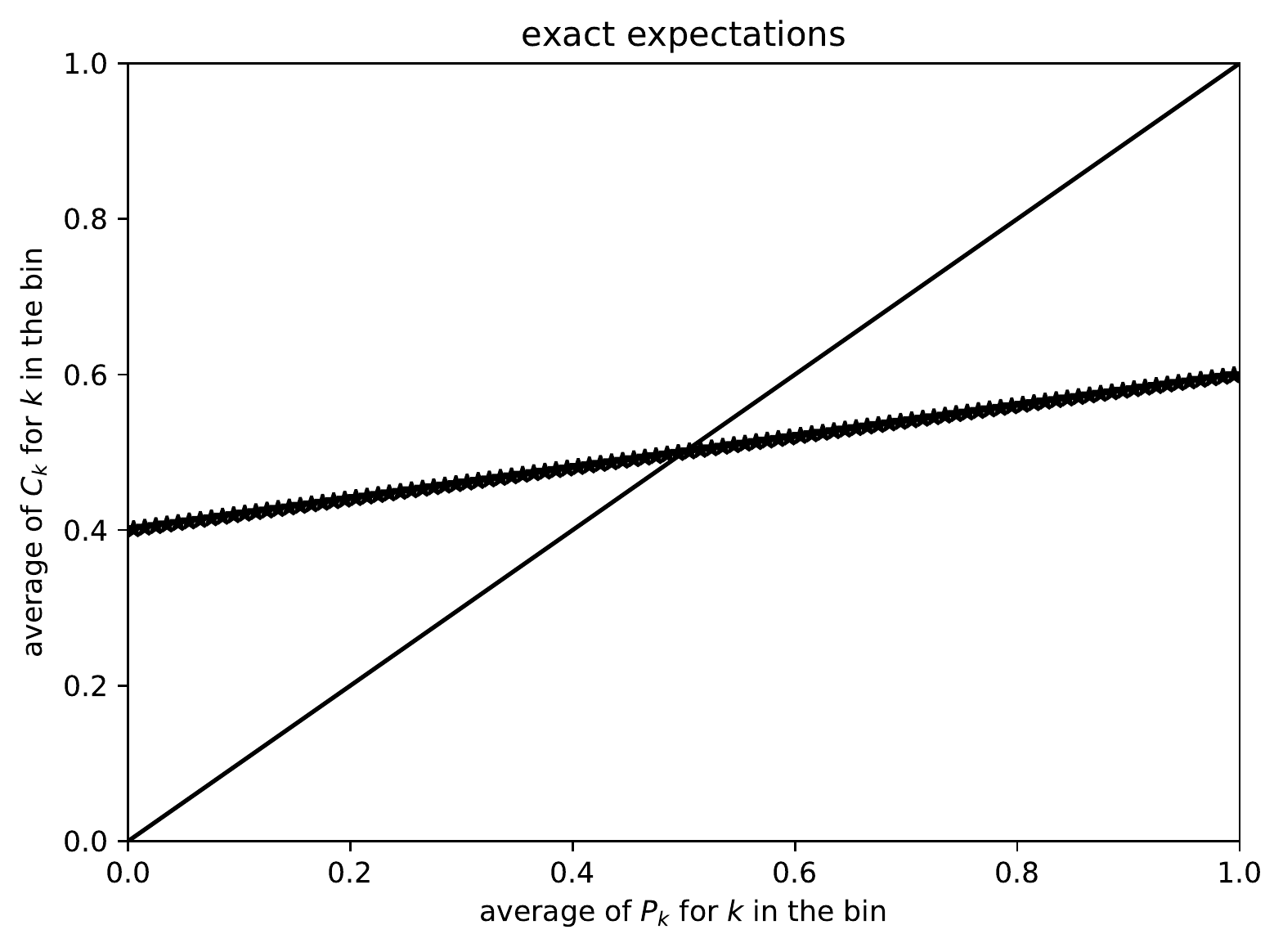}}

\end{centering}
\caption{$n =$ 100; $P_1$, $P_2$, \dots, $P_n$ are equispaced}
\label{100}
\end{figure}

\begin{figure}
\begin{centering}

\parbox{\imsize}{\includegraphics[width=\imsize]
                 {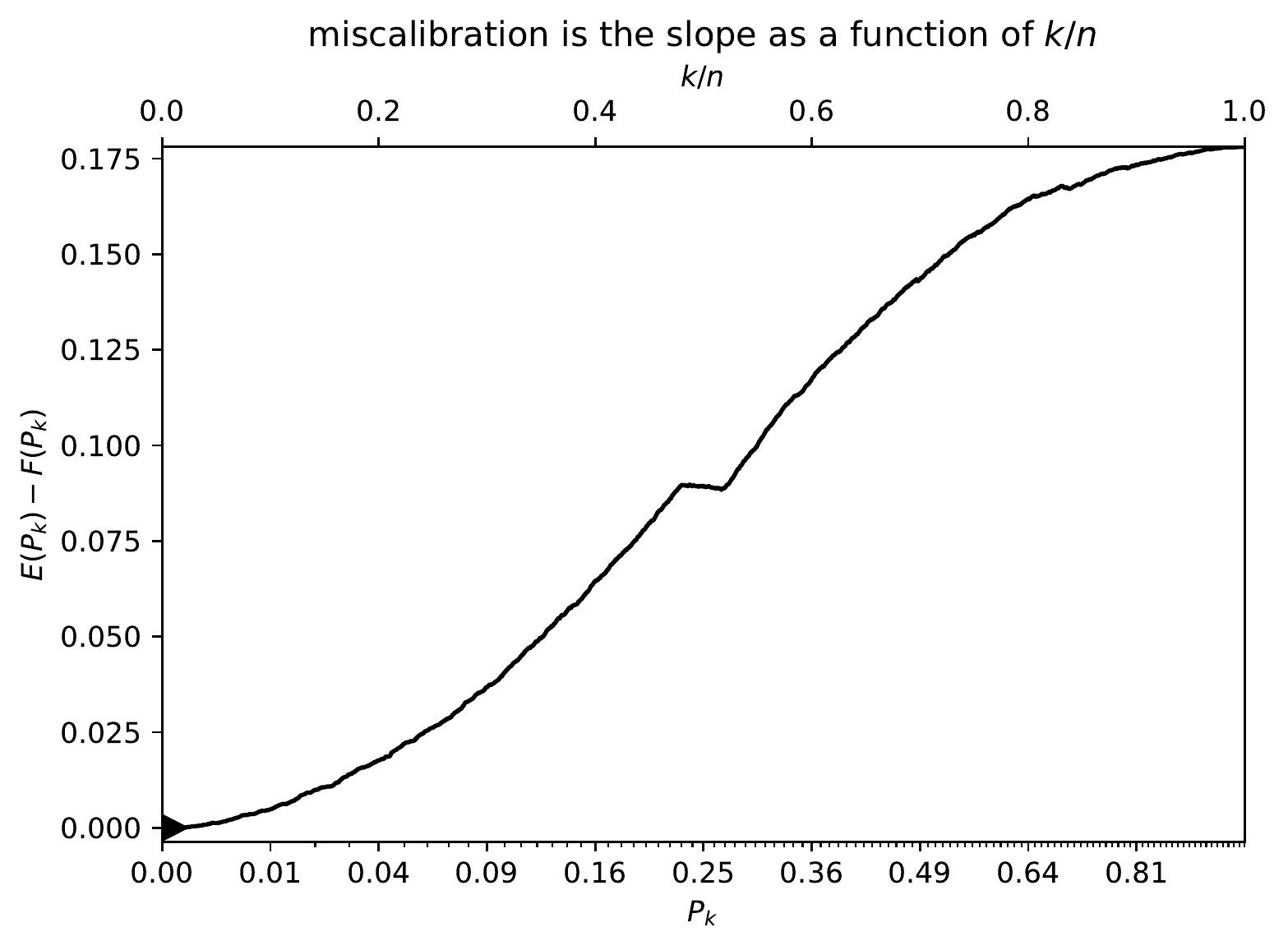}}
\quad\quad
\parbox{\imsize}{\includegraphics[width=\imsize]
                 {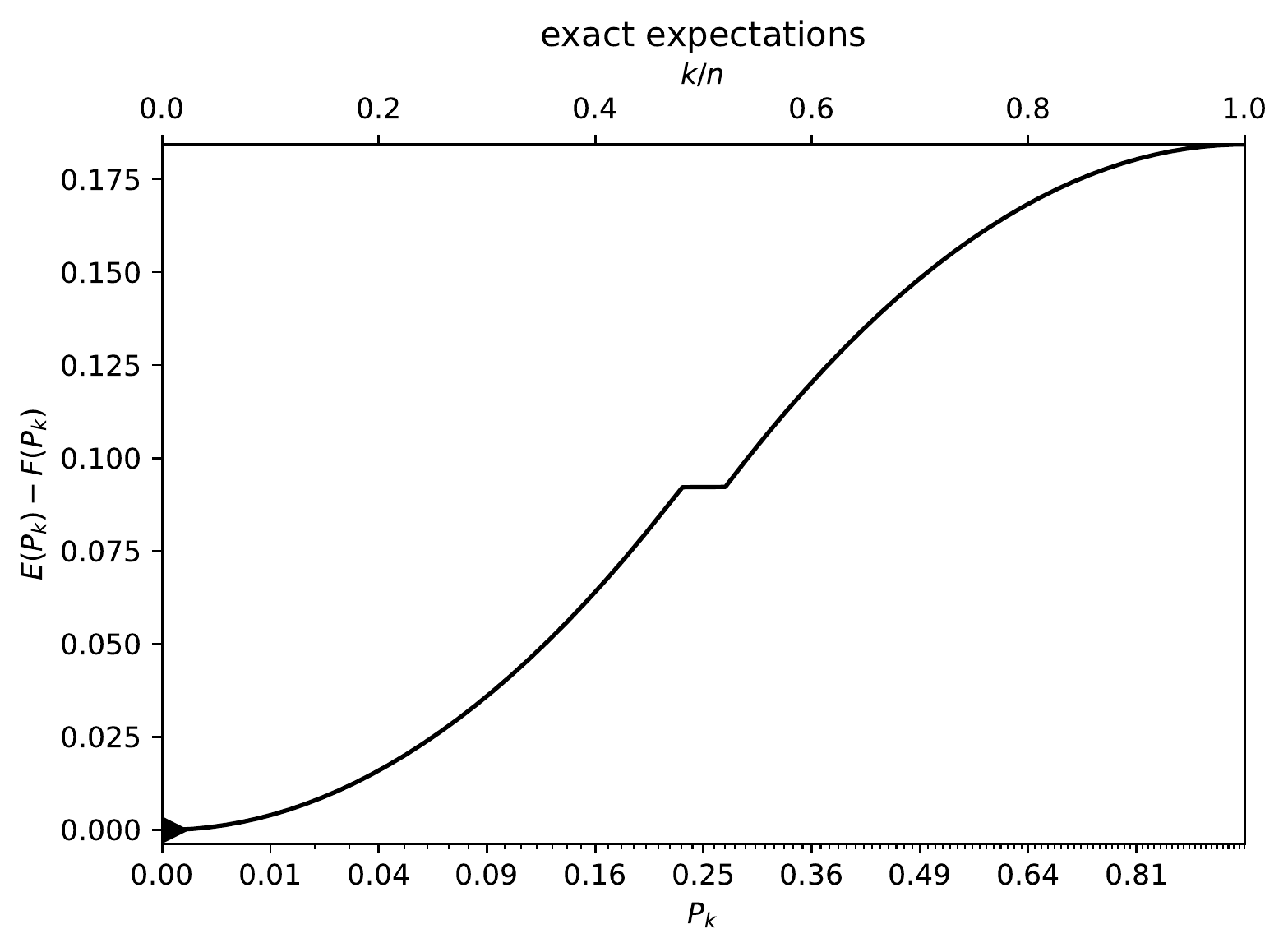}}

\vspace{\vertsep}

\parbox{\imsize}{\includegraphics[width=\imsize]
                 {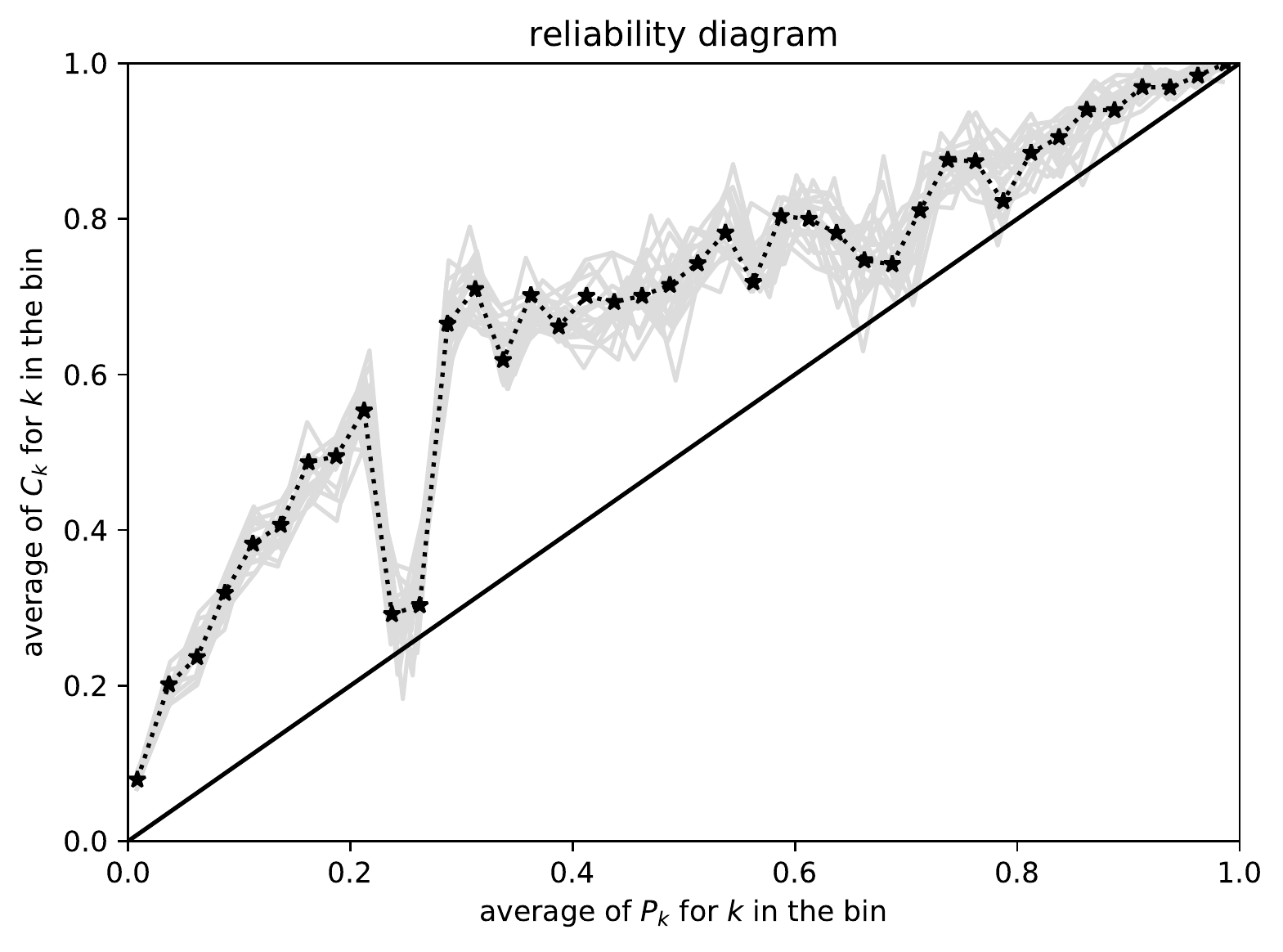}}
\quad\quad
\parbox{\imsize}{\includegraphics[width=\imsize]
                 {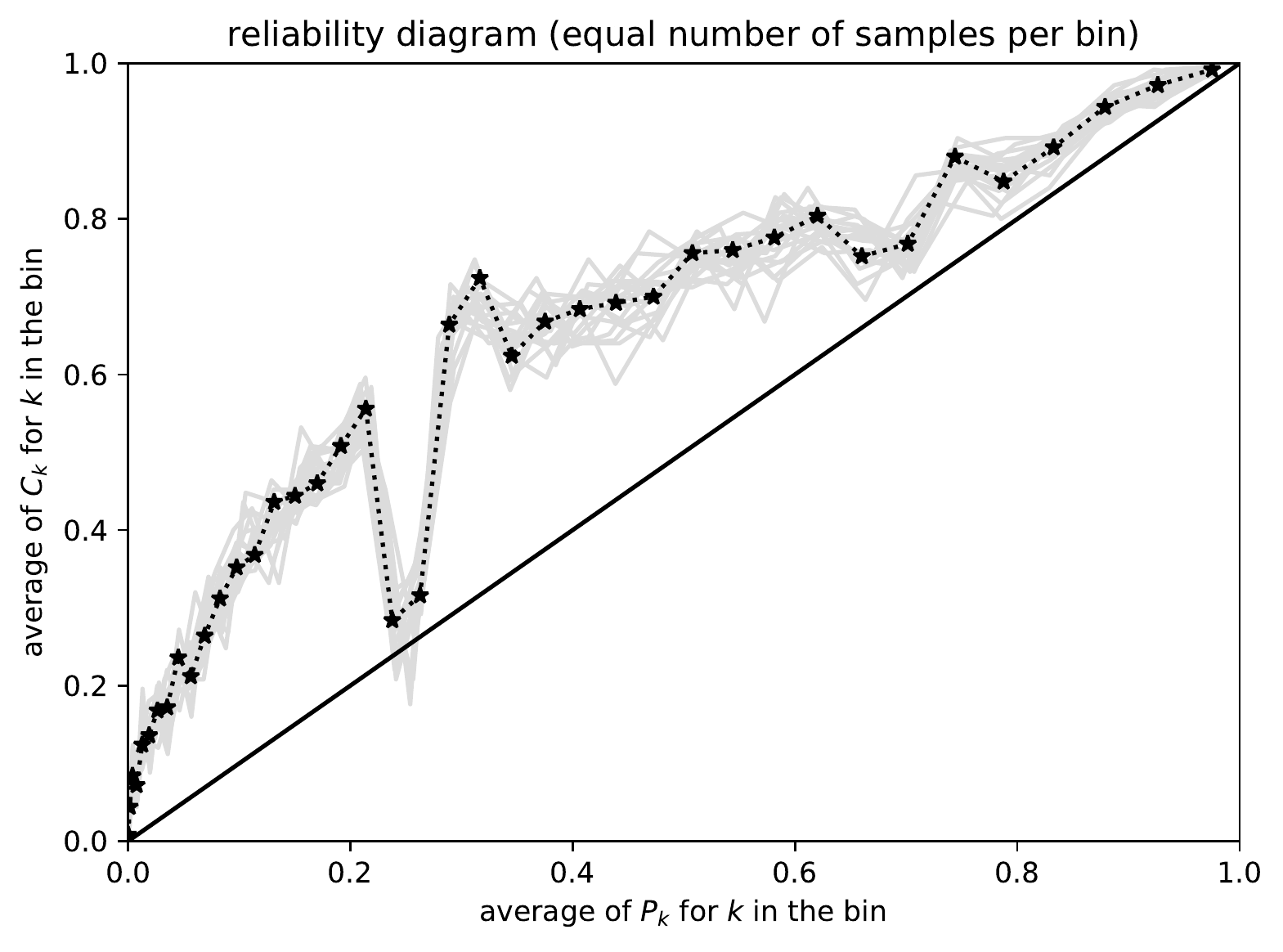}}

\vspace{\vertsep}

\parbox{\imsize}{\includegraphics[width=\imsize]
                 {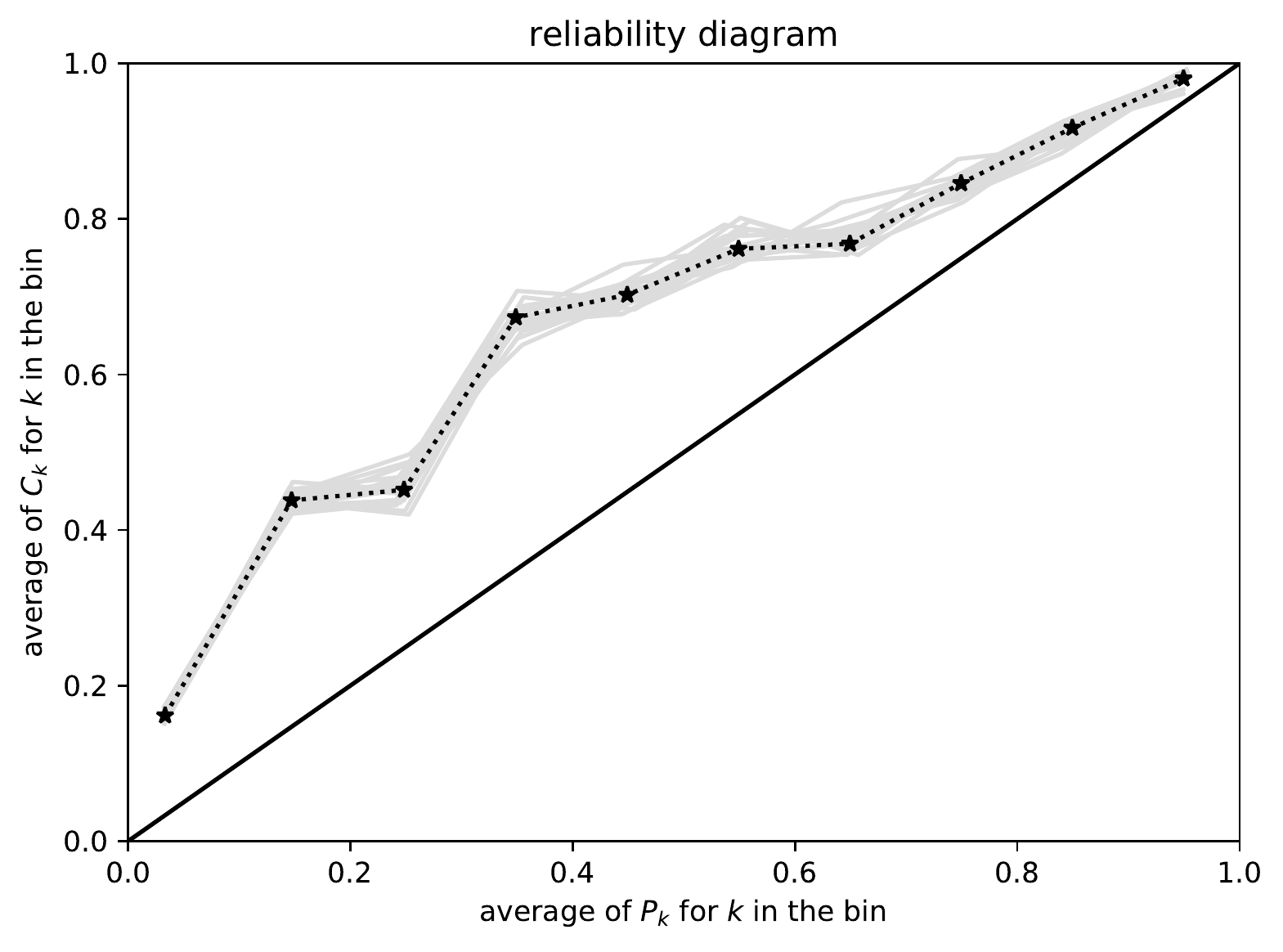}}
\quad\quad
\parbox{\imsize}{\includegraphics[width=\imsize]
                 {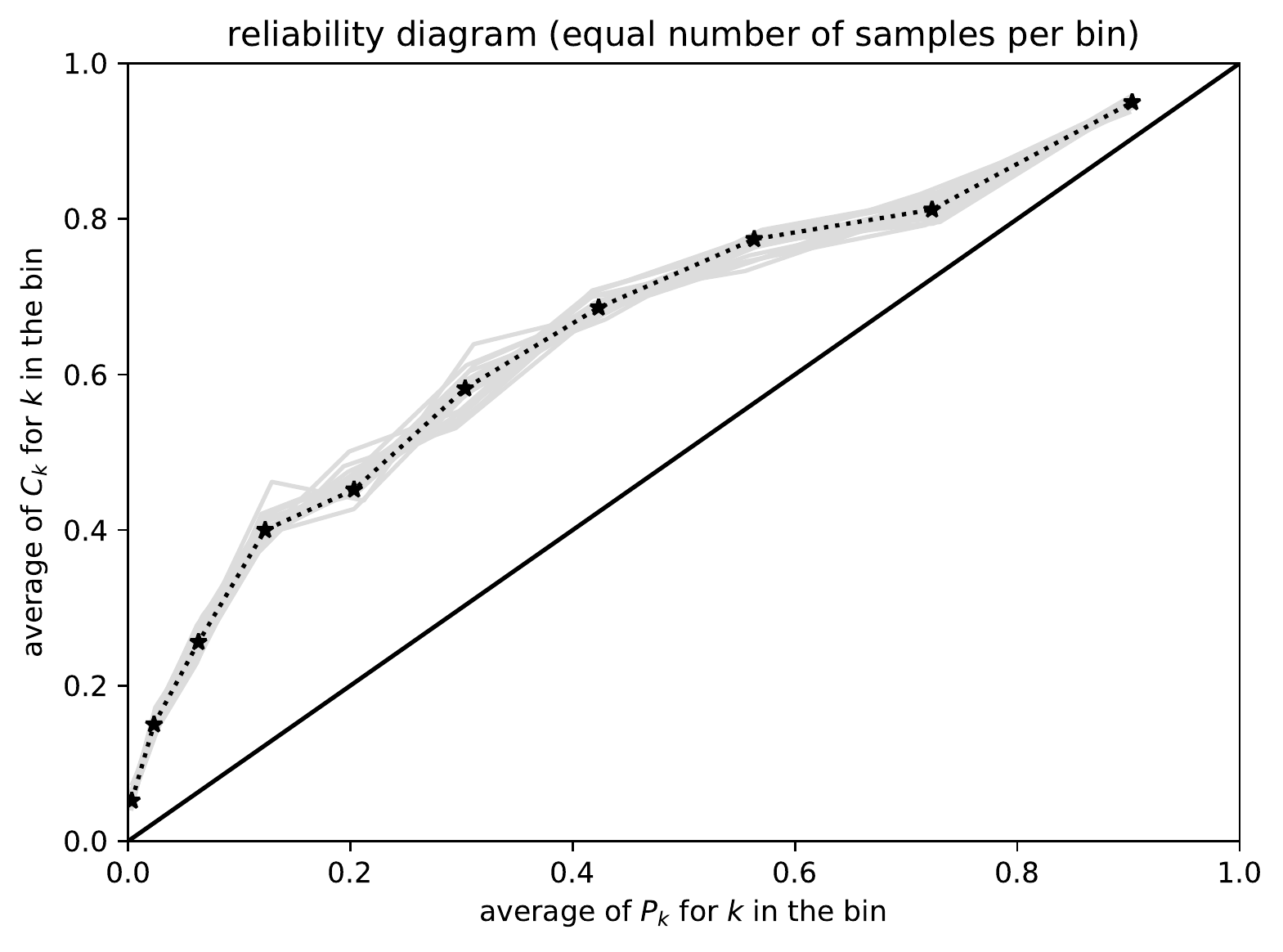}}

\vspace{\vertsep}

\parbox{\imsize}{\includegraphics[width=\imsize]
                 {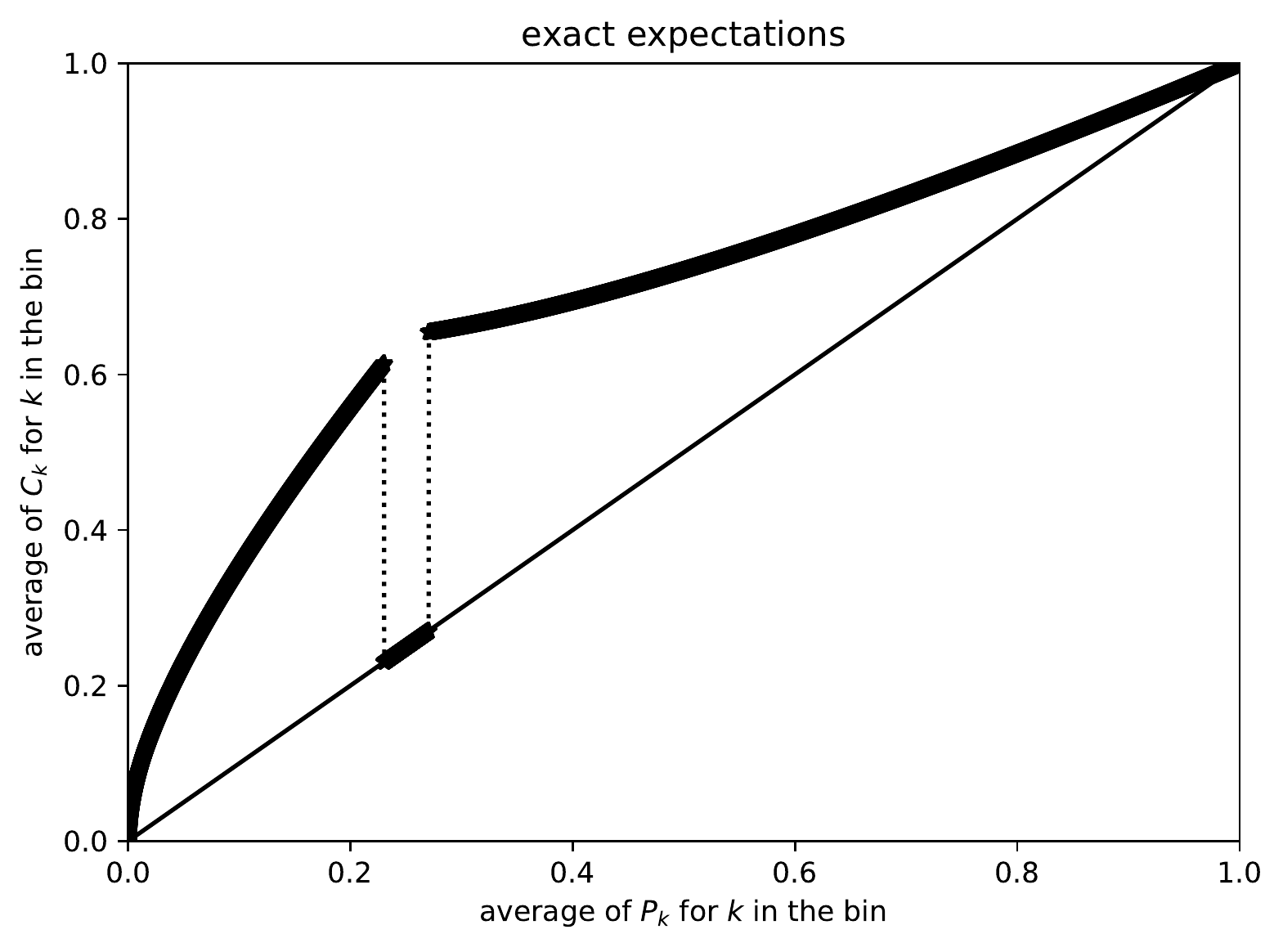}}

\end{centering}
\caption{$n =$ 10,000; $P_1$, $P_2$, \dots, $P_n$ are denser near 0}
\label{10000_0}
\end{figure}

\begin{figure}
\begin{centering}

\parbox{\imsize}{\includegraphics[width=\imsize]
                 {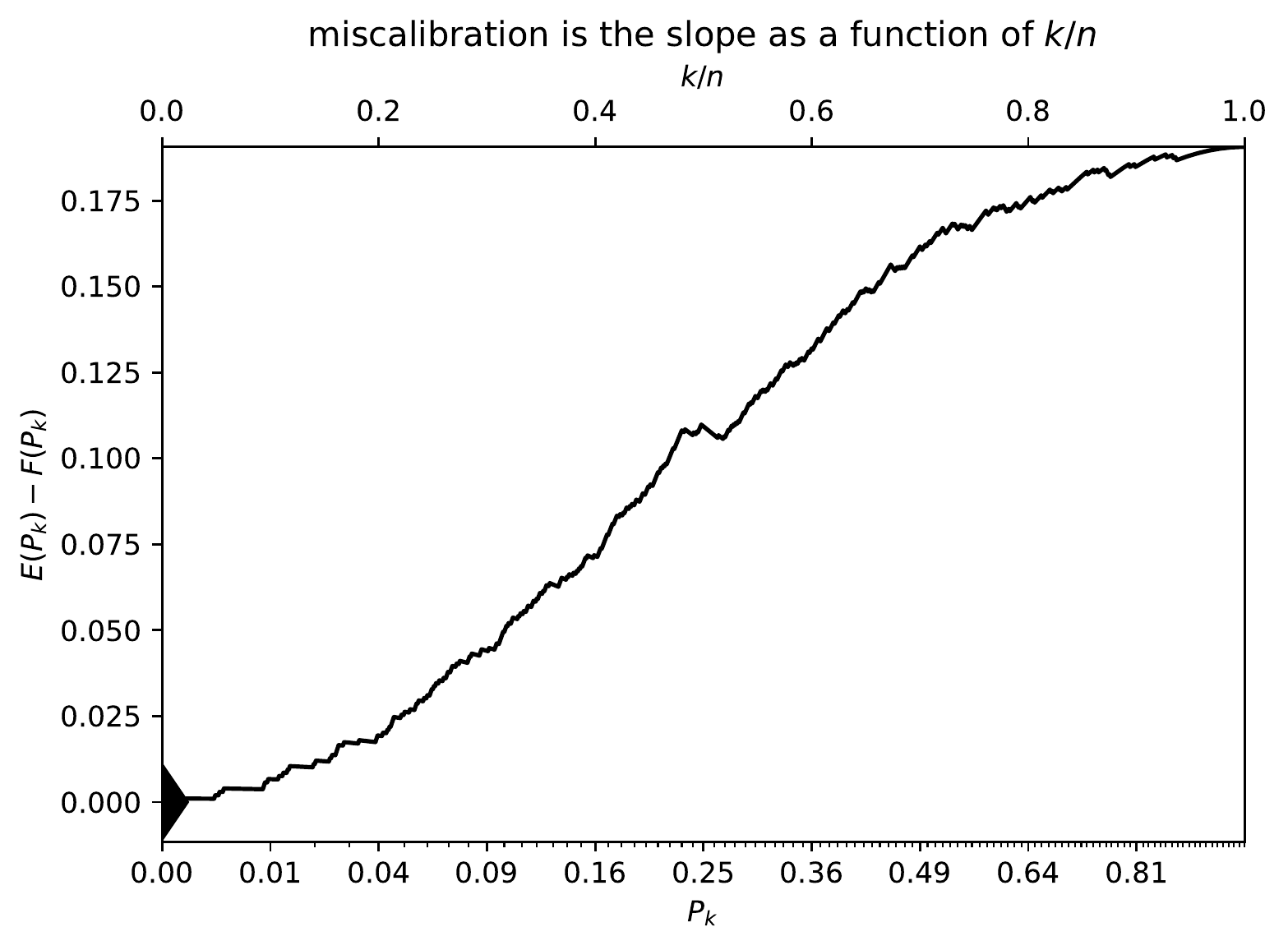}}
\quad\quad
\parbox{\imsize}{\includegraphics[width=\imsize]
                 {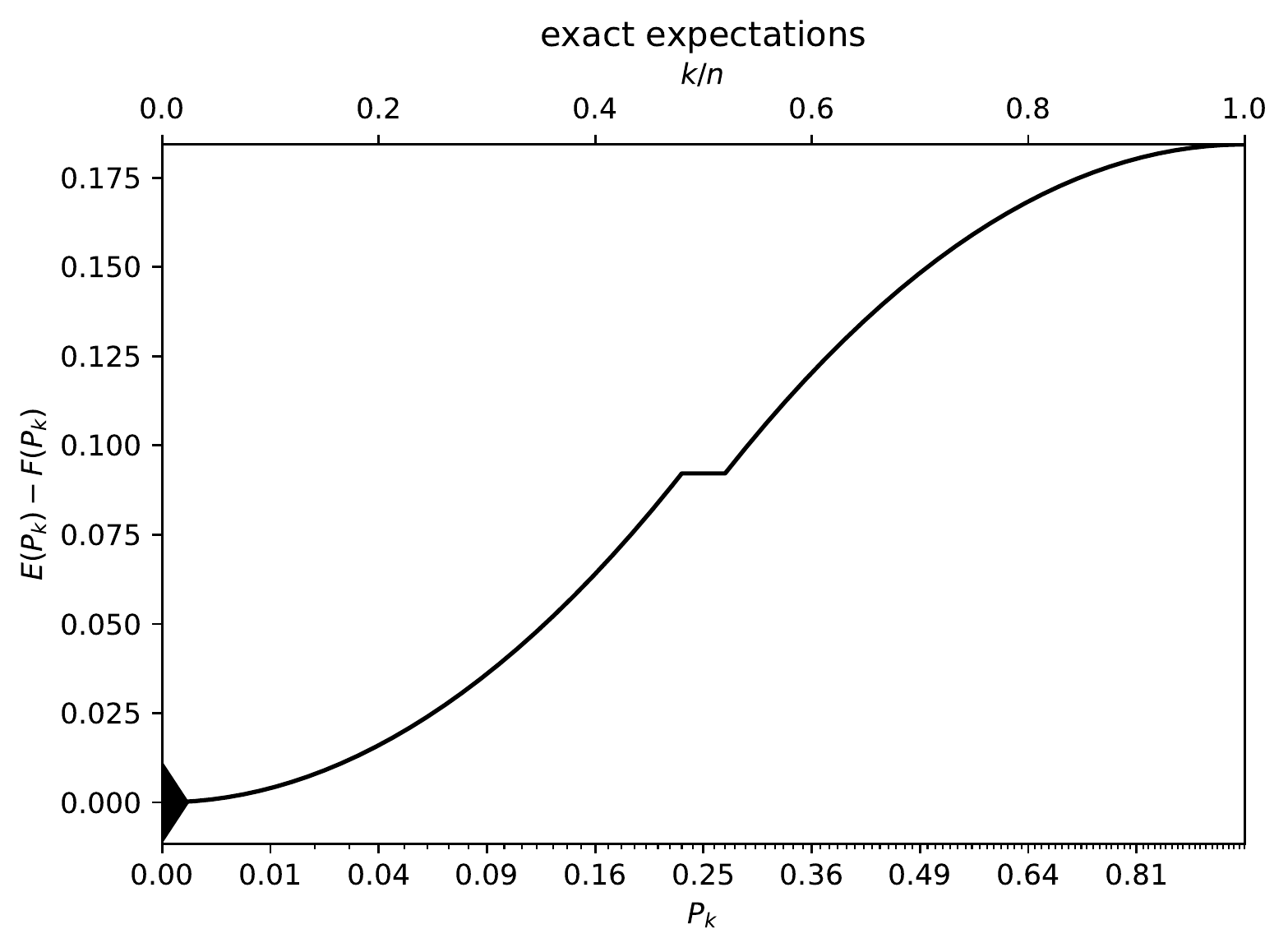}}

\vspace{\vertsep}

\parbox{\imsize}{\includegraphics[width=\imsize]
                 {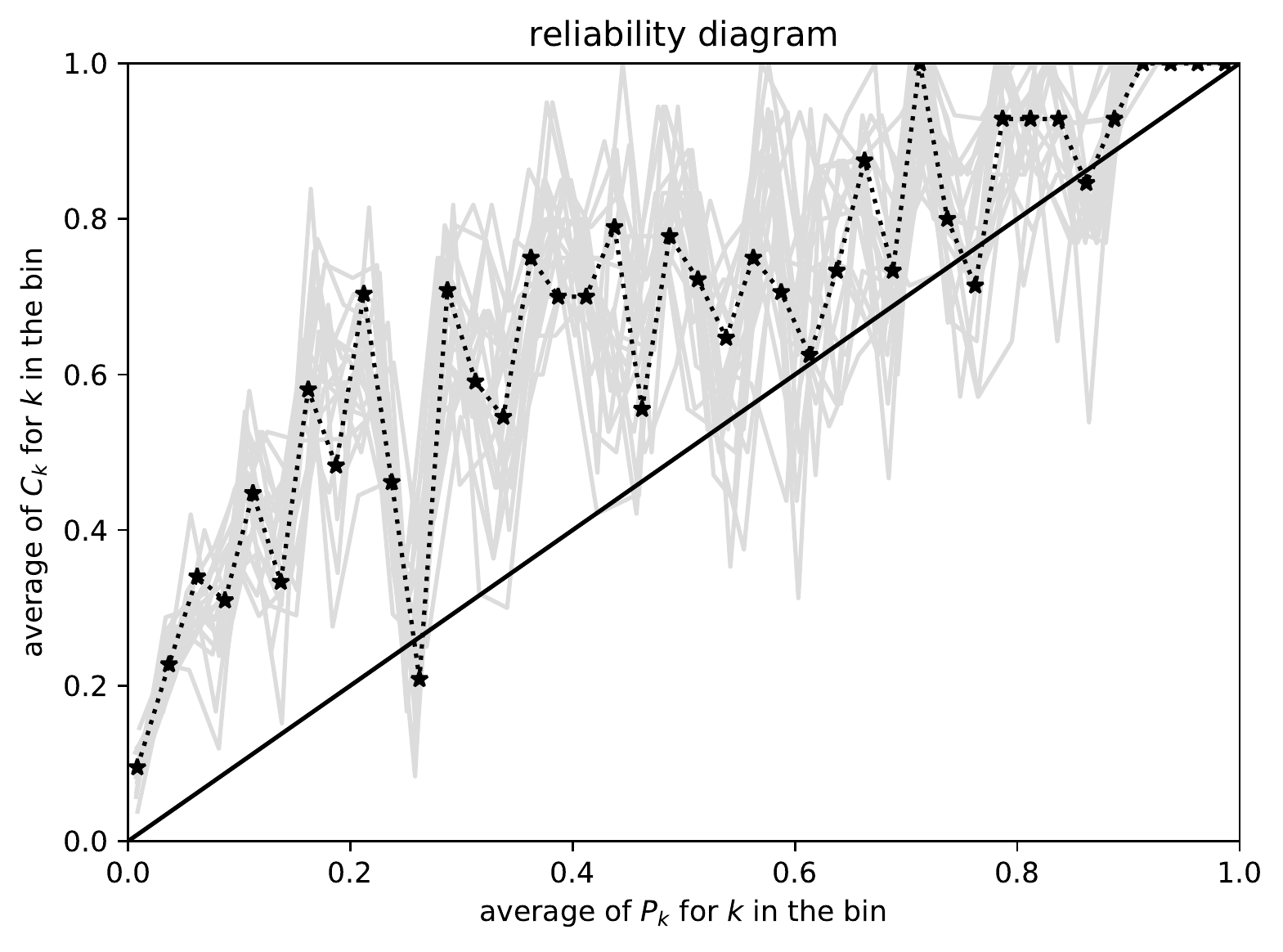}}
\quad\quad
\parbox{\imsize}{\includegraphics[width=\imsize]
                 {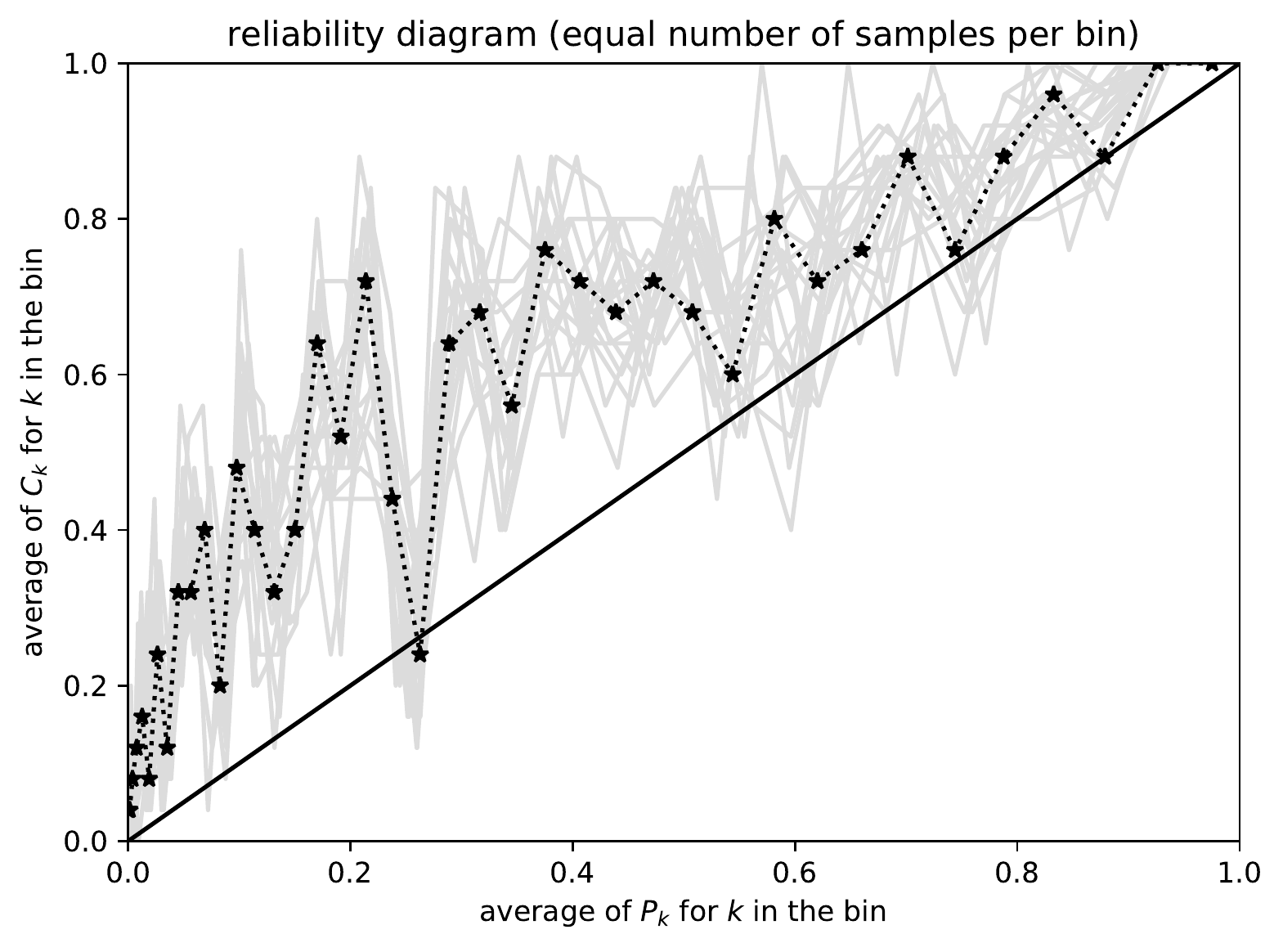}}

\vspace{\vertsep}

\parbox{\imsize}{\includegraphics[width=\imsize]
                 {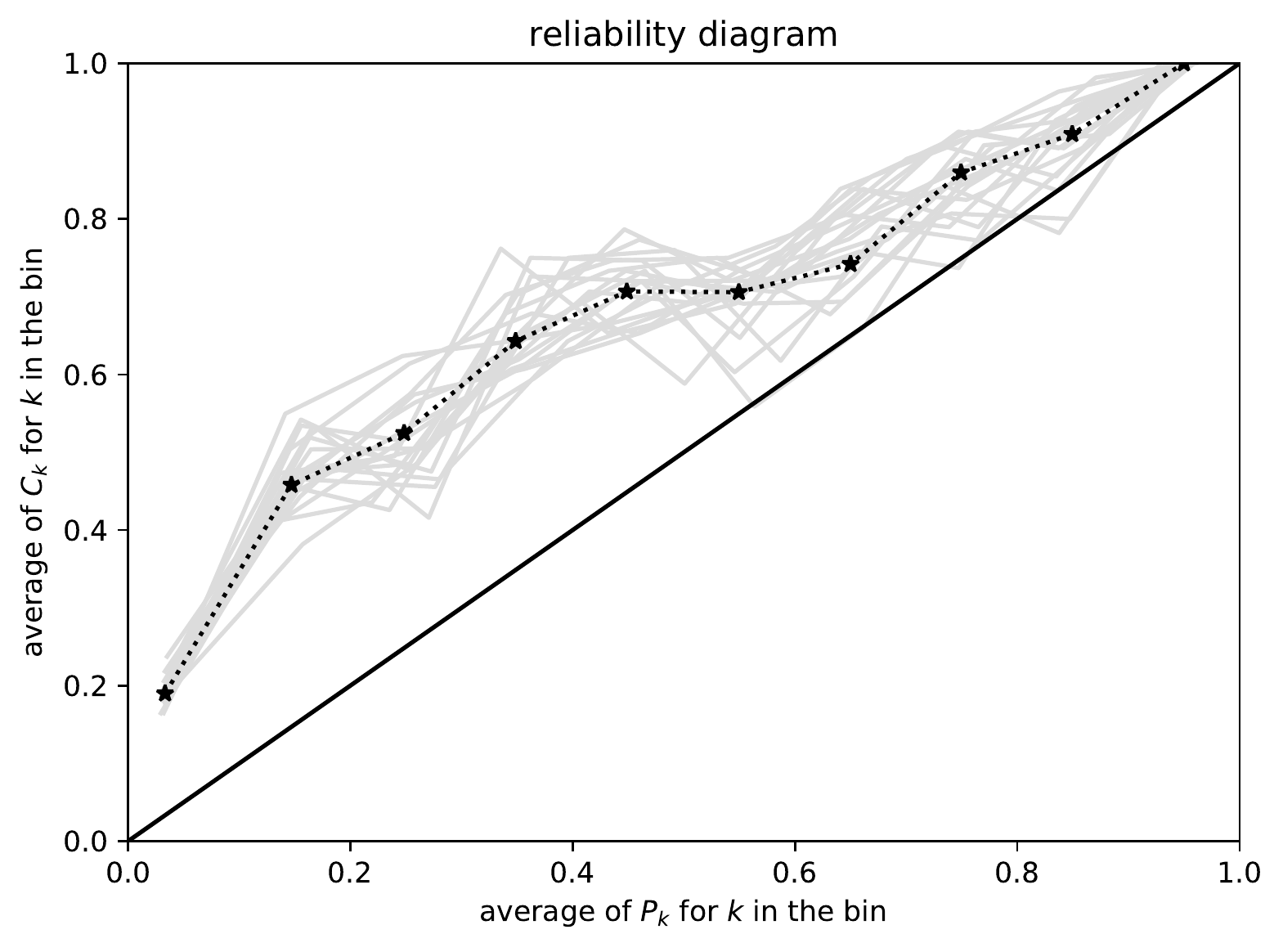}}
\quad\quad
\parbox{\imsize}{\includegraphics[width=\imsize]
                 {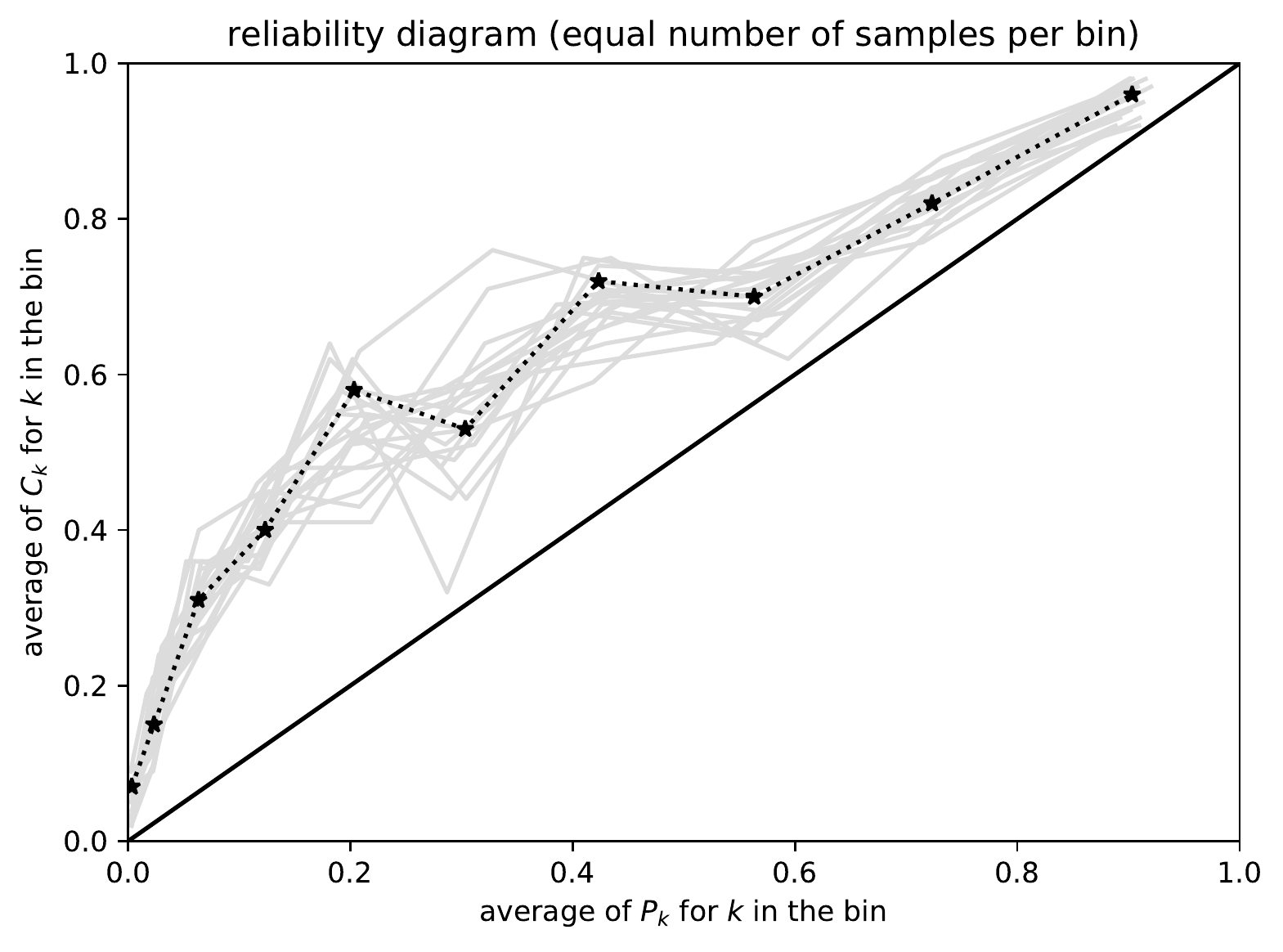}}

\vspace{\vertsep}

\parbox{\imsize}{\includegraphics[width=\imsize]
                 {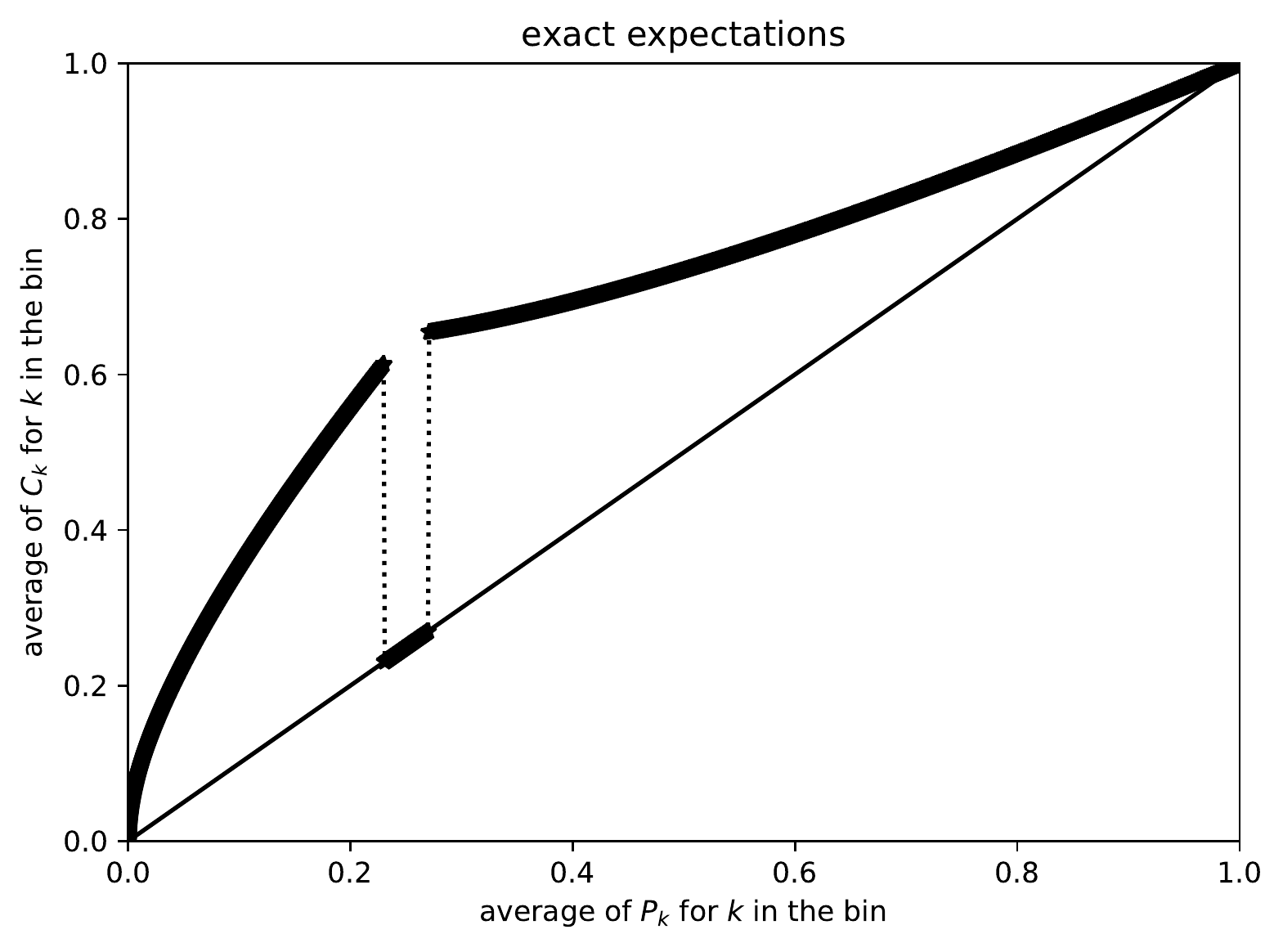}}

\end{centering}
\caption{$n =$ 1,000; $P_1$, $P_2$, \dots, $P_n$ are denser near 0}
\label{1000_0}
\end{figure}

\begin{figure}
\begin{centering}

\parbox{\imsize}{\includegraphics[width=\imsize]
                 {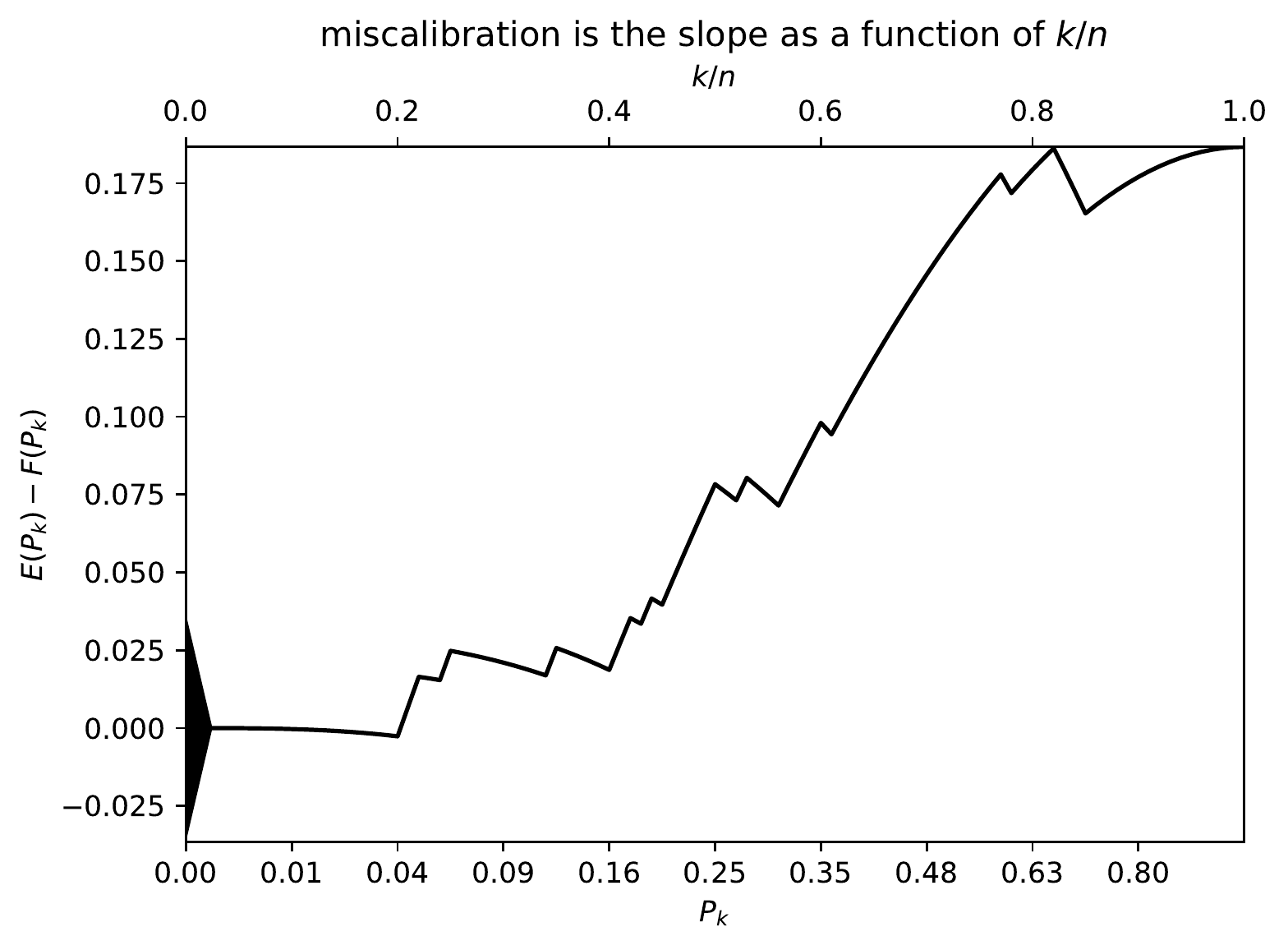}}
\quad\quad
\parbox{\imsize}{\includegraphics[width=\imsize]
                 {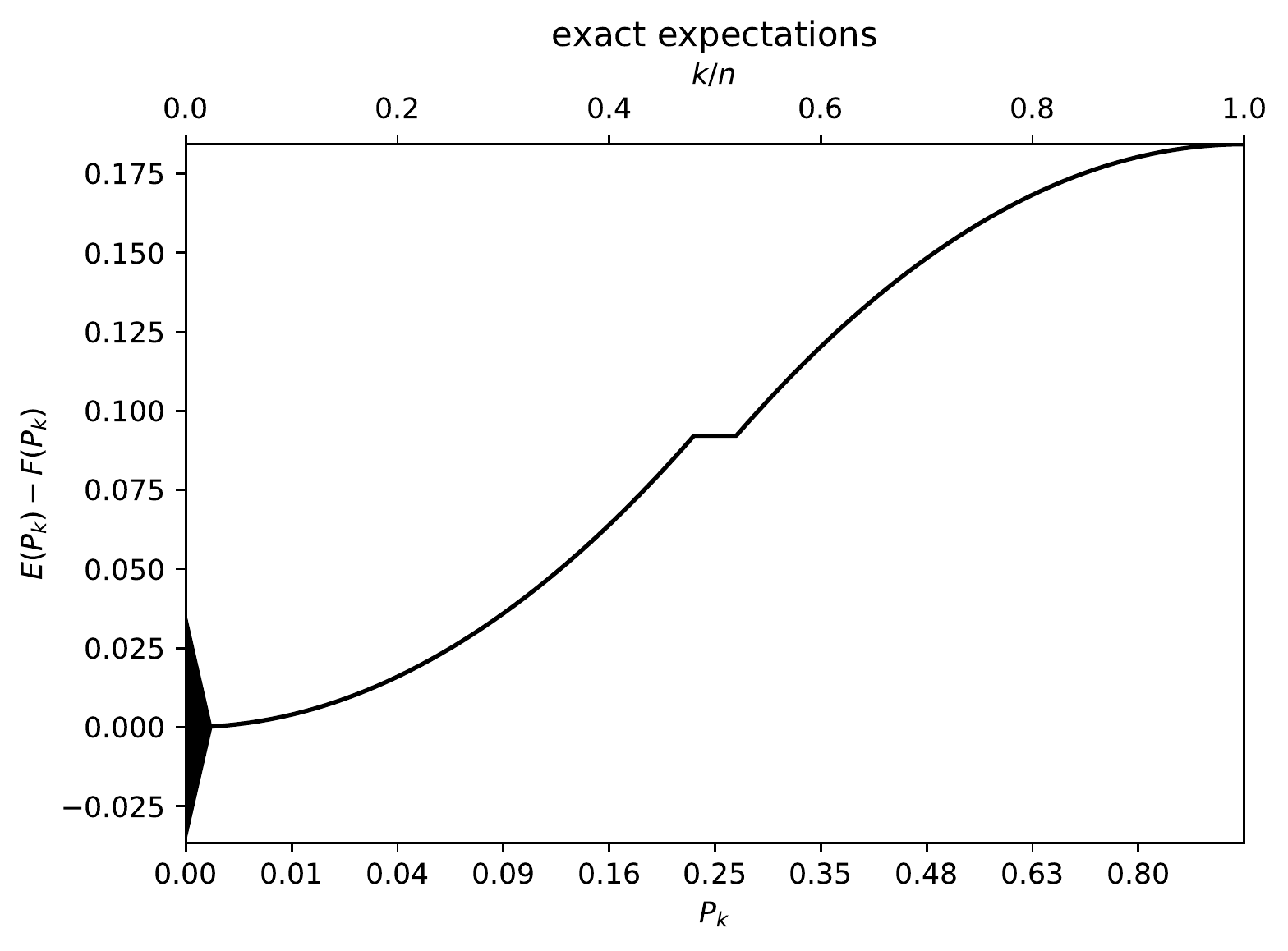}}

\vspace{\vertsep}

\parbox{\imsize}{\includegraphics[width=\imsize]
                 {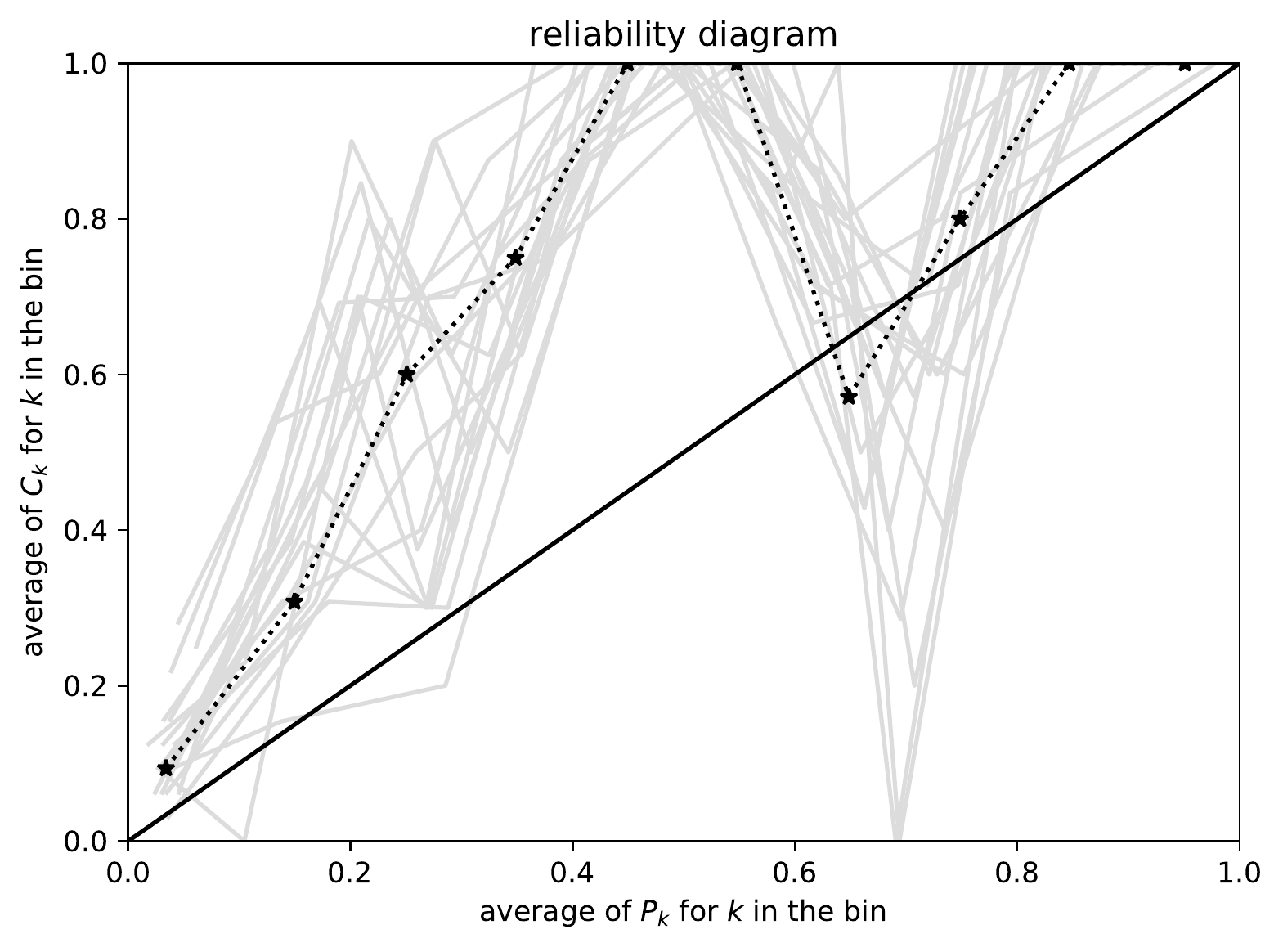}}
\quad\quad
\parbox{\imsize}{\includegraphics[width=\imsize]
                 {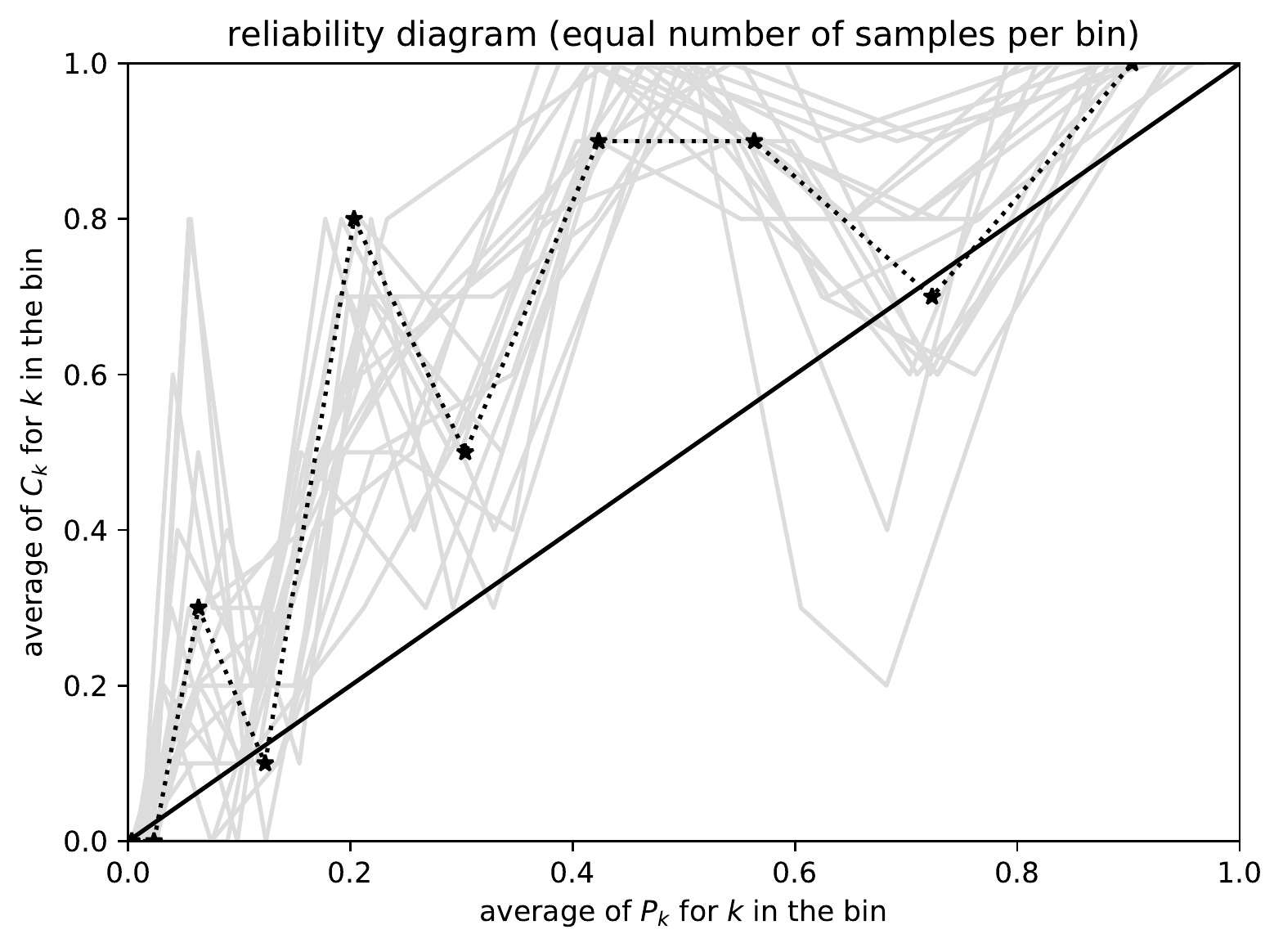}}

\vspace{\vertsep}

\parbox{\imsize}{\includegraphics[width=\imsize]
                 {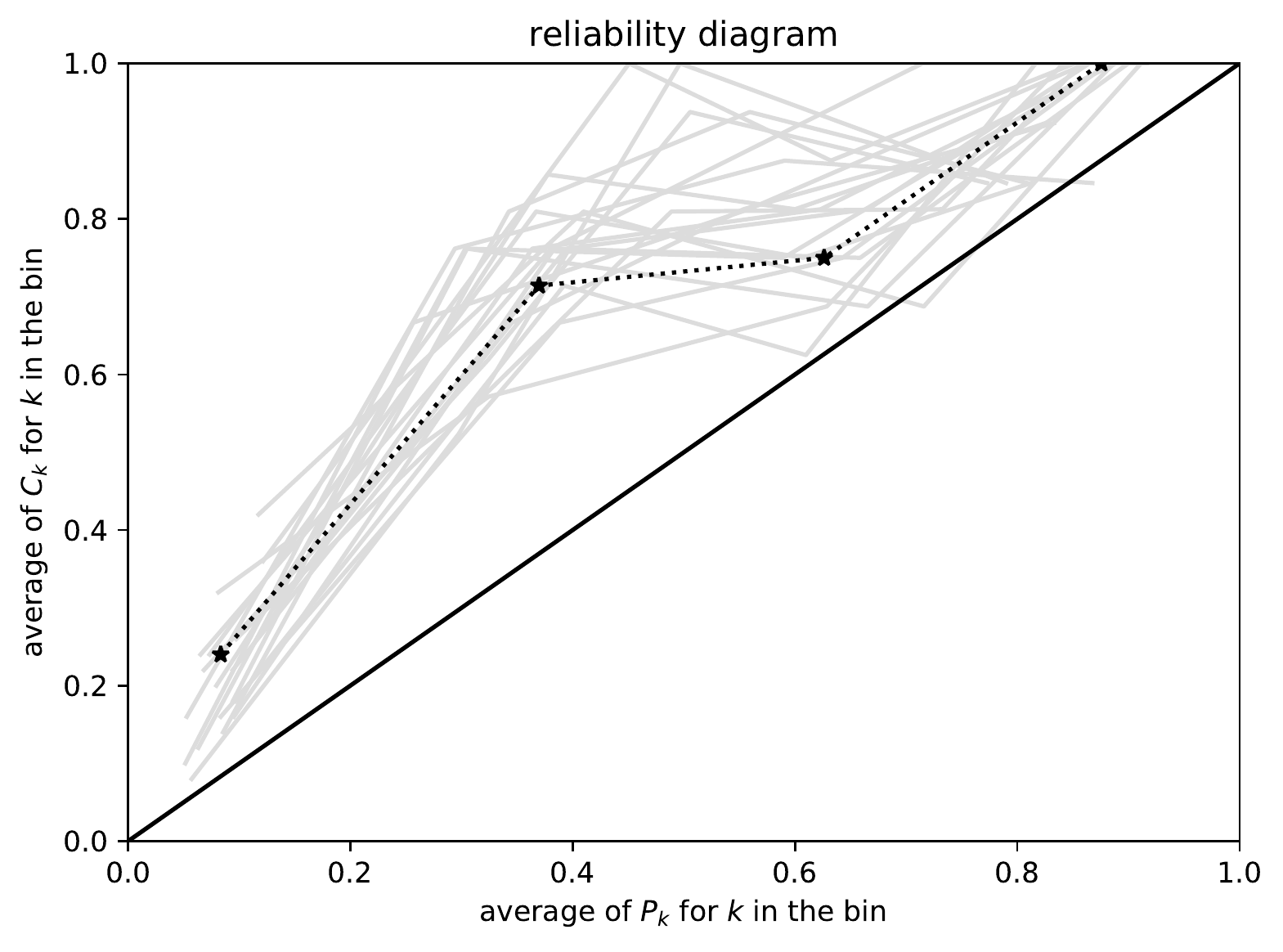}}
\quad\quad
\parbox{\imsize}{\includegraphics[width=\imsize]
                 {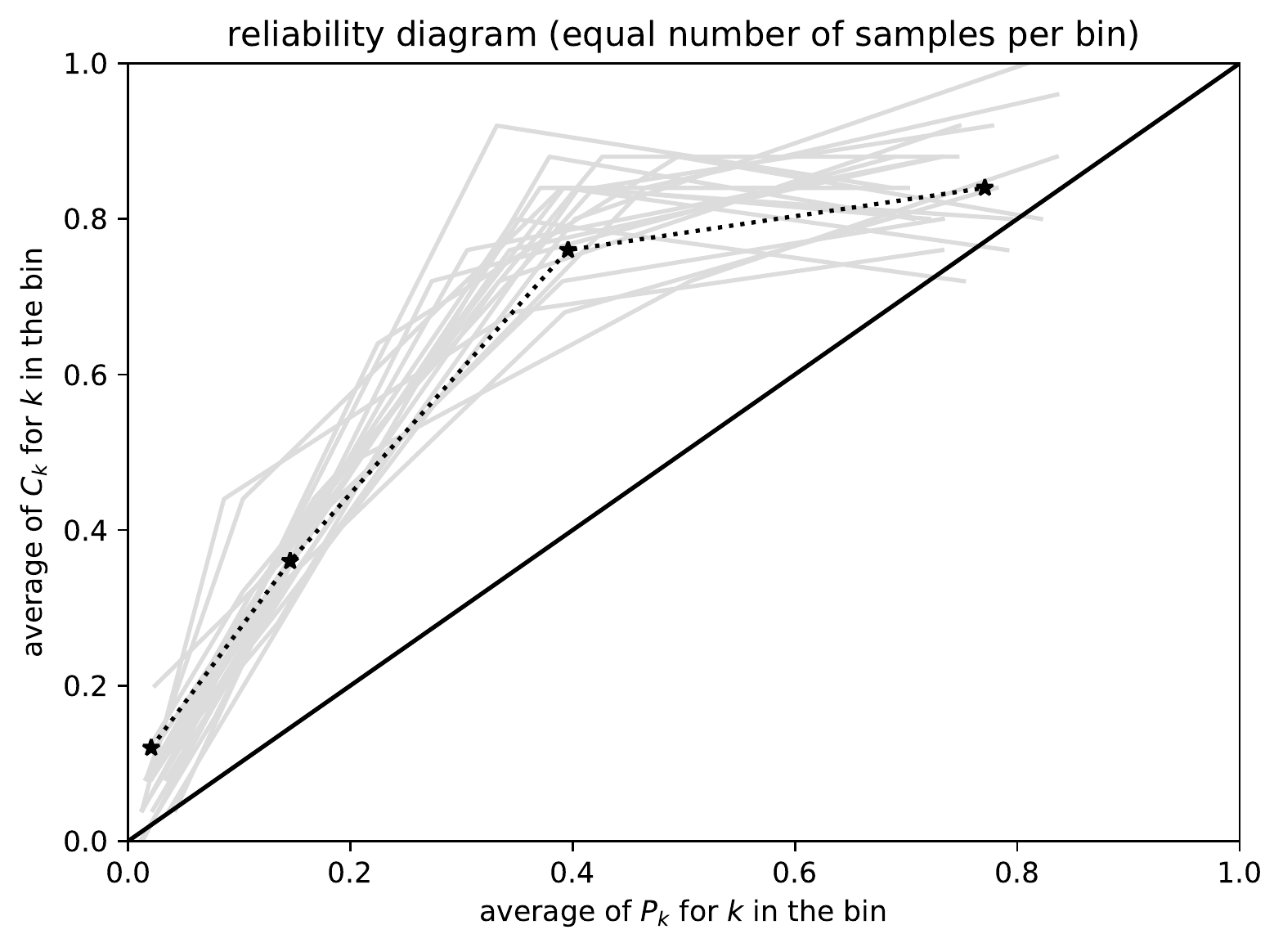}}

\vspace{\vertsep}

\parbox{\imsize}{\includegraphics[width=\imsize]
                 {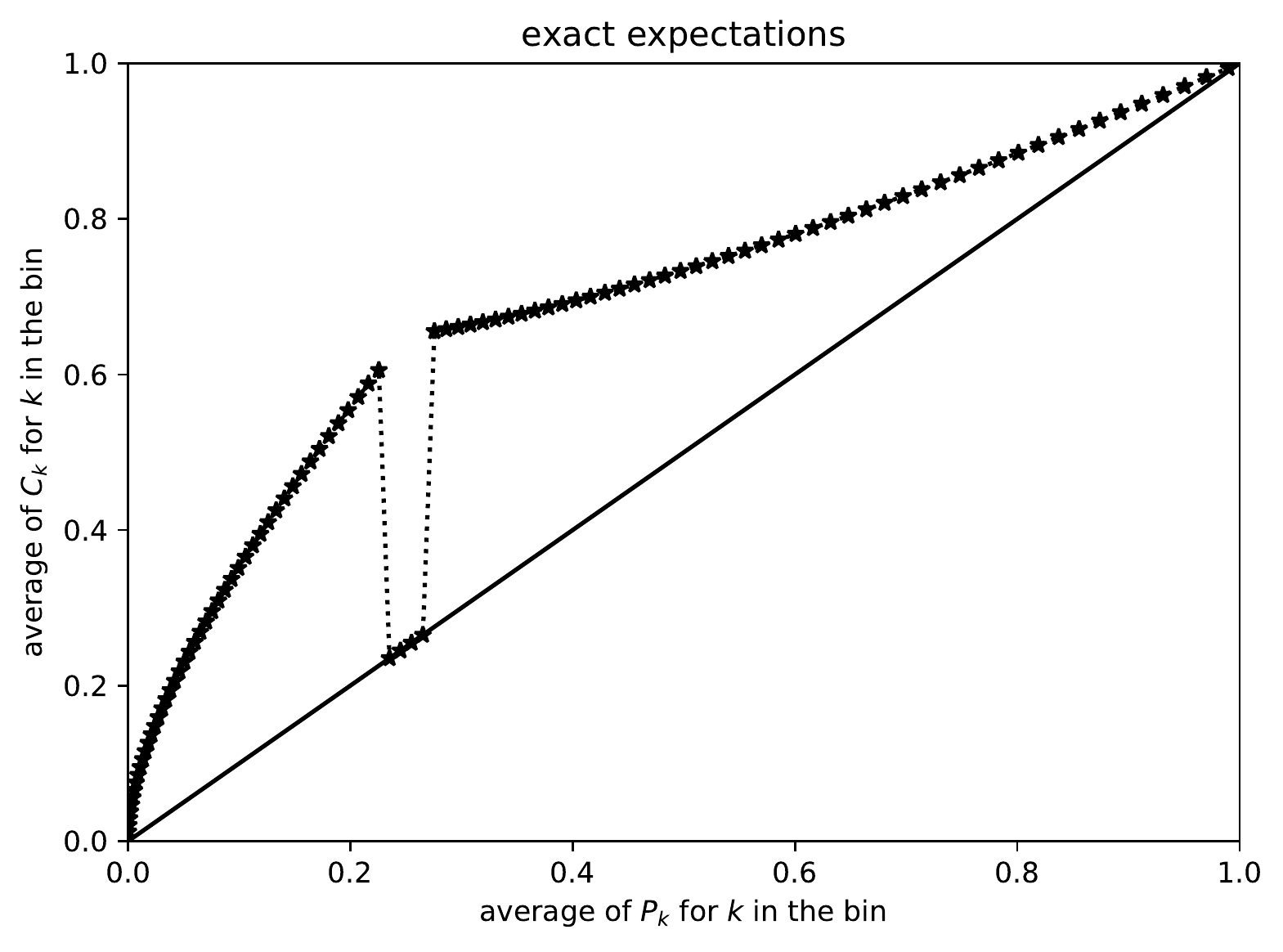}}

\end{centering}
\caption{$n =$ 100; $P_1$, $P_2$, \dots, $P_n$ are denser near 0}
\label{100_0}
\end{figure}

\begin{figure}
\begin{centering}

\parbox{\imsize}{\includegraphics[width=\imsize]
                 {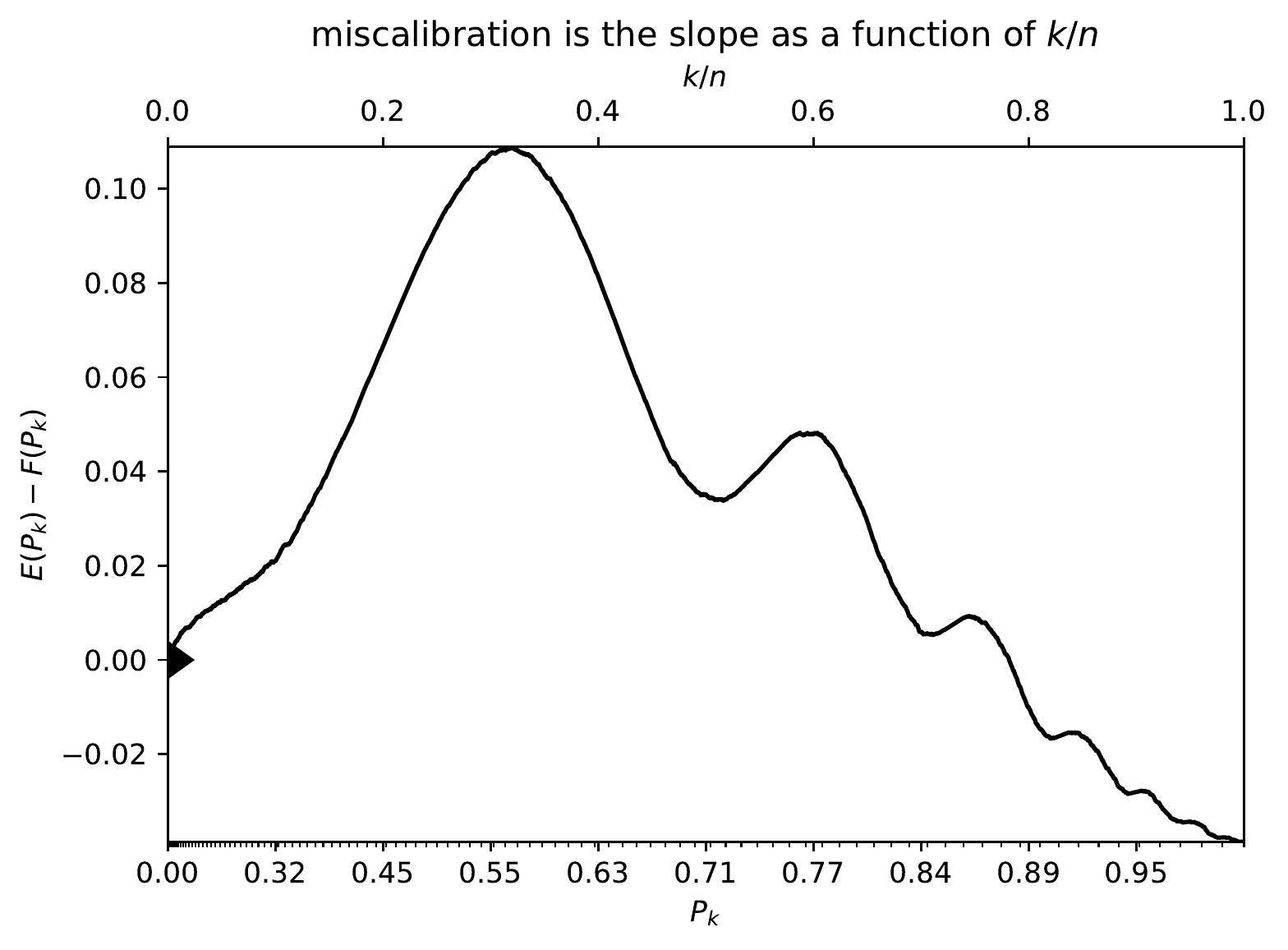}}
\quad\quad
\parbox{\imsize}{\includegraphics[width=\imsize]
                 {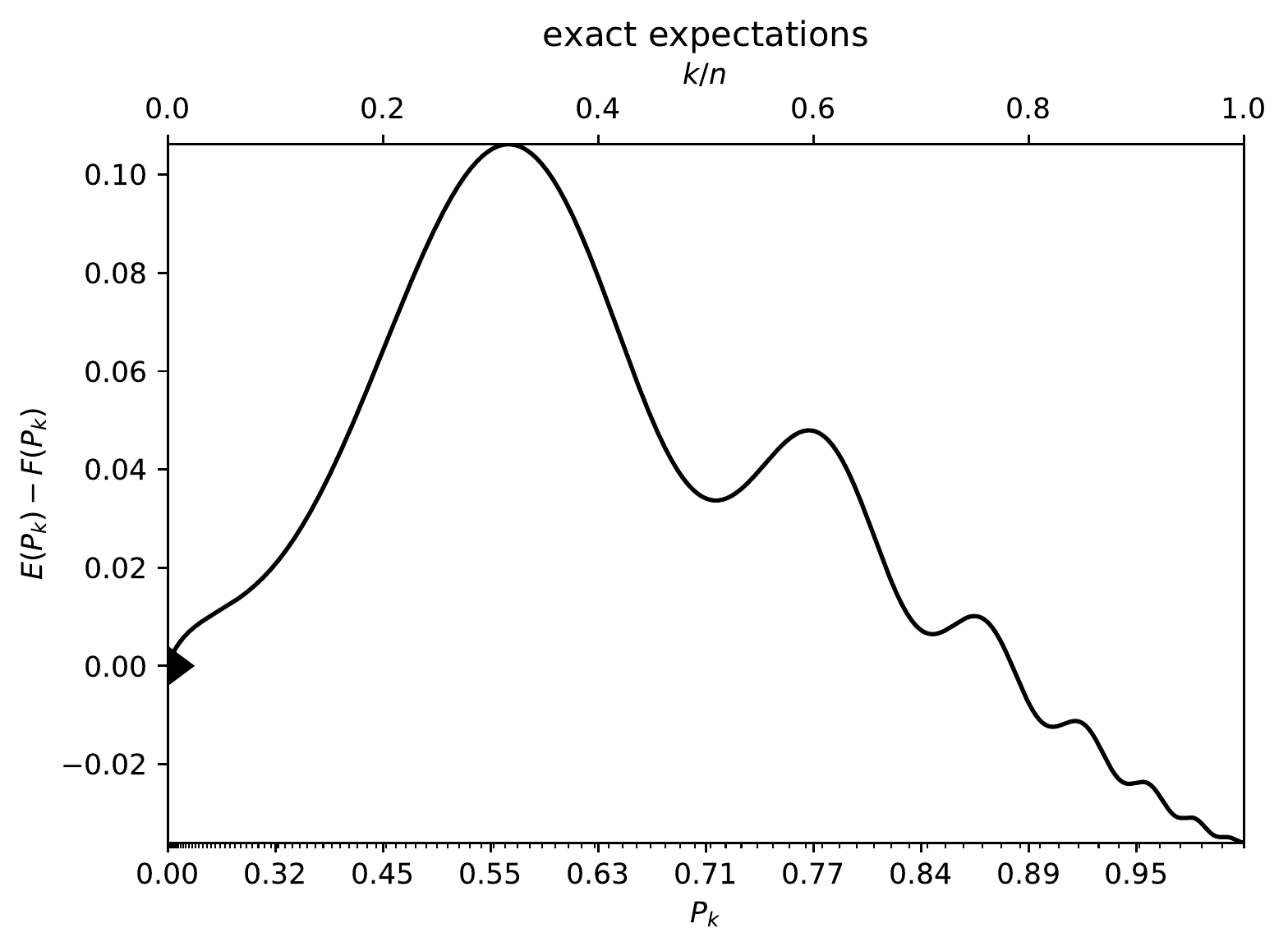}}

\vspace{\vertsep}

\parbox{\imsize}{\includegraphics[width=\imsize]
                 {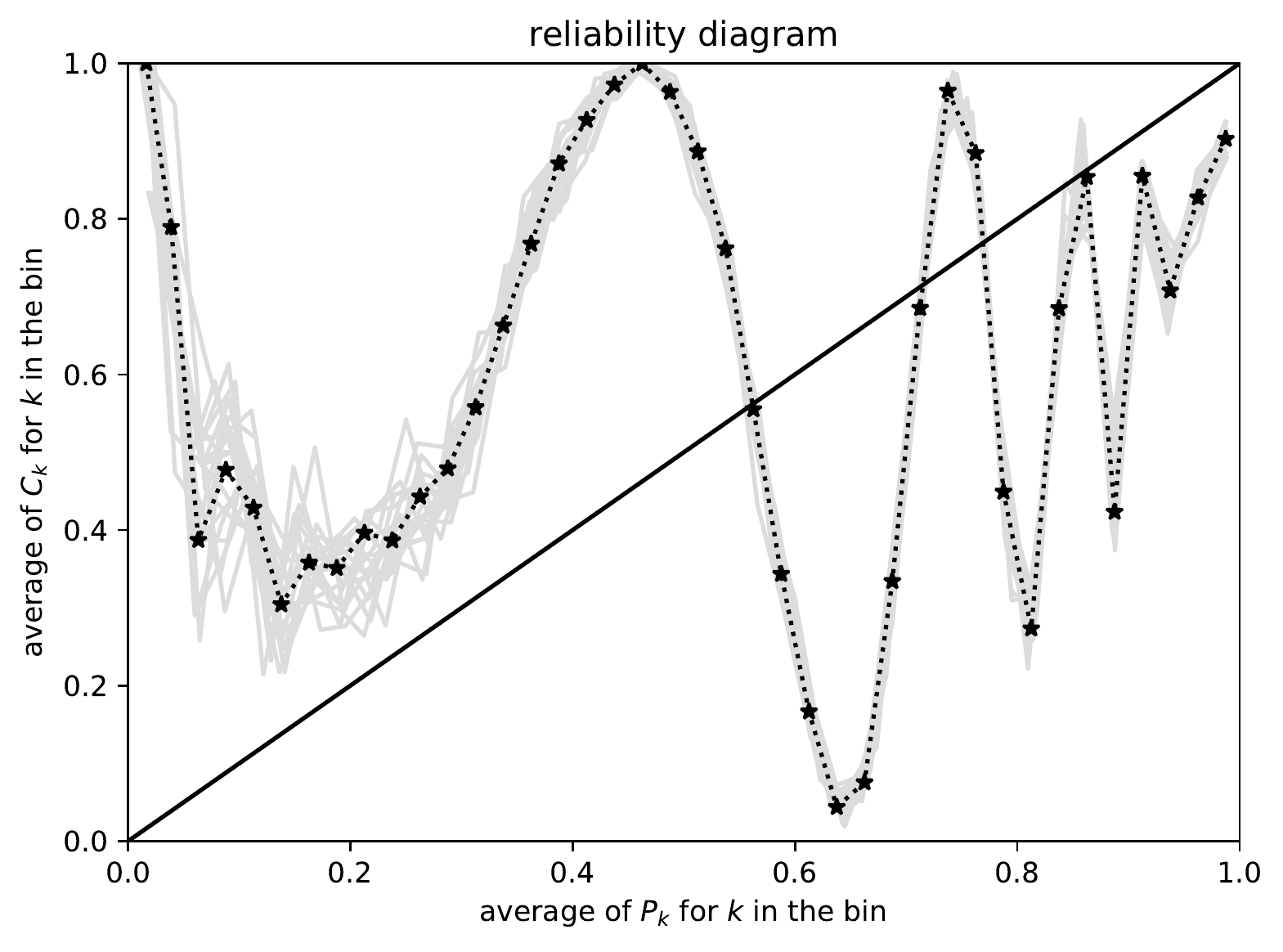}}
\quad\quad
\parbox{\imsize}{\includegraphics[width=\imsize]
                 {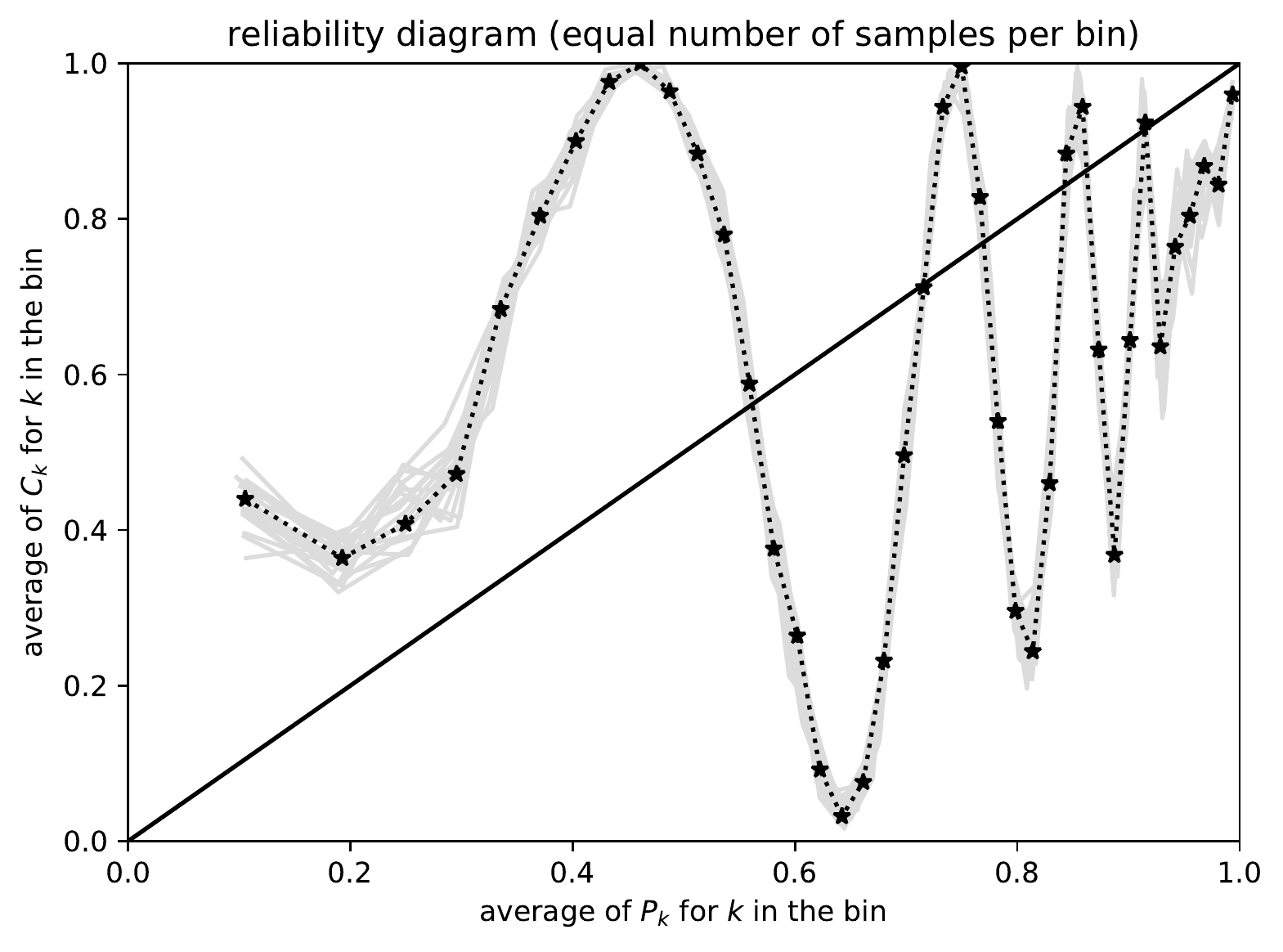}}

\vspace{\vertsep}

\parbox{\imsize}{\includegraphics[width=\imsize]
                 {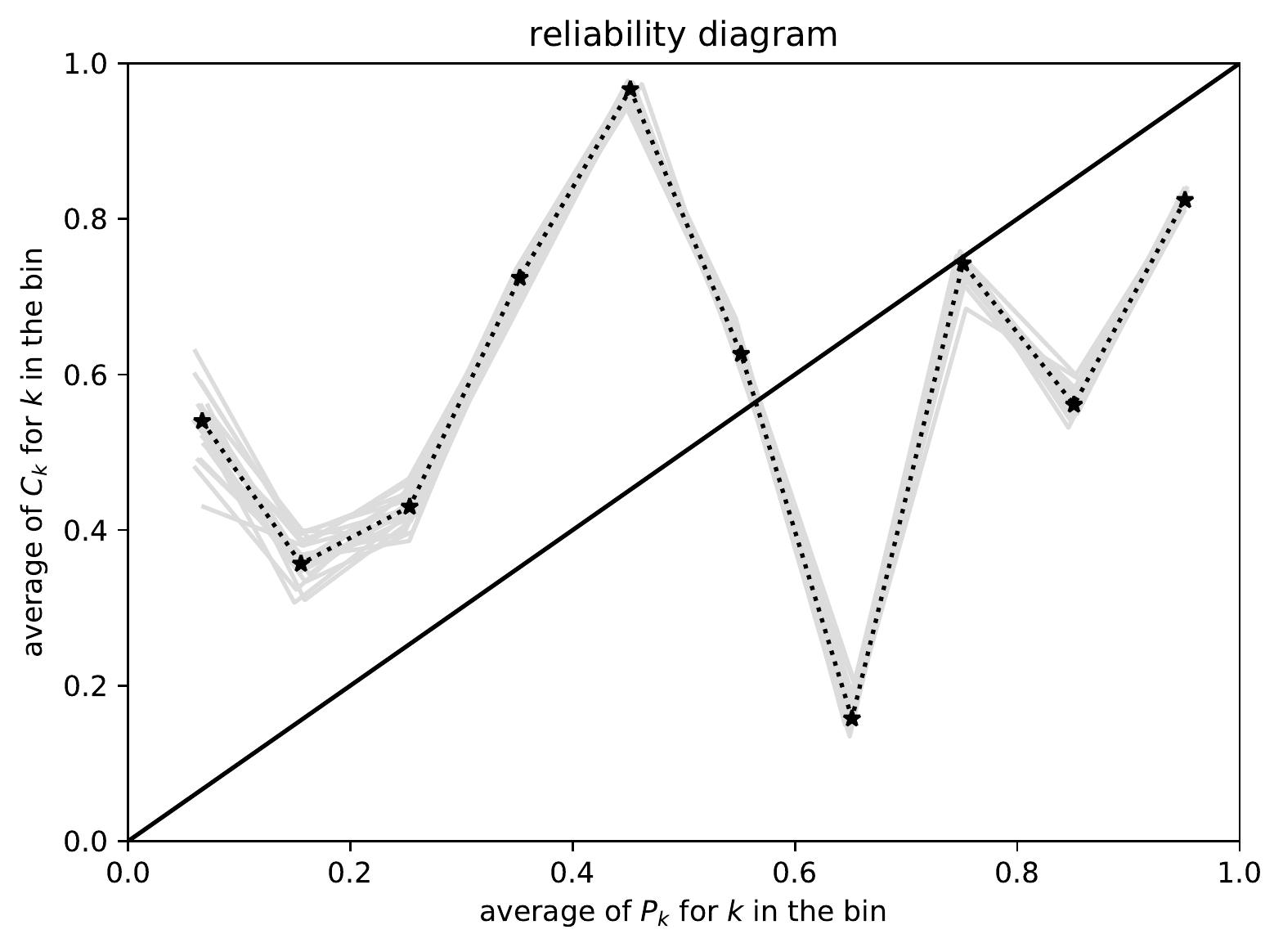}}
\quad\quad
\parbox{\imsize}{\includegraphics[width=\imsize]
                 {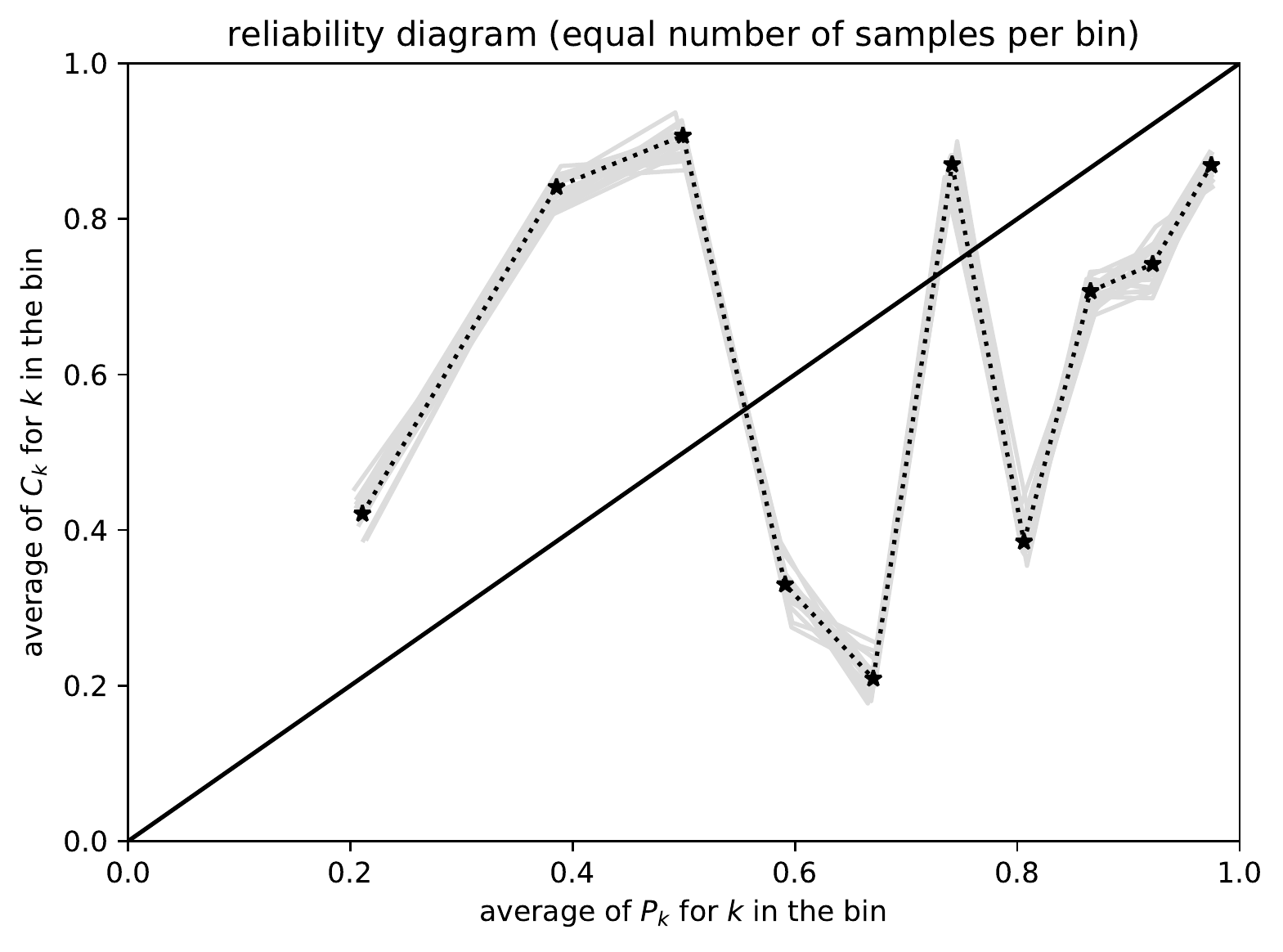}}

\vspace{\vertsep}

\parbox{\imsize}{\includegraphics[width=\imsize]
                 {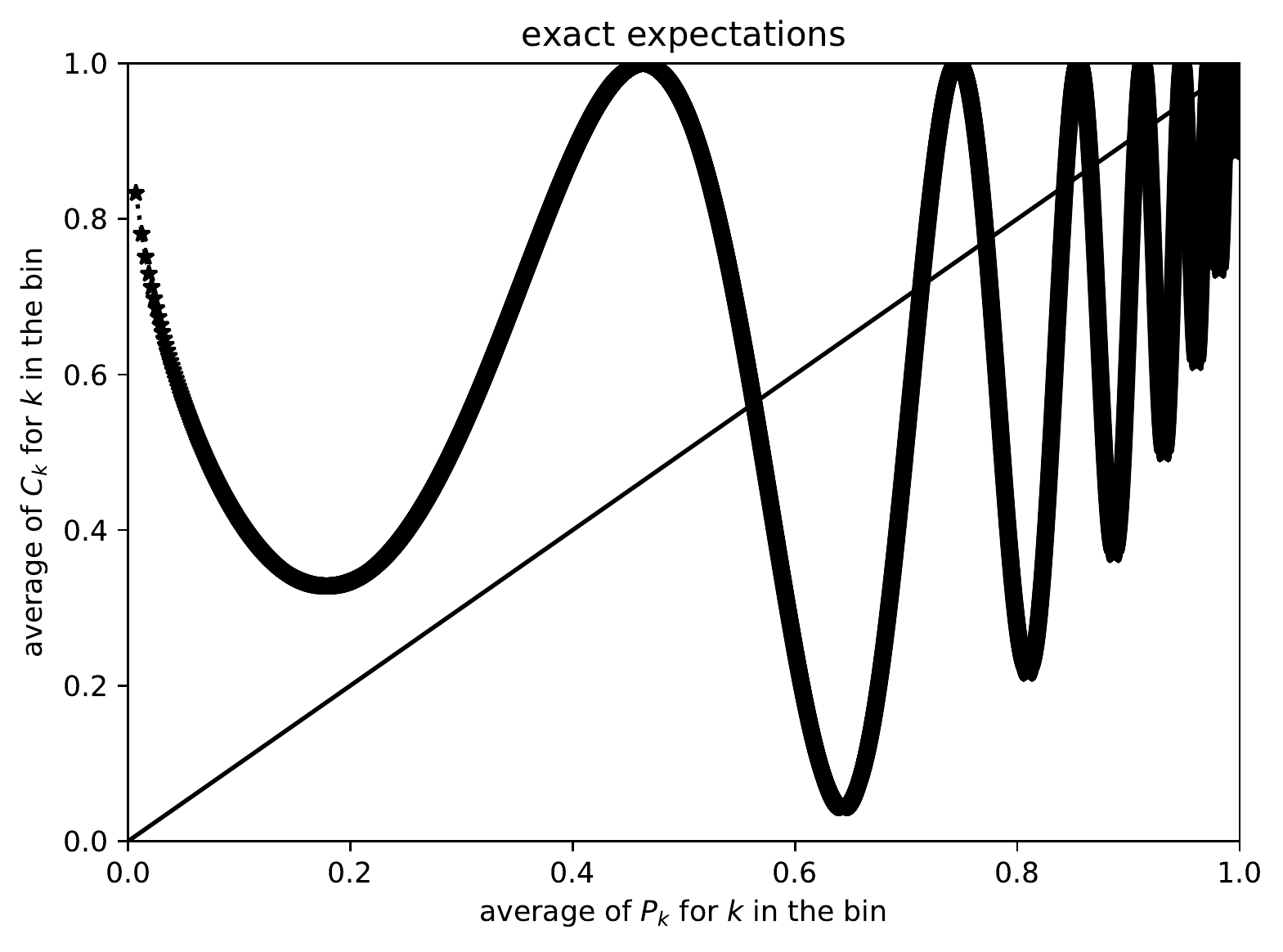}}

\end{centering}
\caption{$n =$ 10,000; $P_1$, $P_2$, \dots, $P_n$ are denser near 1}
\label{10000_1}
\end{figure}

\begin{figure}
\begin{centering}

\parbox{\imsize}{\includegraphics[width=\imsize]
                 {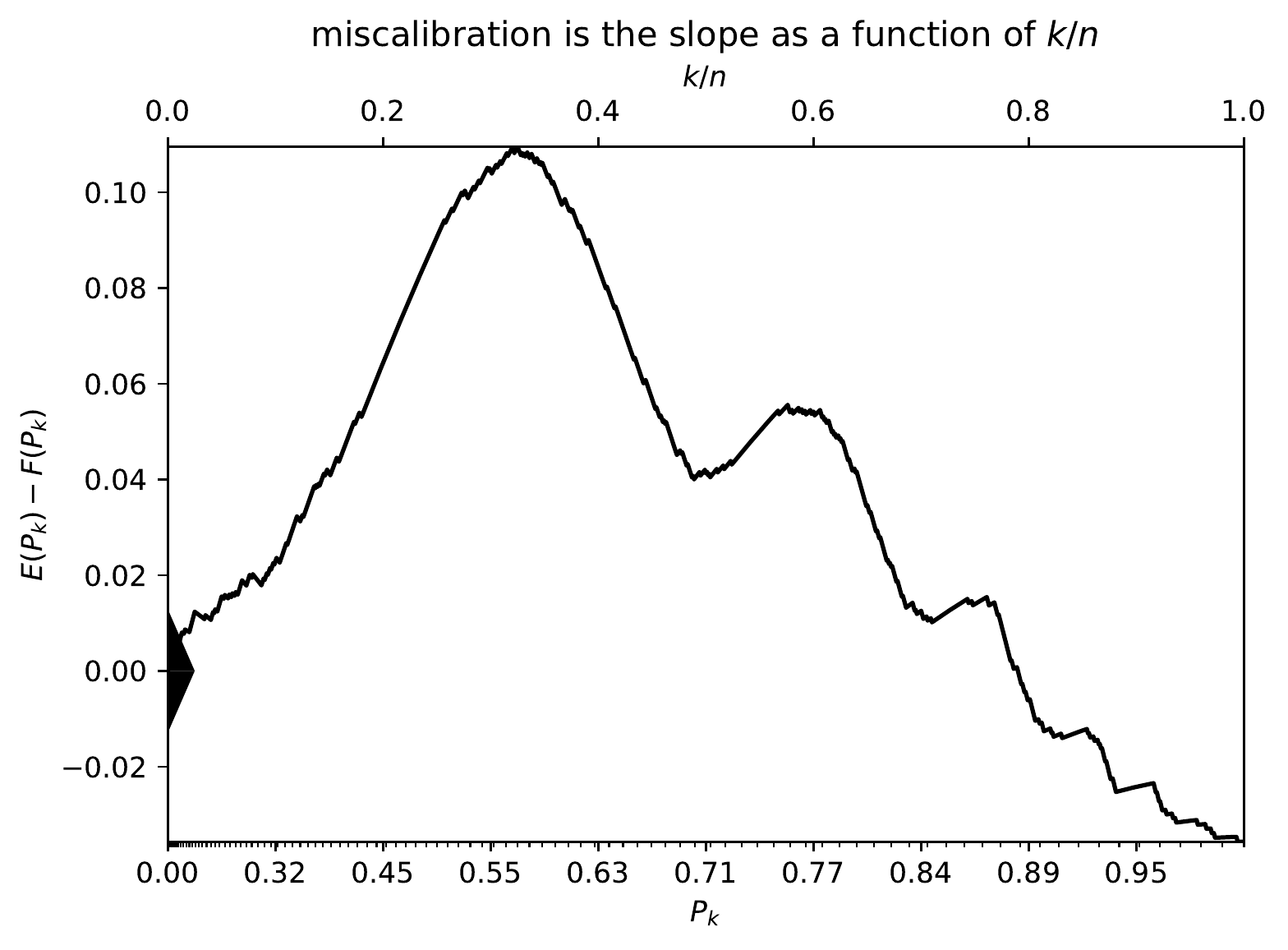}}
\quad\quad
\parbox{\imsize}{\includegraphics[width=\imsize]
                 {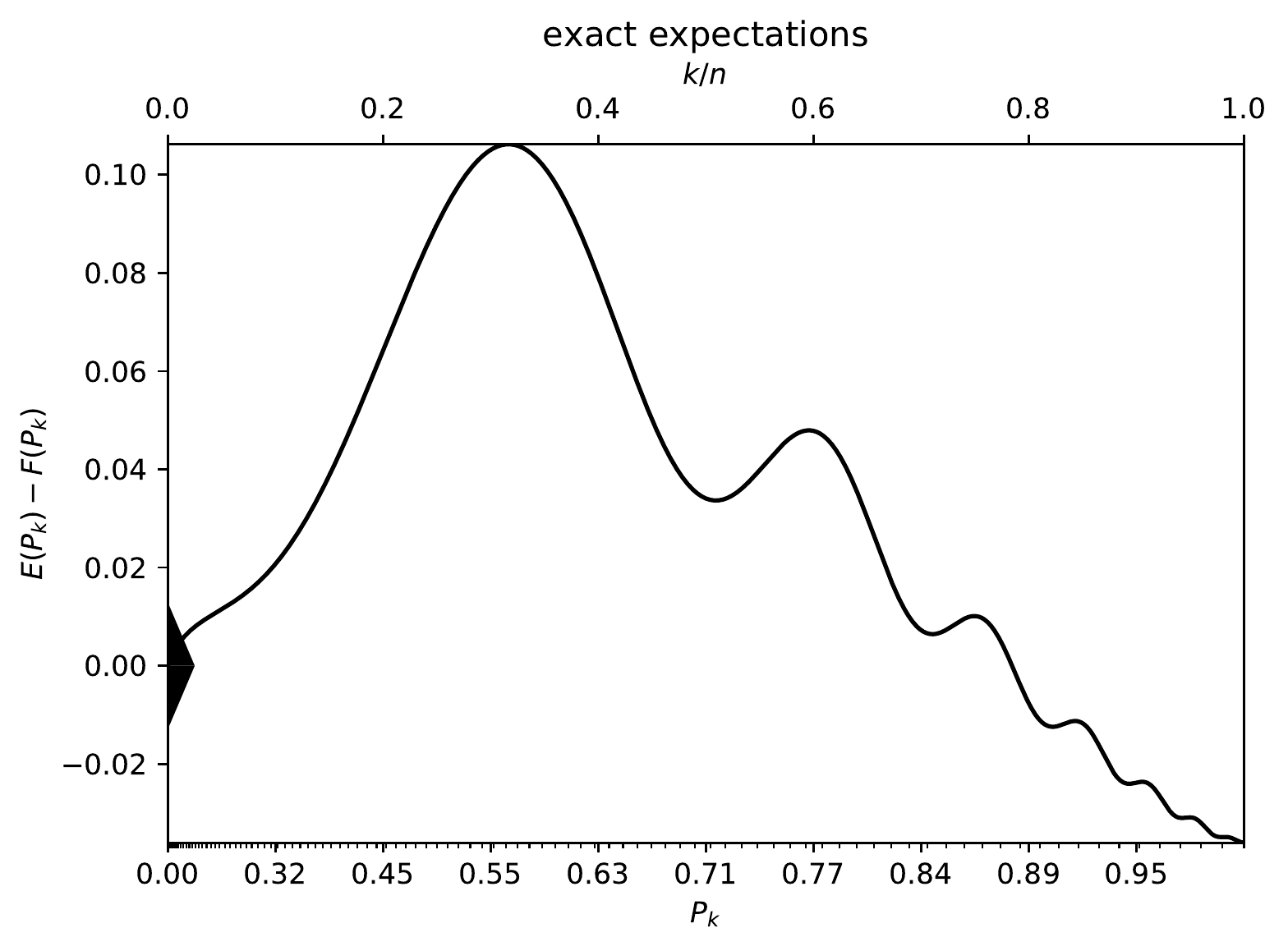}}

\vspace{\vertsep}

\parbox{\imsize}{\includegraphics[width=\imsize]
                 {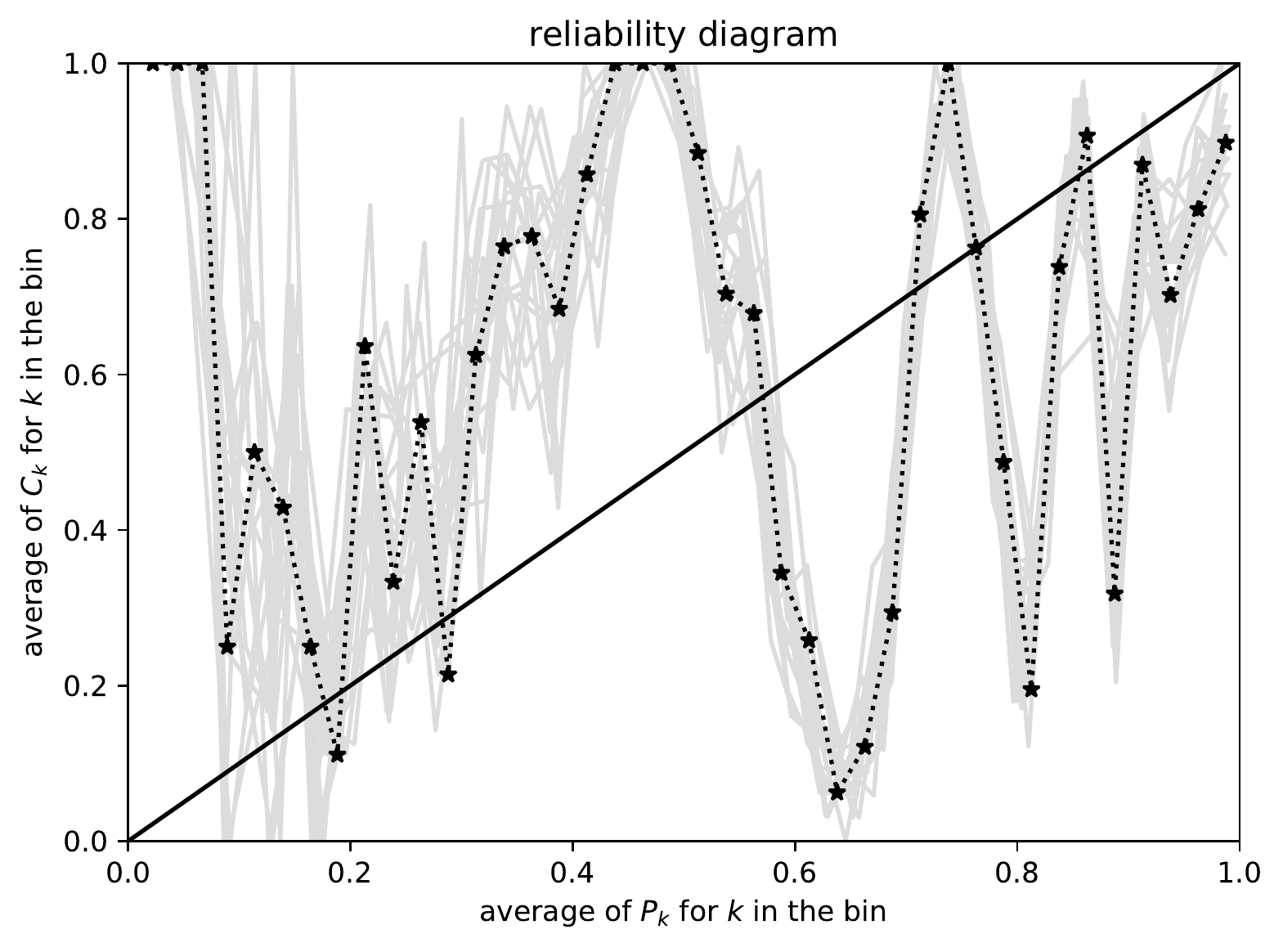}}
\quad\quad
\parbox{\imsize}{\includegraphics[width=\imsize]
                 {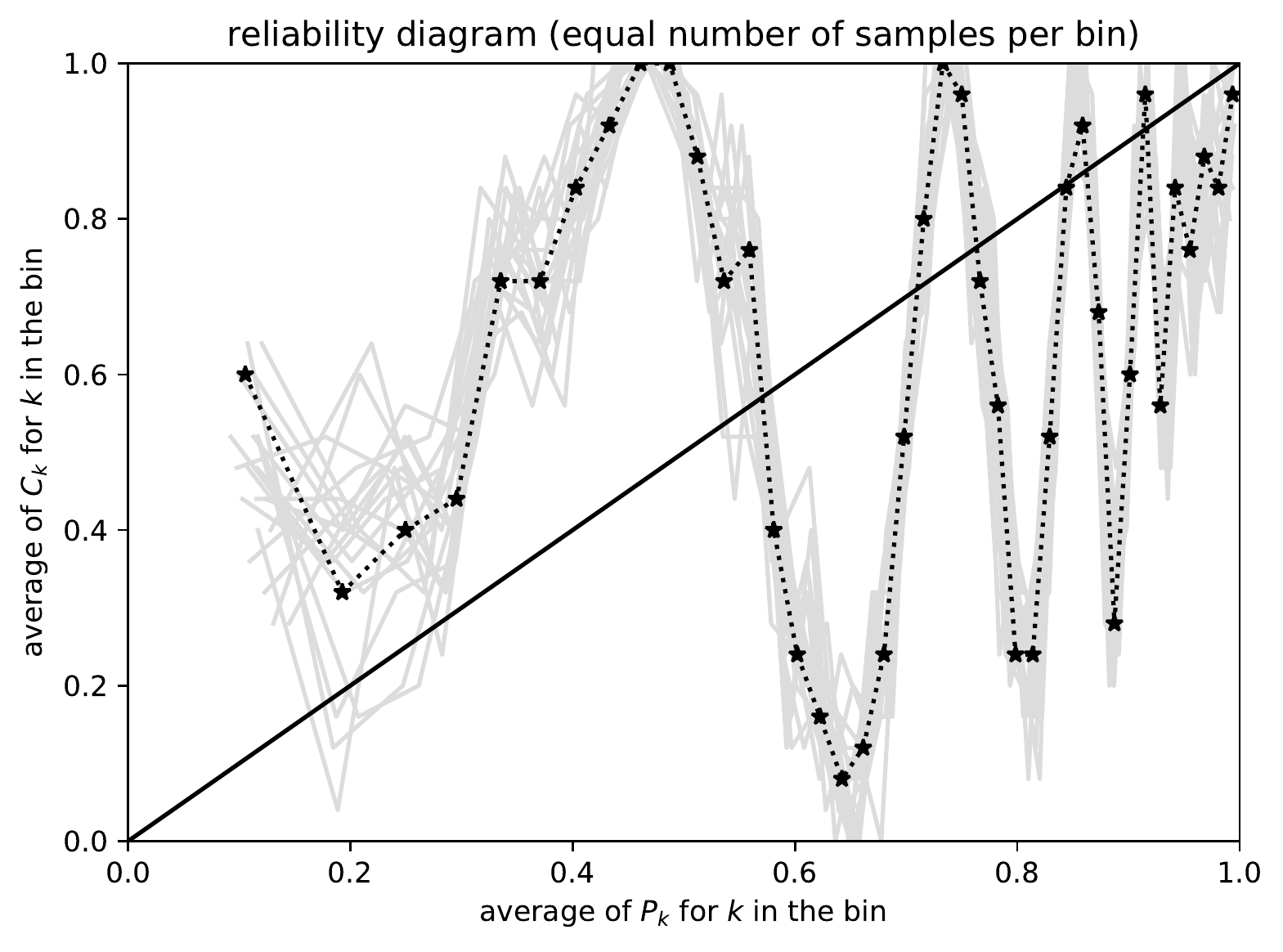}}

\vspace{\vertsep}

\parbox{\imsize}{\includegraphics[width=\imsize]
                 {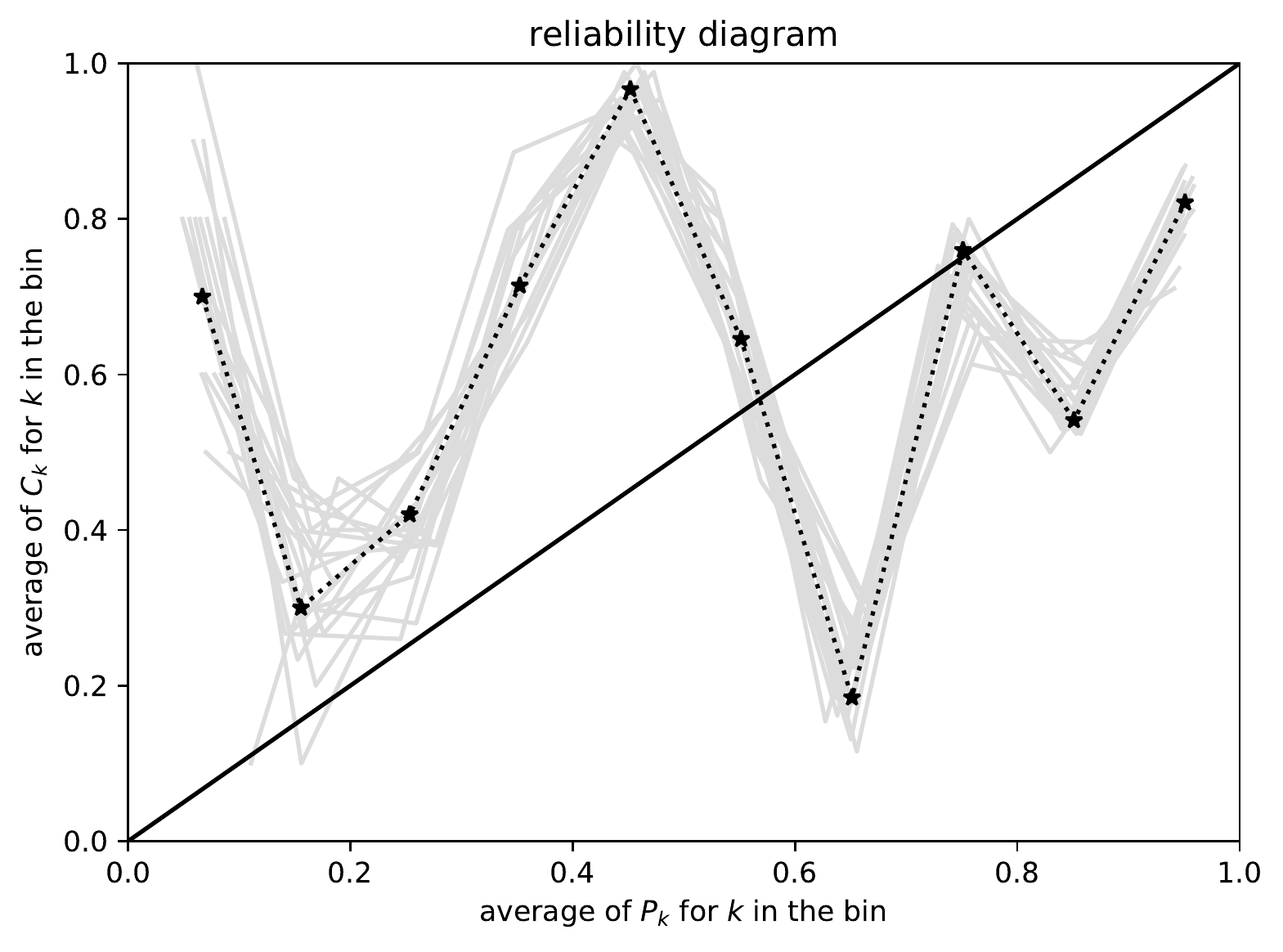}}
\quad\quad
\parbox{\imsize}{\includegraphics[width=\imsize]
                 {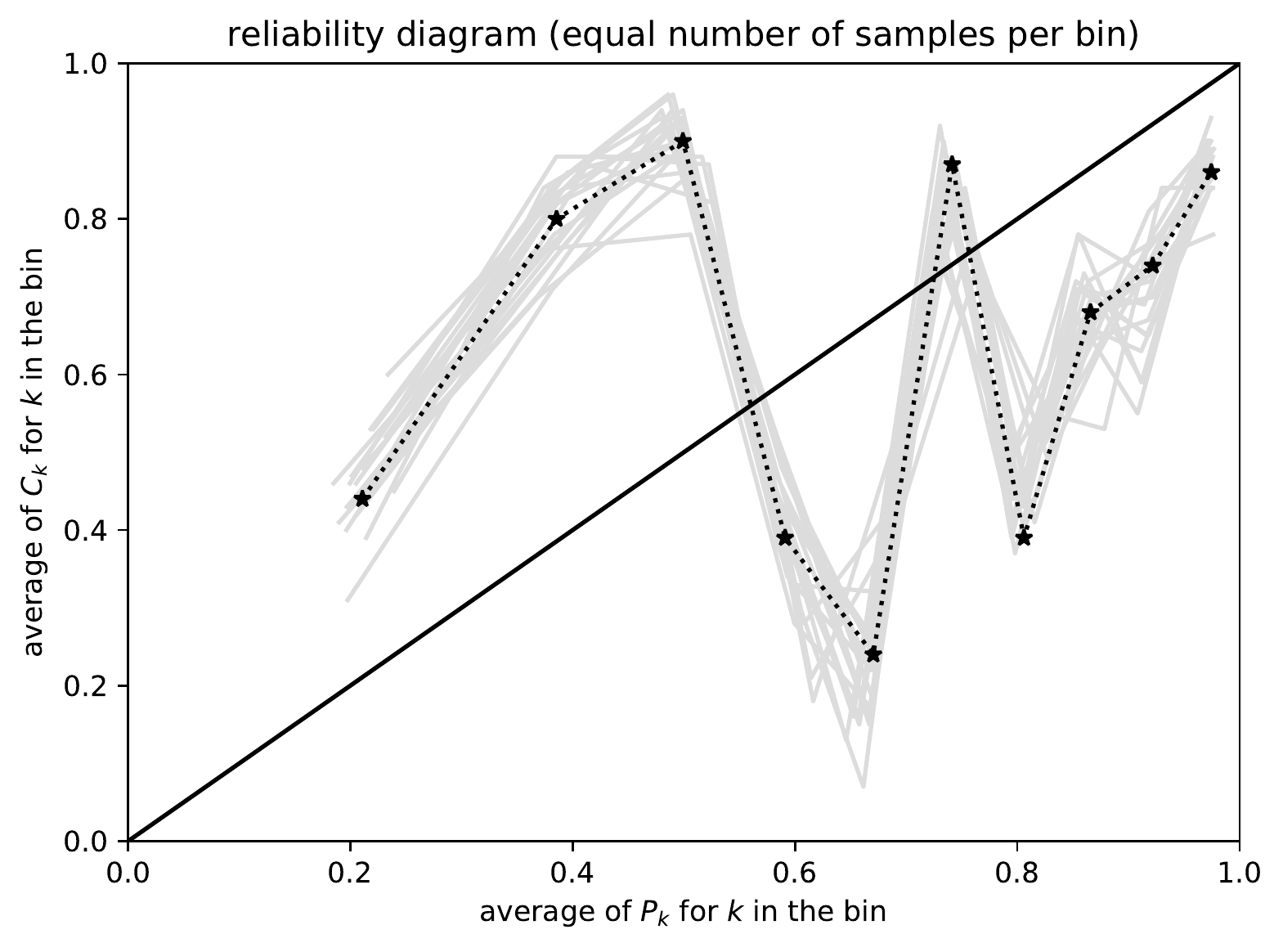}}

\vspace{\vertsep}

\parbox{\imsize}{\includegraphics[width=\imsize]
                 {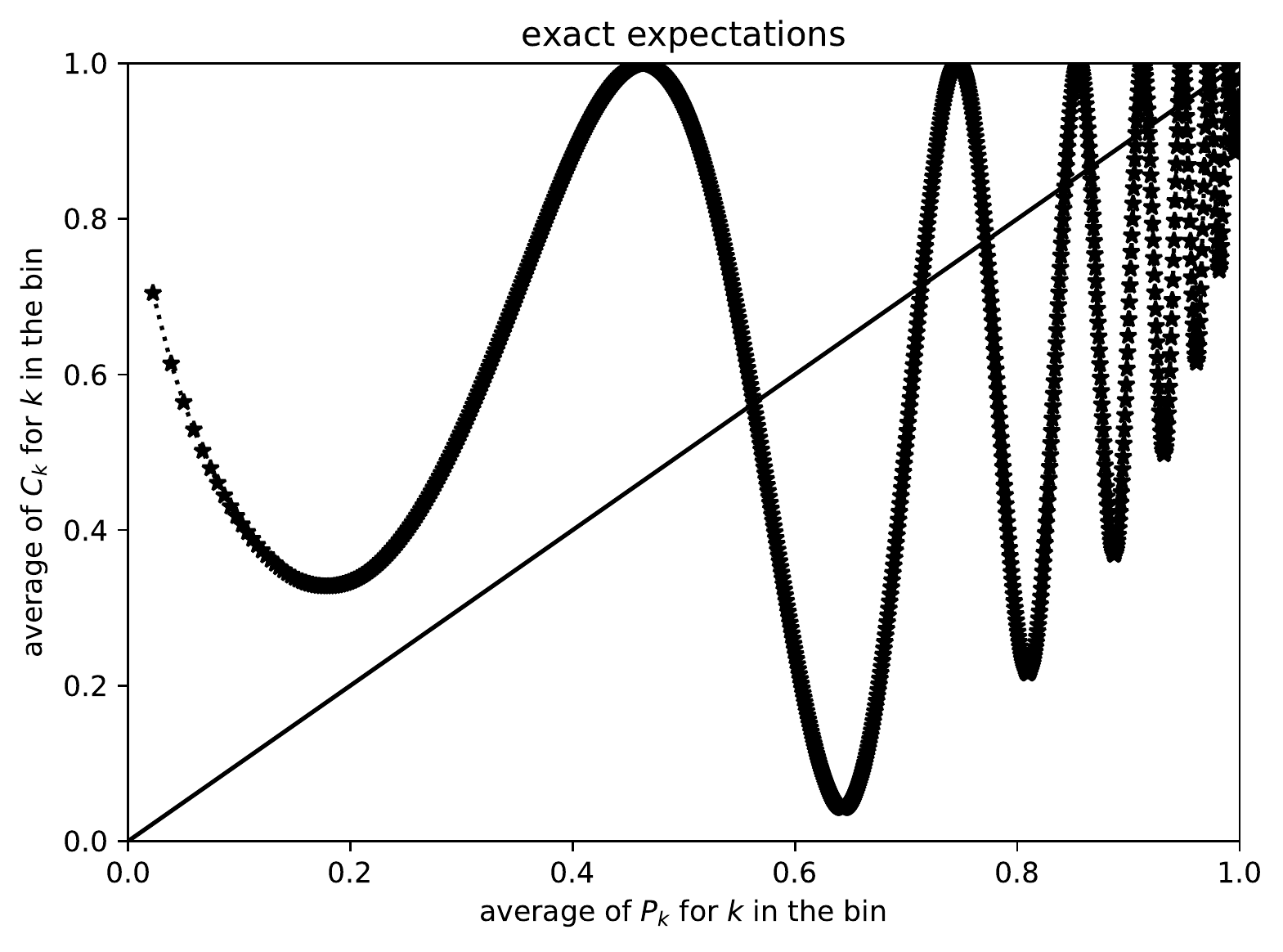}}

\end{centering}
\caption{$n =$ 1,000; $P_1$, $P_2$, \dots, $P_n$ are denser near 1}
\label{1000_1}
\end{figure}

\begin{figure}
\begin{centering}

\parbox{\imsize}{\includegraphics[width=\imsize]
                 {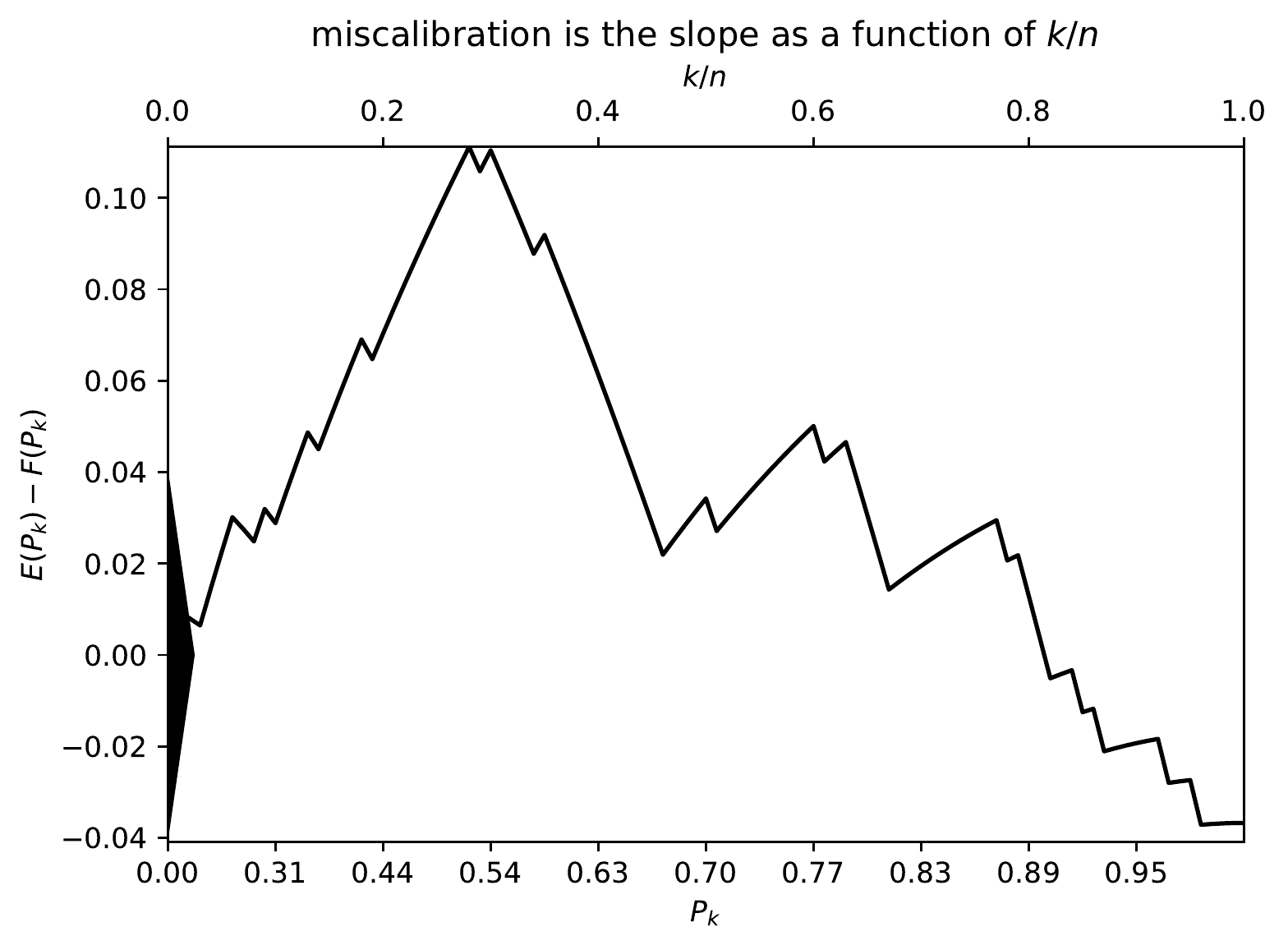}}
\quad\quad
\parbox{\imsize}{\includegraphics[width=\imsize]
                 {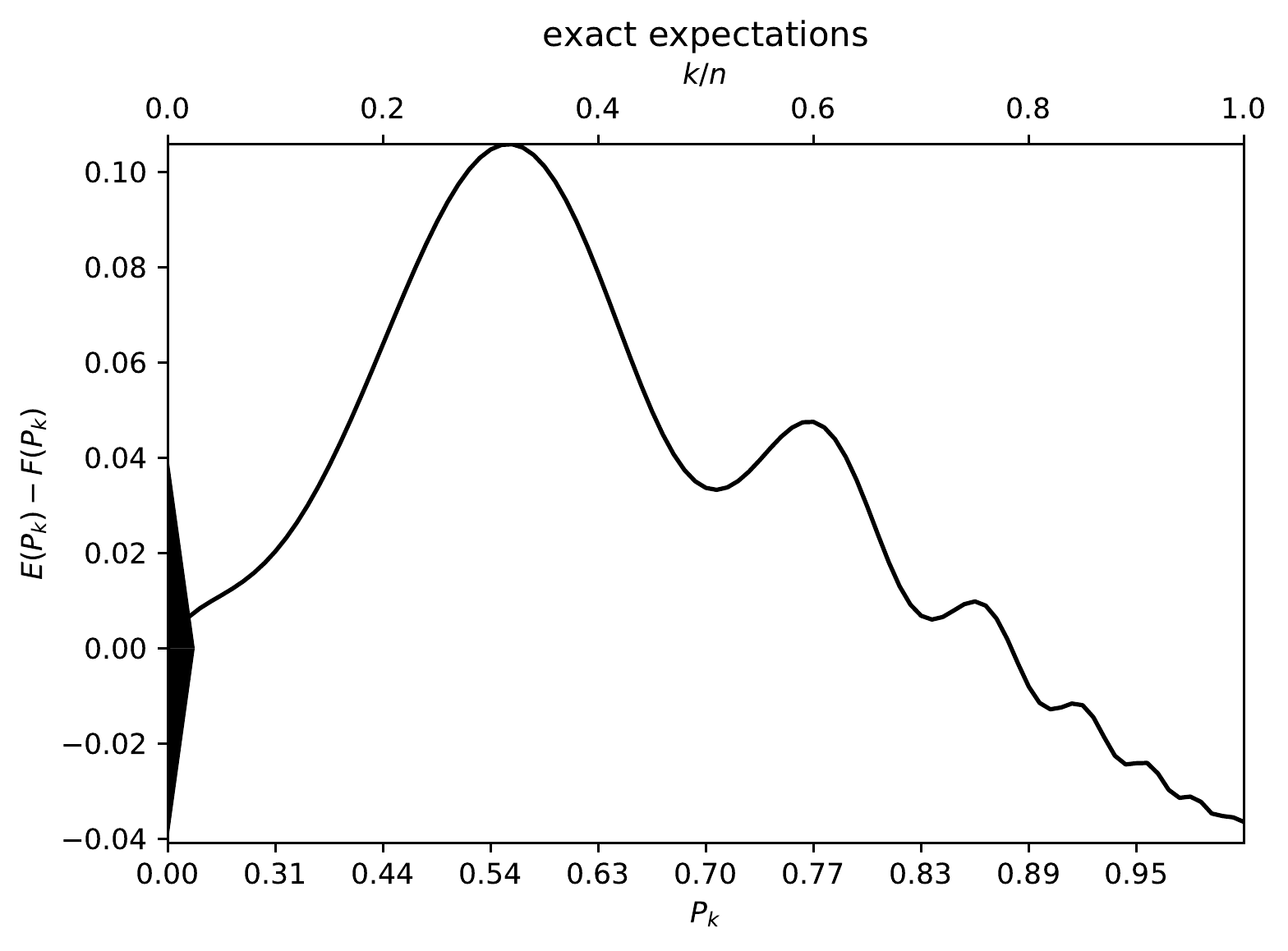}}

\vspace{\vertsep}

\parbox{\imsize}{\includegraphics[width=\imsize]
                 {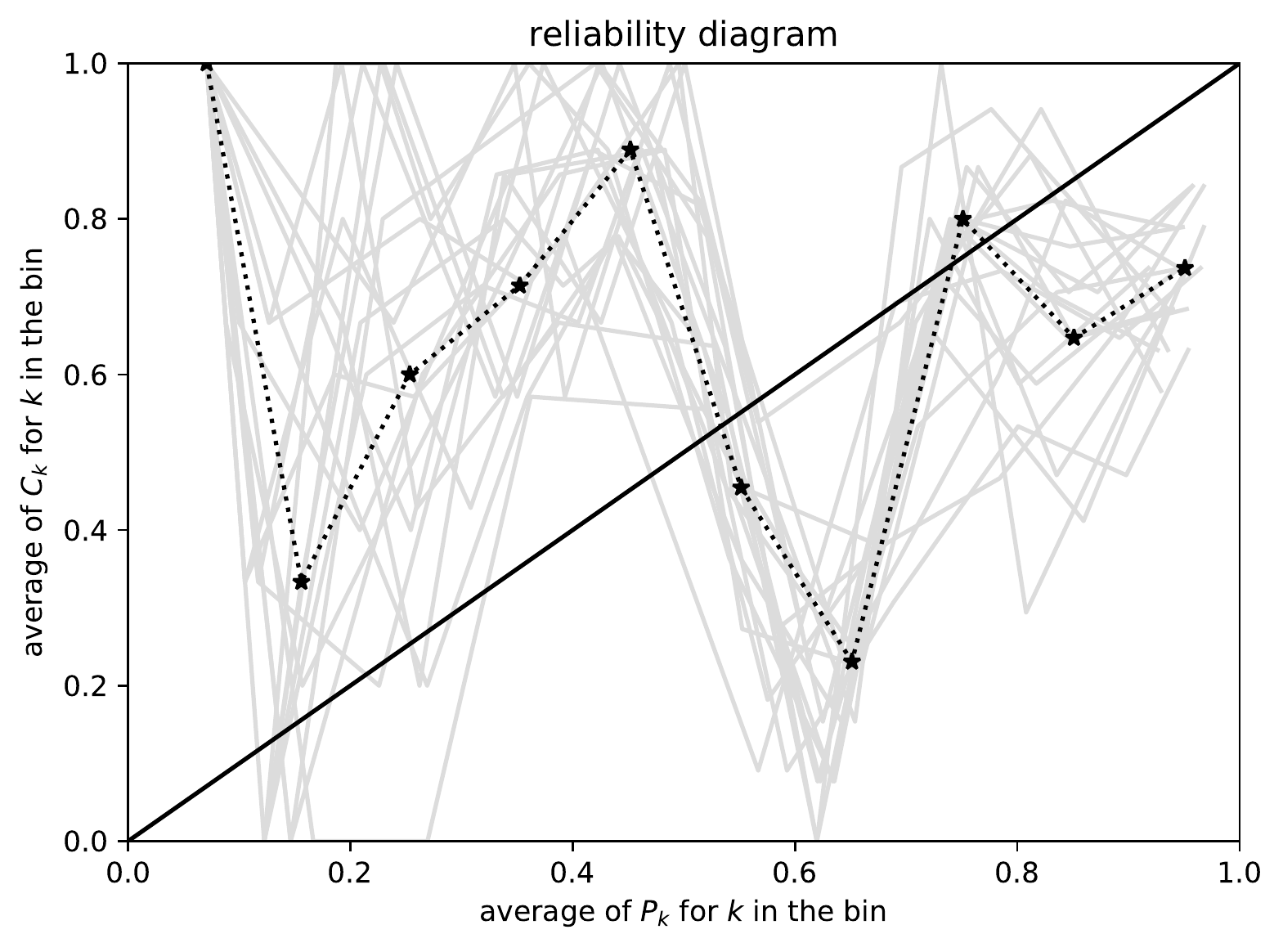}}
\quad\quad
\parbox{\imsize}{\includegraphics[width=\imsize]
                 {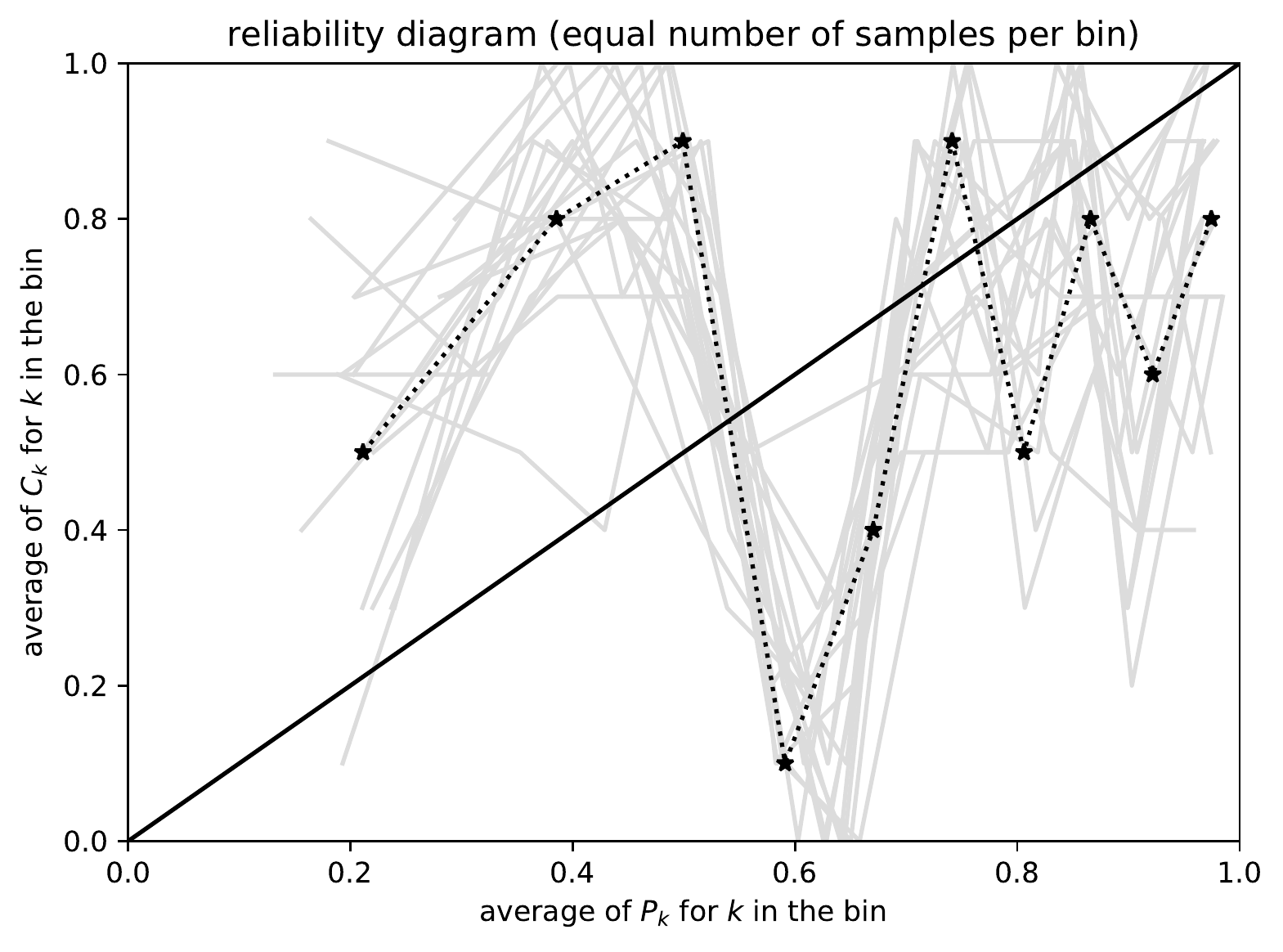}}

\vspace{\vertsep}

\parbox{\imsize}{\includegraphics[width=\imsize]
                 {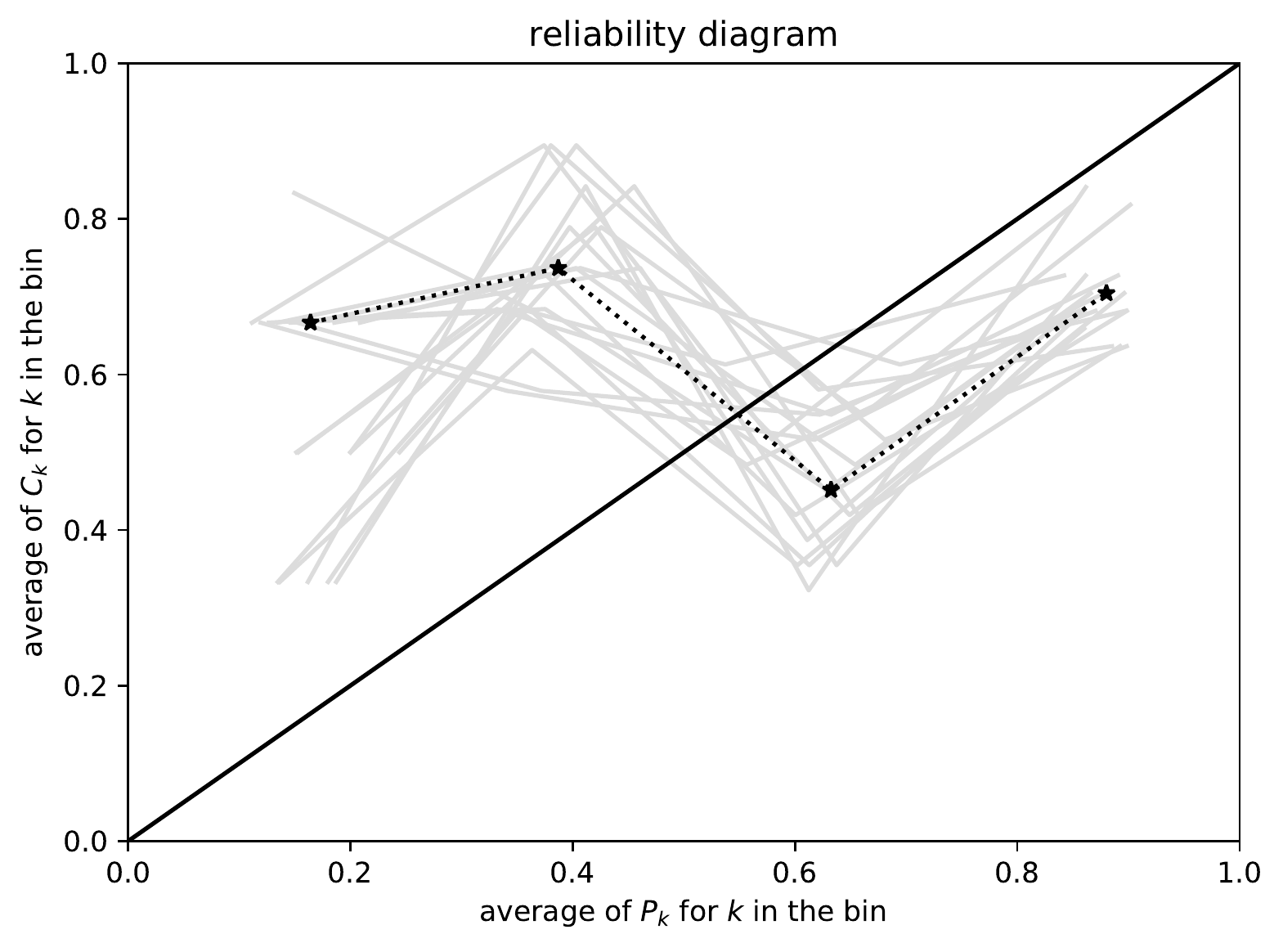}}
\quad\quad
\parbox{\imsize}{\includegraphics[width=\imsize]
                 {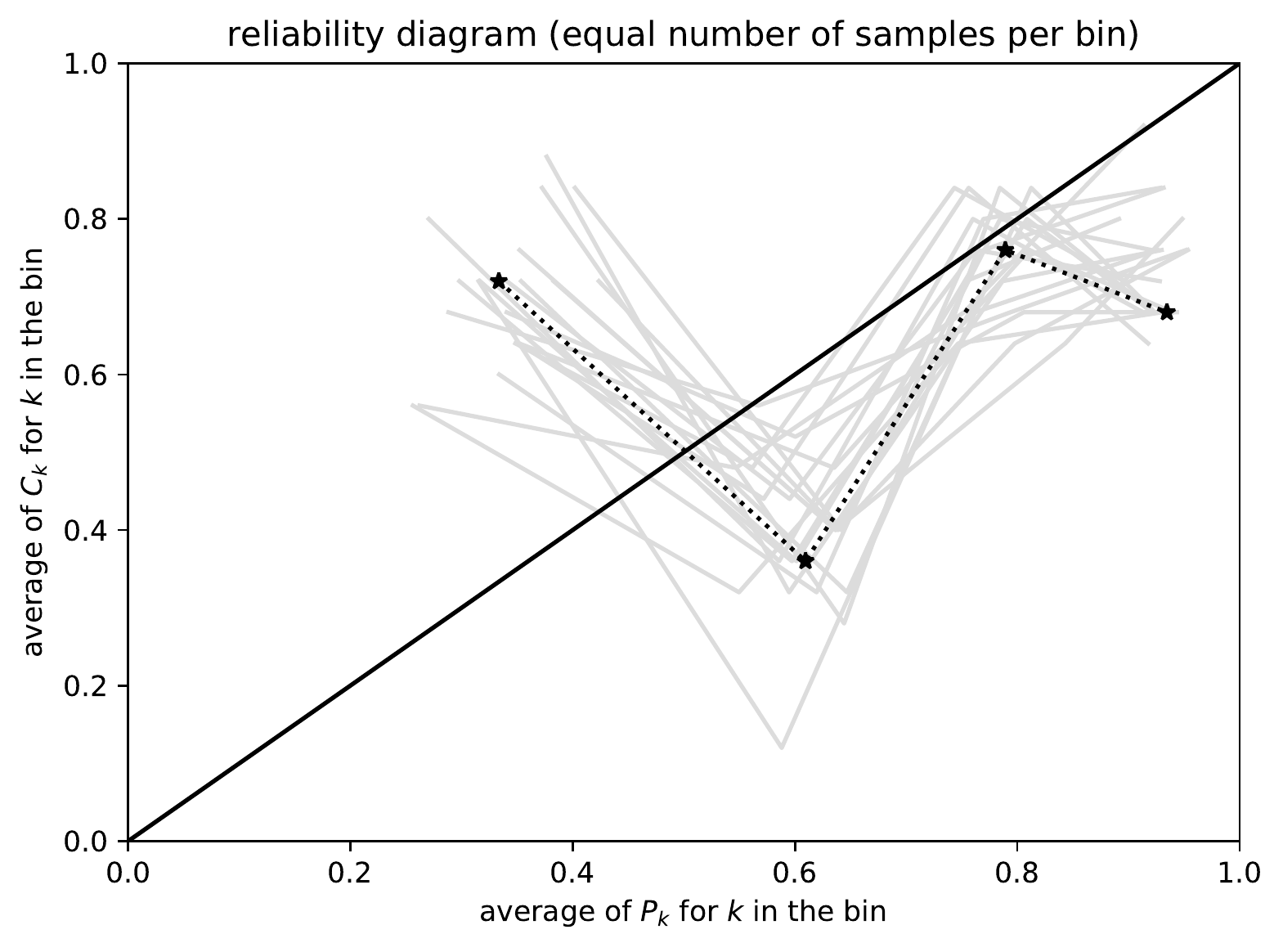}}

\vspace{\vertsep}

\parbox{\imsize}{\includegraphics[width=\imsize]
                 {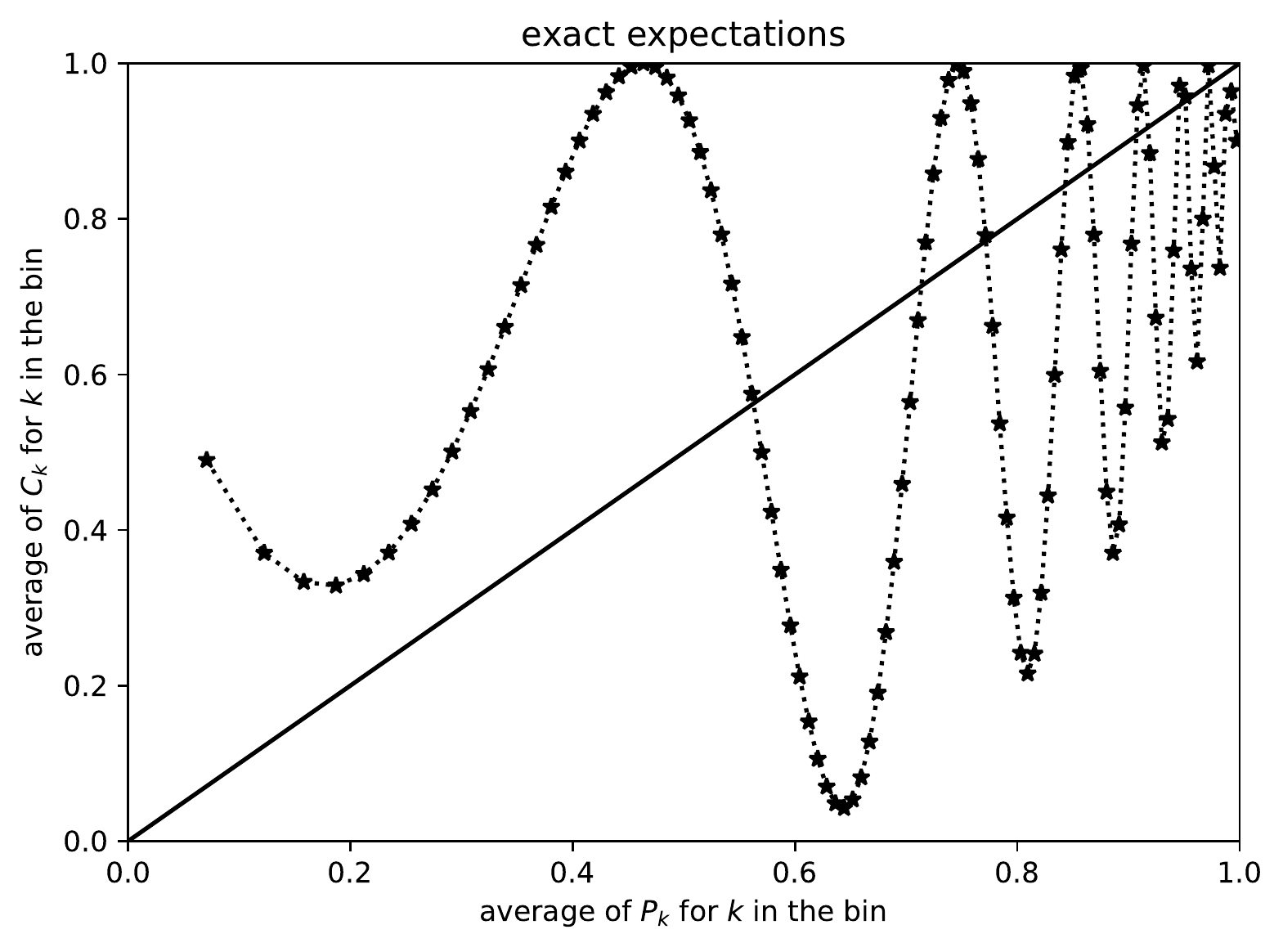}}

\end{centering}
\caption{$n =$ 100; $P_1$, $P_2$, \dots, $P_n$ are denser near 1}
\label{100_1}
\end{figure}

\section{Discussion and conclusion}
\label{conclusion}

Graphing the cumulative differences between observed and expected values
of the sorted predictions sidesteps having to make an arbitrary choice
of widths for bins or convolutional kernels --- a choice which is necessary
in the canonical reliability diagrams and their variants.
As reviewed above, the selection can also be made somewhat less arbitrary
by constructing multiple plots with varying numbers of bins 
or by adding estimates of errors with resampling schemes such as the bootstrap.
Choosing between the cumulative plot
and the more complicated conventional reliability diagrams
may be merely a matter of convenience and personal preference.
The plot of cumulative differences encodes miscalibration directly
as the slope of secant lines for the graph, and such slope is easy to perceive
independent of any irrelevant constant offset of a secant line;
the graph of cumulative differences very directly enables
detection and quantification of miscalibration,
along with identification of the ranges of miscalibrated probabilities.
The cumulative differences estimate the distribution
of miscalibration fully nonparametrically, letting the data samples
speak for themselves (or nearly for themselves --- the triangle at the origin
helps convey the scale of a driftless random walk's
expected random fluctuations).
As seen in the figures, the graph of cumulative differences automatically
adapts its resolving power to the distribution of miscalibration and sampling,
not imposing any artificial grid of bins or set-width smoothing kernel,
unlike the conventional reliability diagrams and calibration plots.

\section*{Acknowledgements}

We would like to thank Tiffany Cai, Joaquin Candela, Kenneth Hung, Mike Rabbat,
and Adina Williams.

\appendix
\section{Appendix: Random walks}

\begin{figure}
\begin{centering}

\parbox{\imsize}{\includegraphics[width=\imsize]
                 {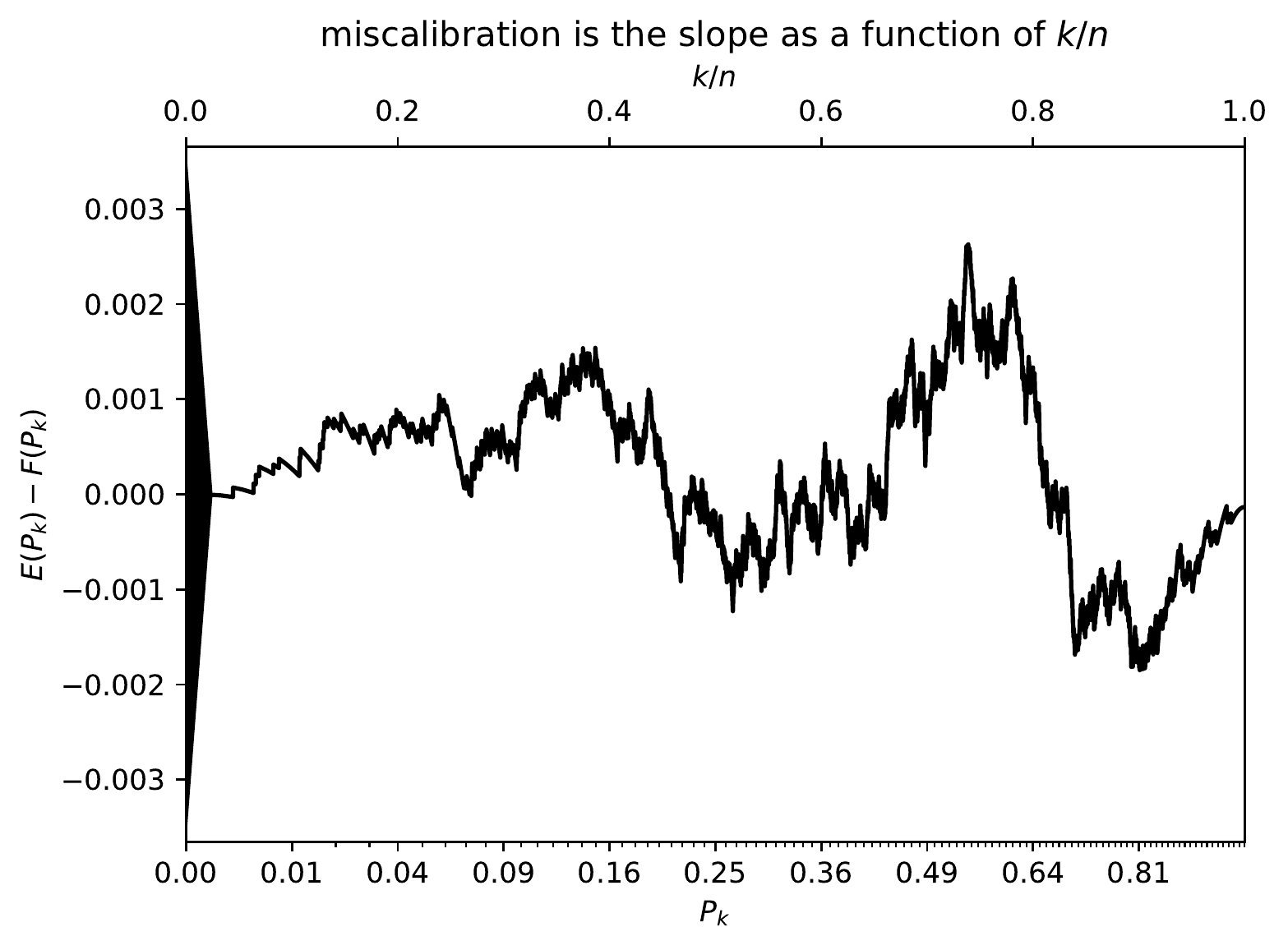}}
\quad\quad
\parbox{\imsize}{\includegraphics[width=\imsize]
                 {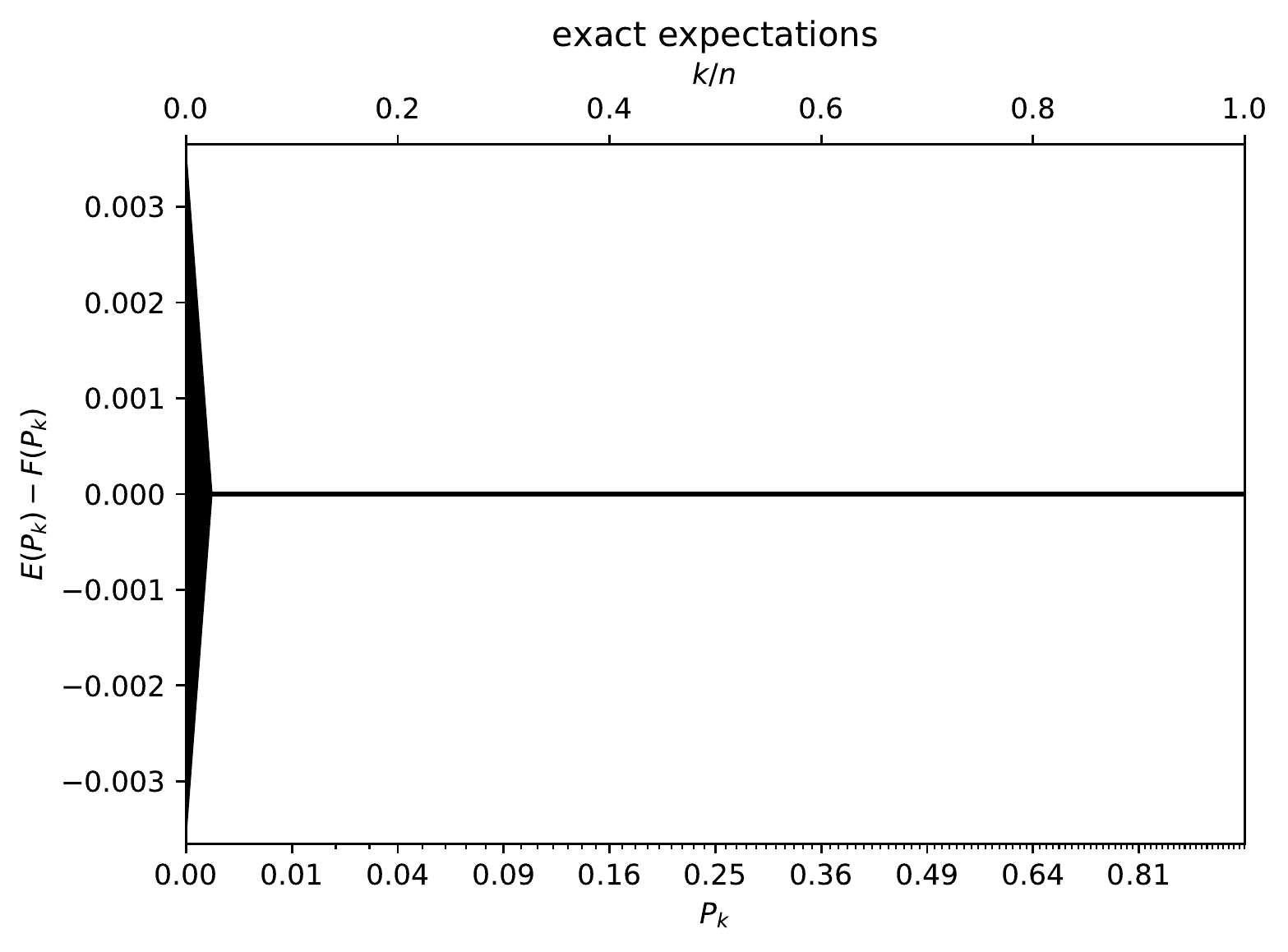}}

\vspace{\vertsep}

\parbox{\imsize}{\includegraphics[width=\imsize]
                 {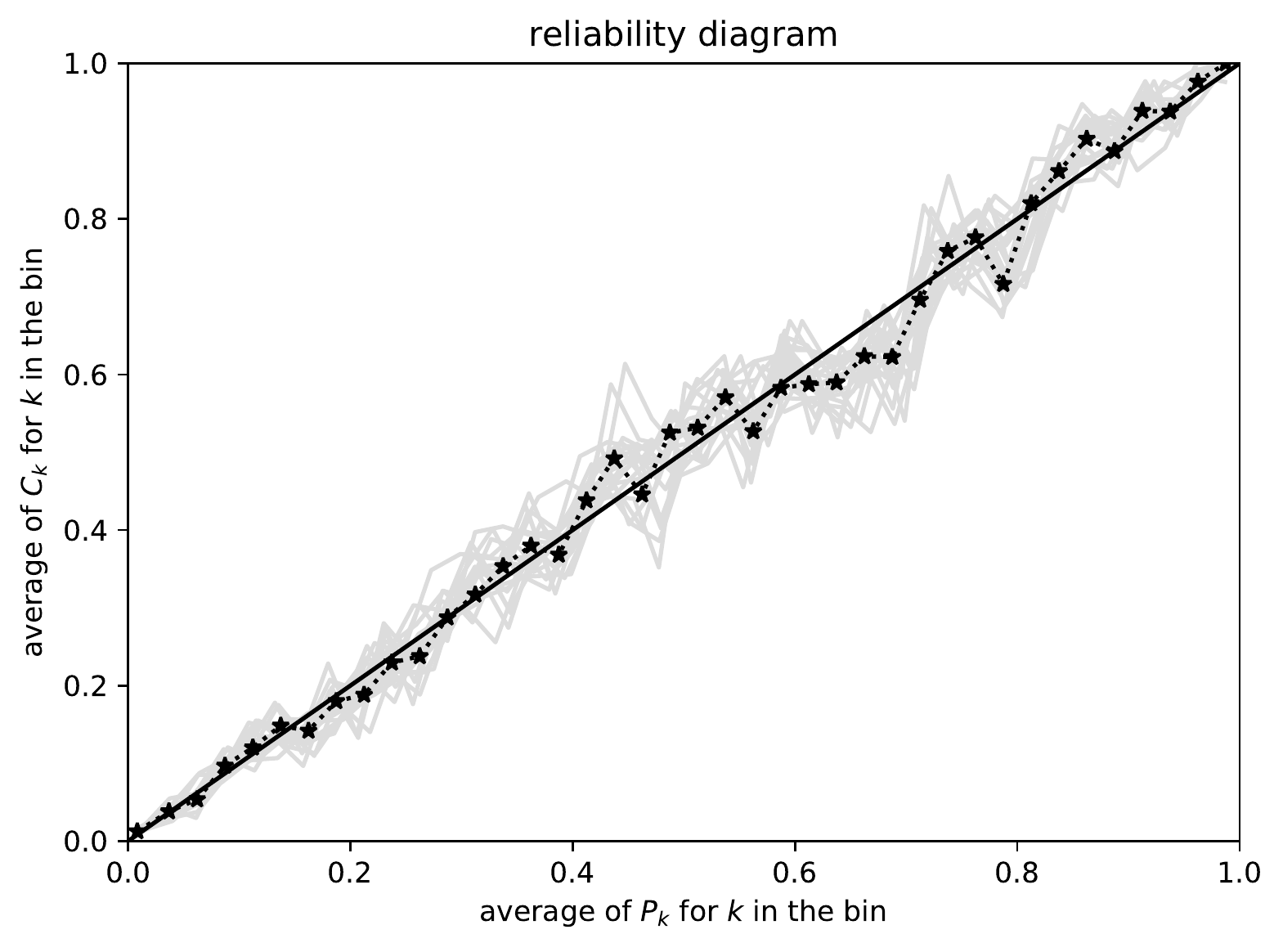}}
\quad\quad
\parbox{\imsize}{\includegraphics[width=\imsize]
                 {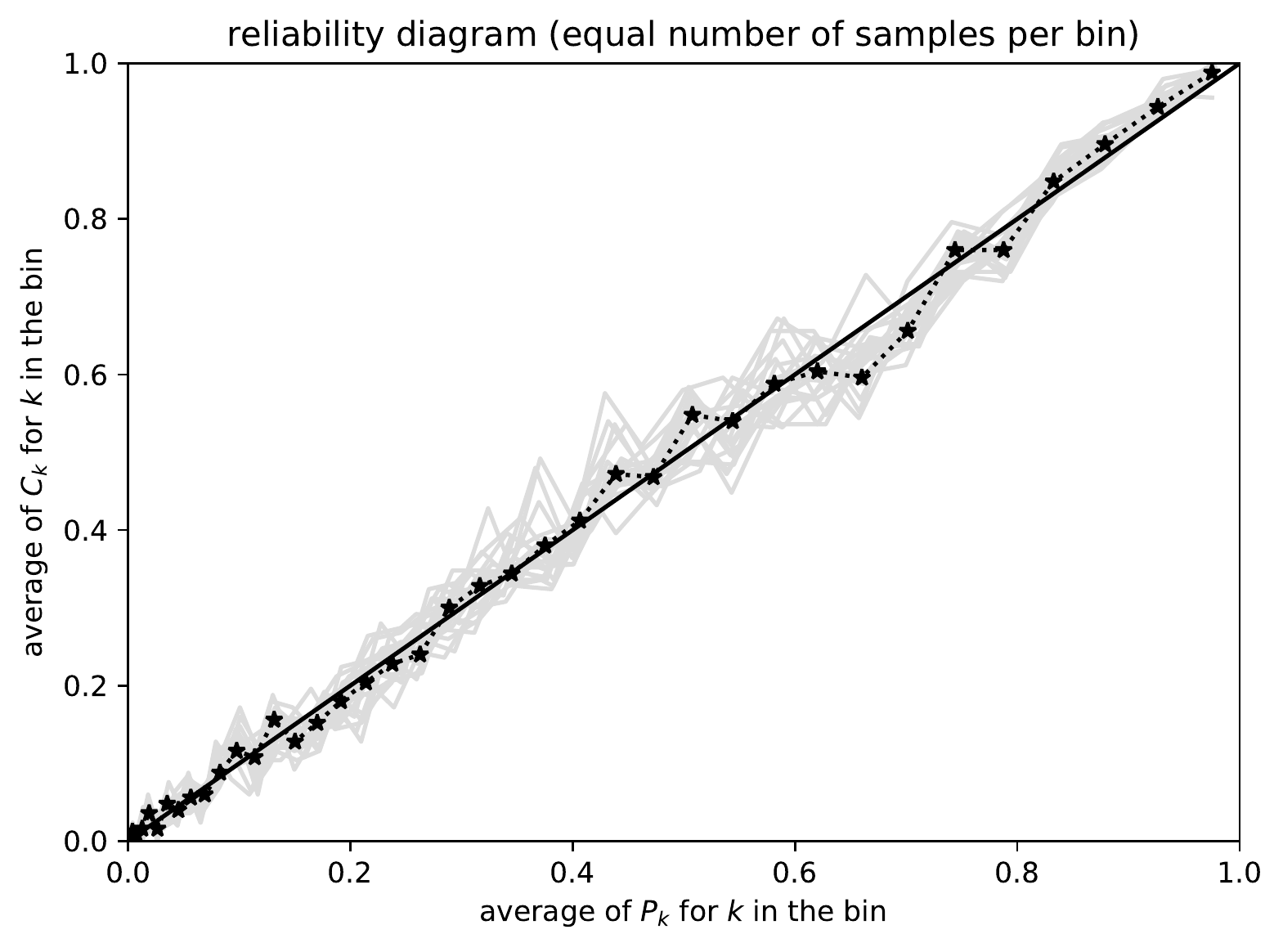}}

\vspace{\vertsep}

\parbox{\imsize}{\includegraphics[width=\imsize]
                 {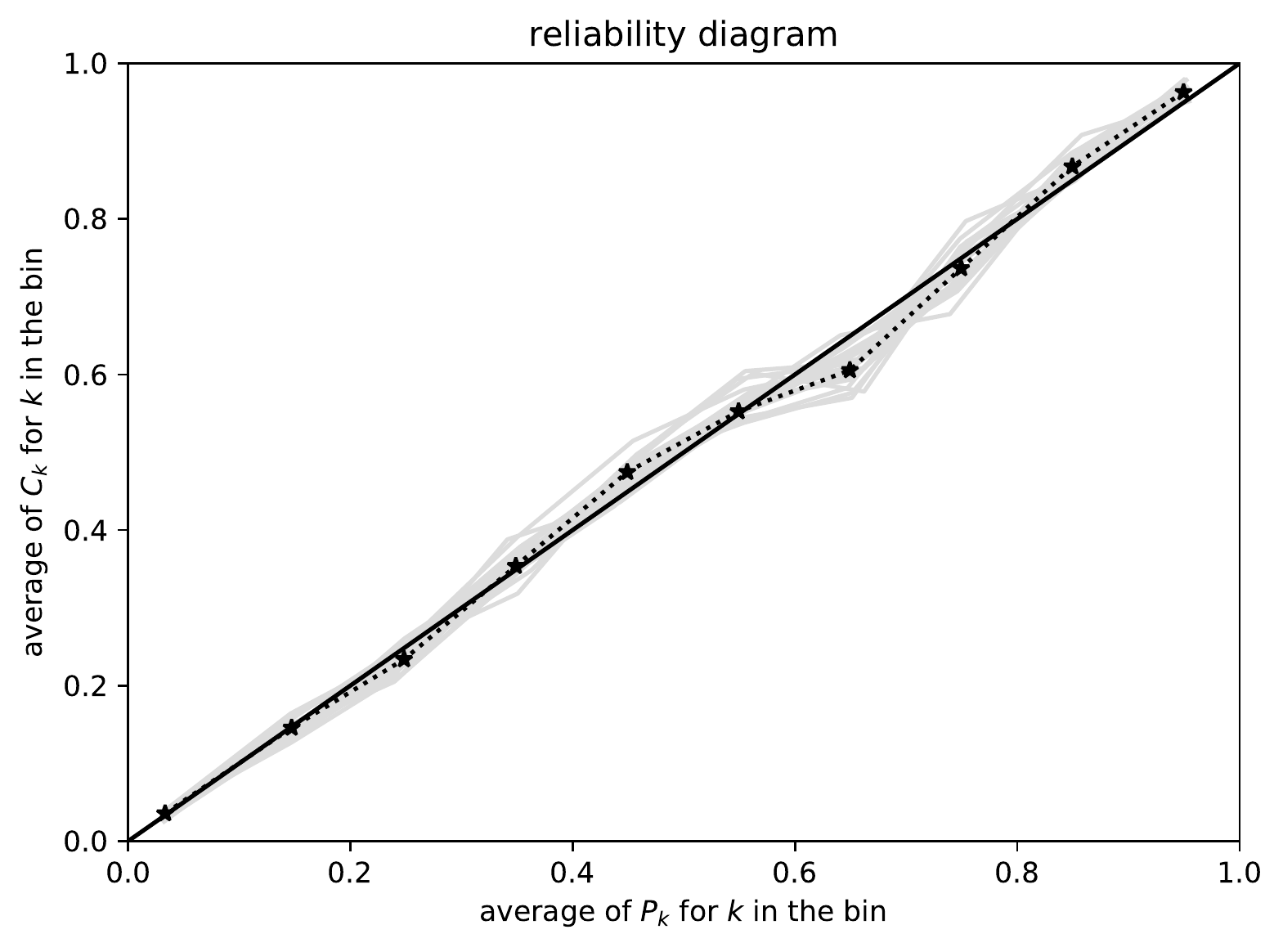}}
\quad\quad
\parbox{\imsize}{\includegraphics[width=\imsize]
                 {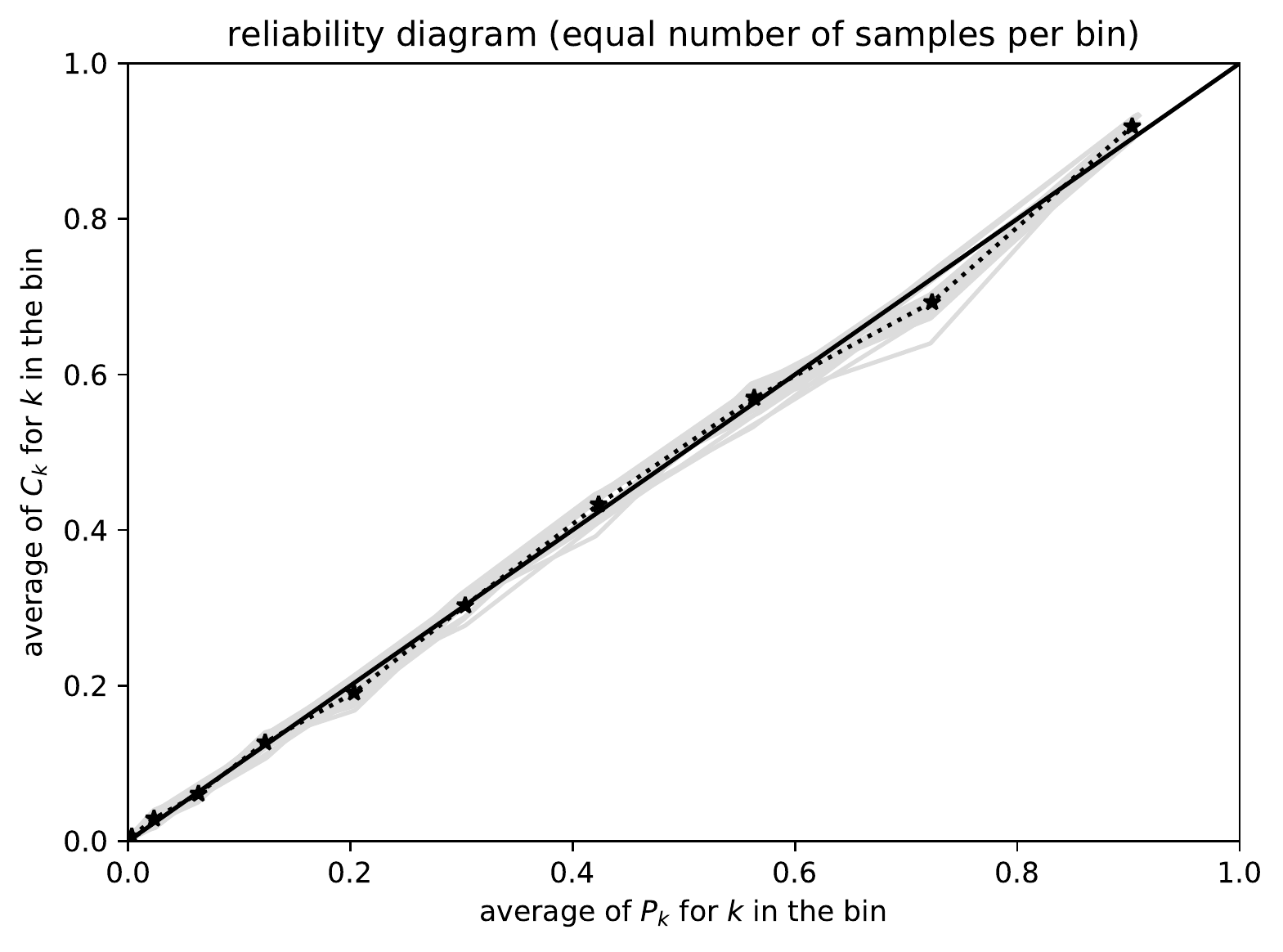}}

\vspace{\vertsep}

\parbox{\imsize}{\includegraphics[width=\imsize]
                 {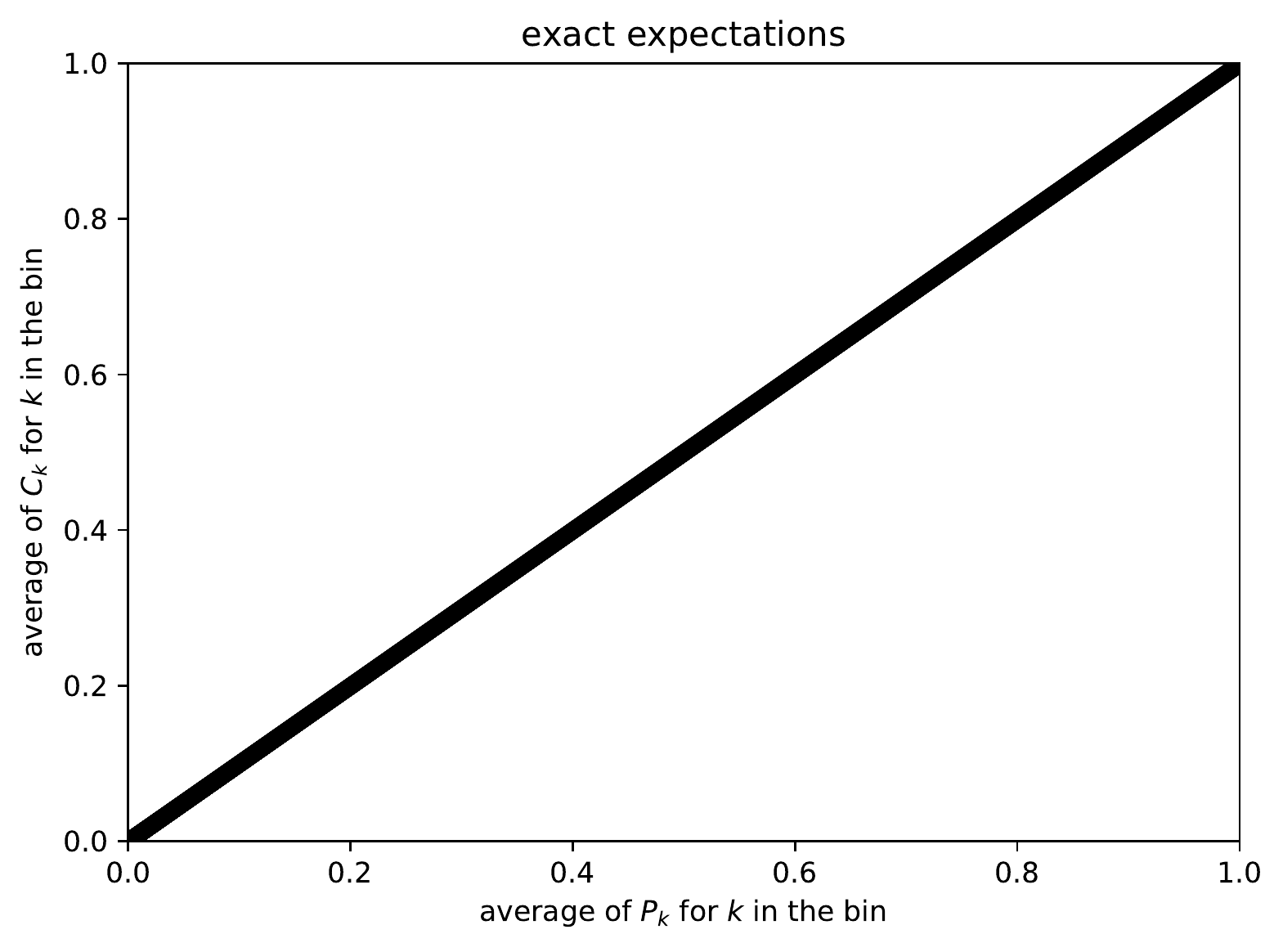}}

\end{centering}
\caption{$n =$ 10,000; $P_1$, $P_2$, \dots, $P_n$ are denser near 0}
\label{10000_00}
\end{figure}

\begin{figure}
\begin{centering}

\parbox{\imsize}{\includegraphics[width=\imsize]
                 {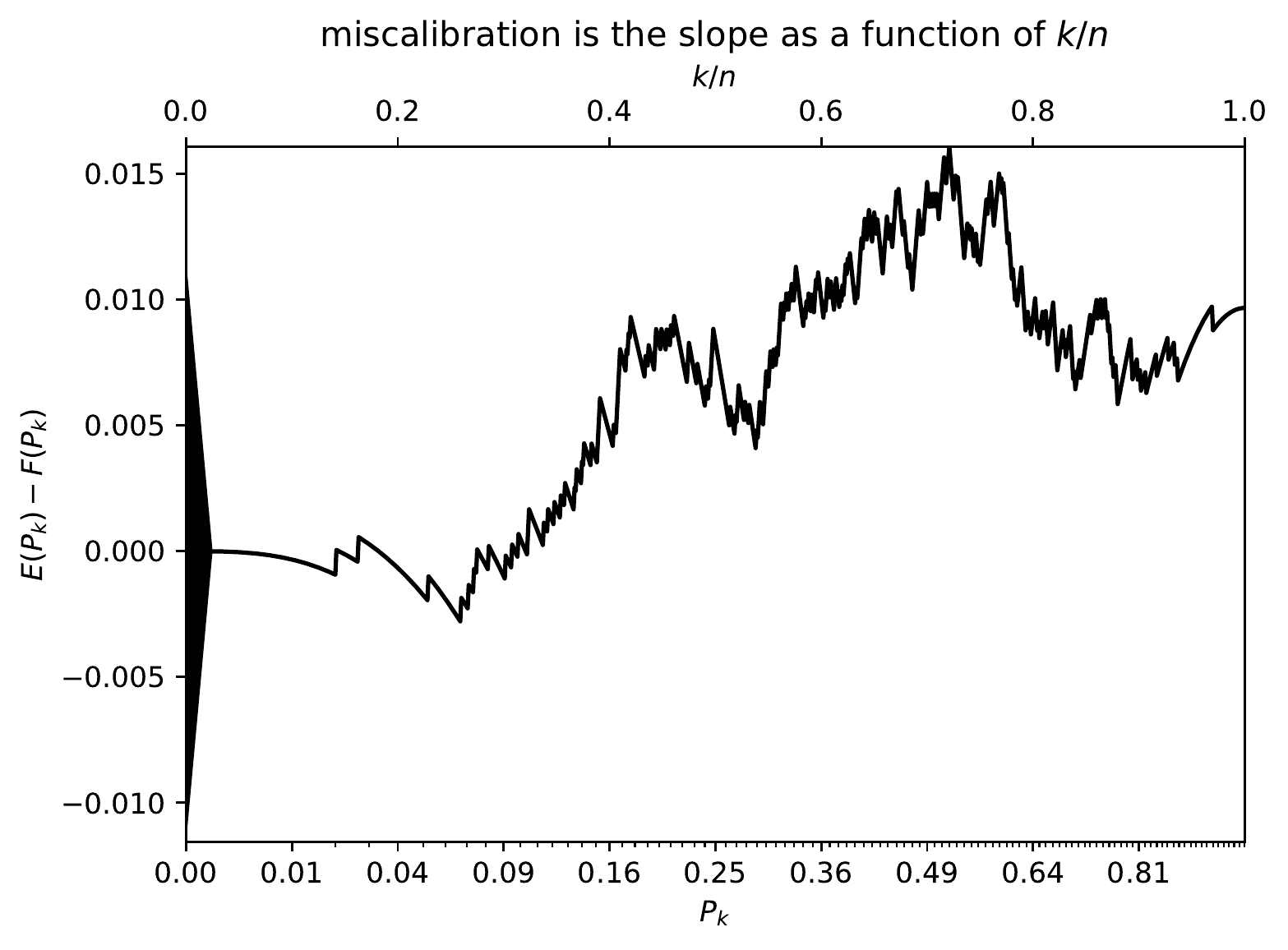}}
\quad\quad
\parbox{\imsize}{\includegraphics[width=\imsize]
                 {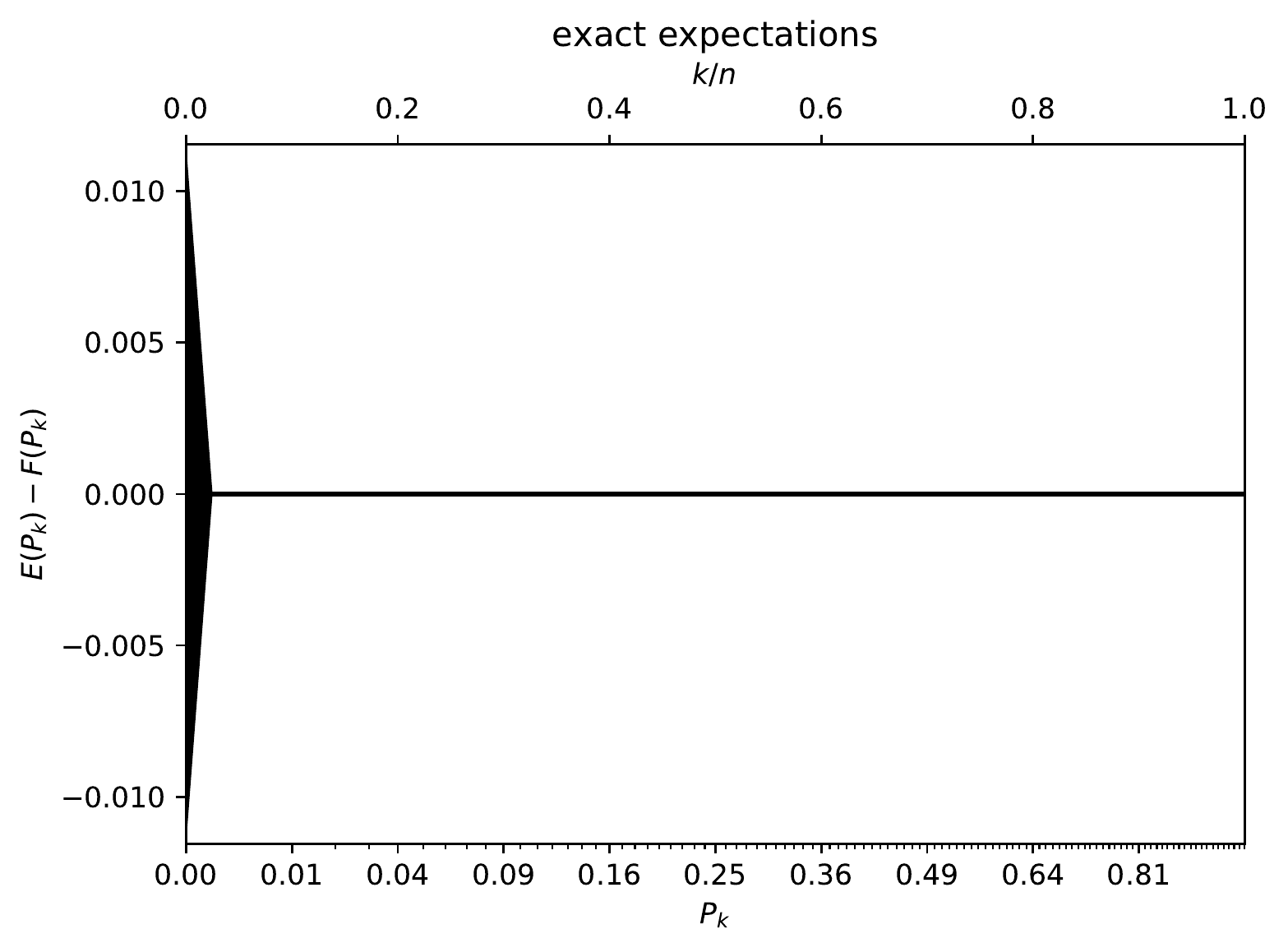}}

\vspace{\vertsep}

\parbox{\imsize}{\includegraphics[width=\imsize]
                 {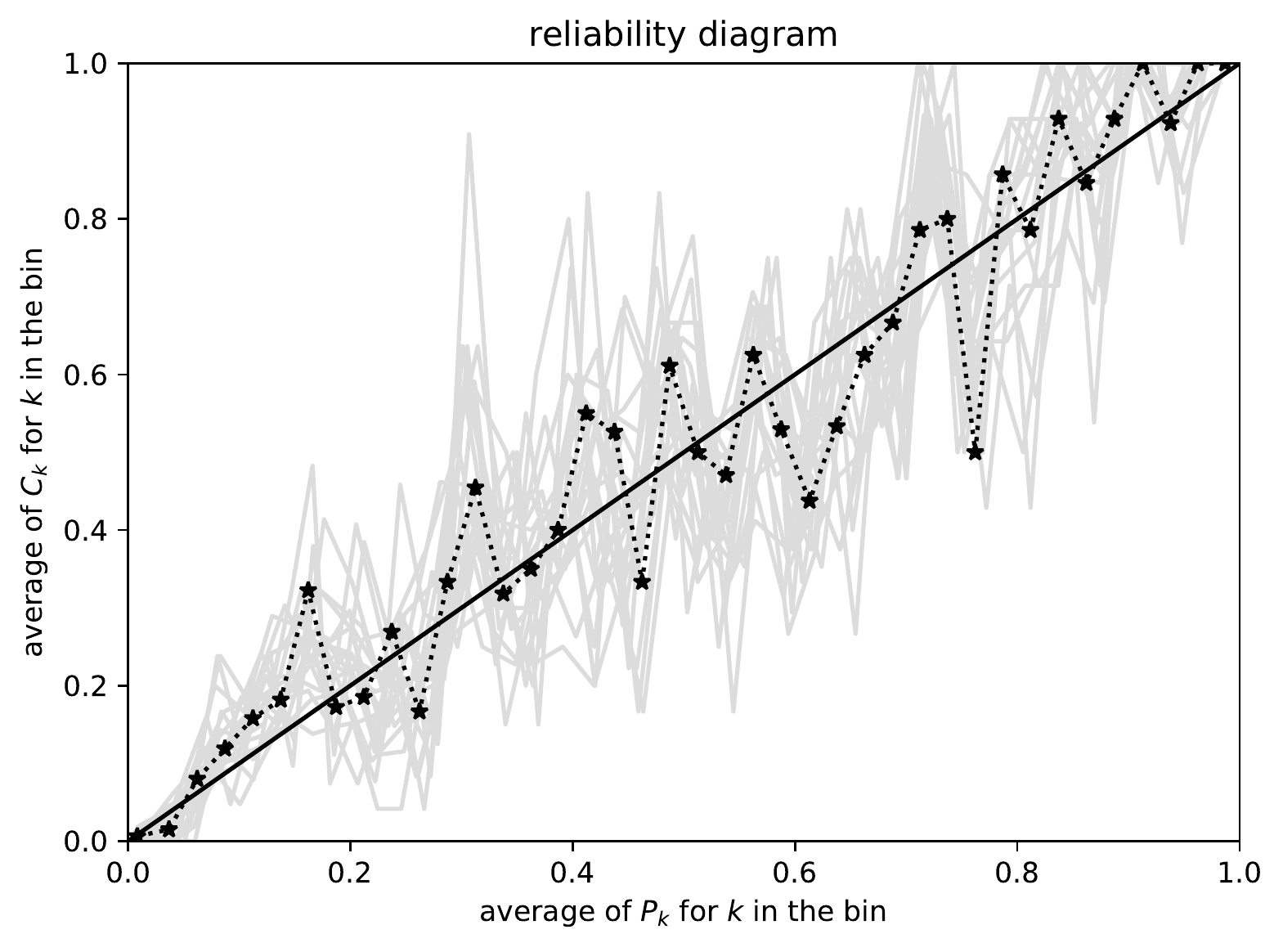}}
\quad\quad
\parbox{\imsize}{\includegraphics[width=\imsize]
                 {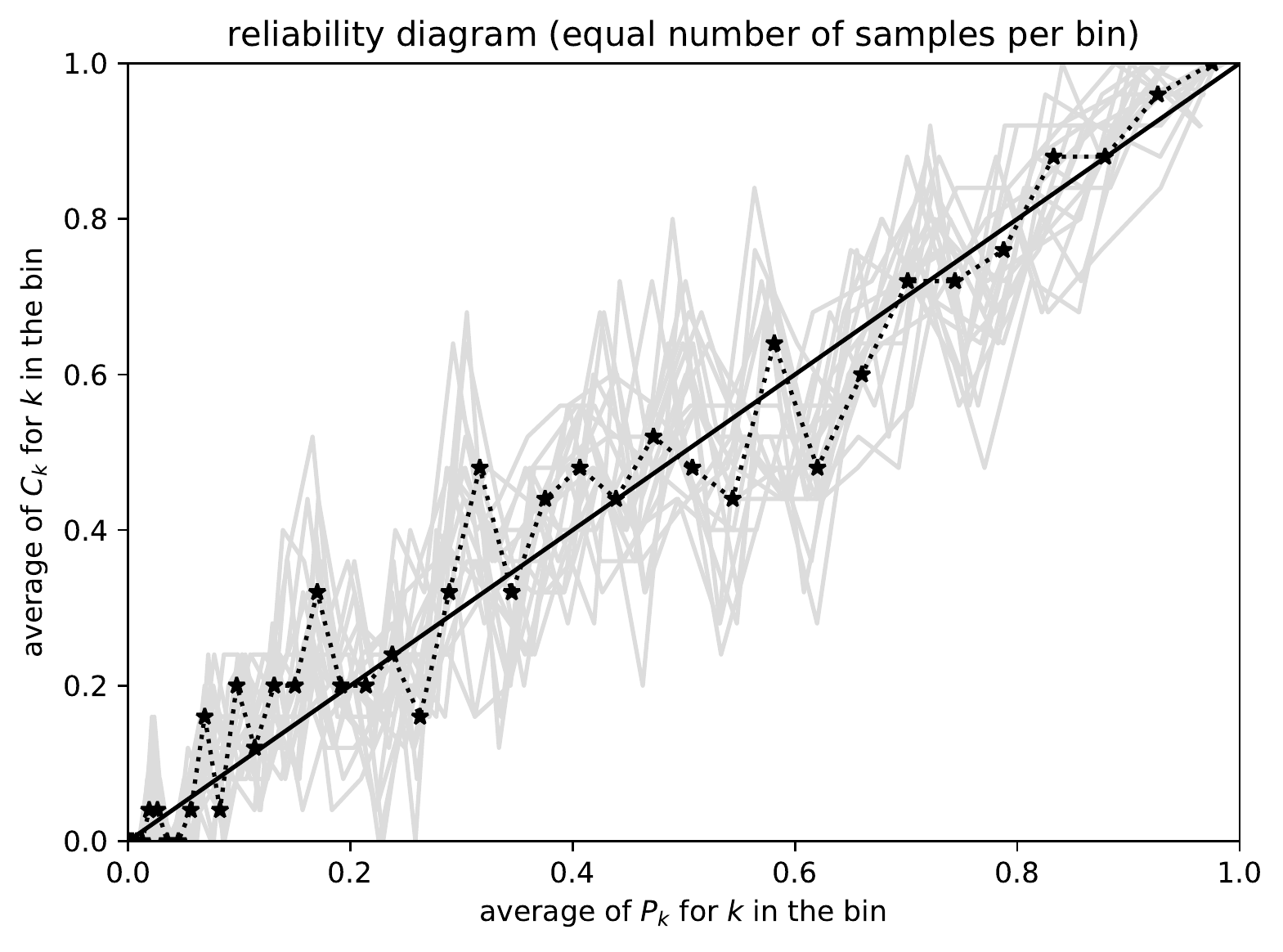}}

\vspace{\vertsep}

\parbox{\imsize}{\includegraphics[width=\imsize]
                 {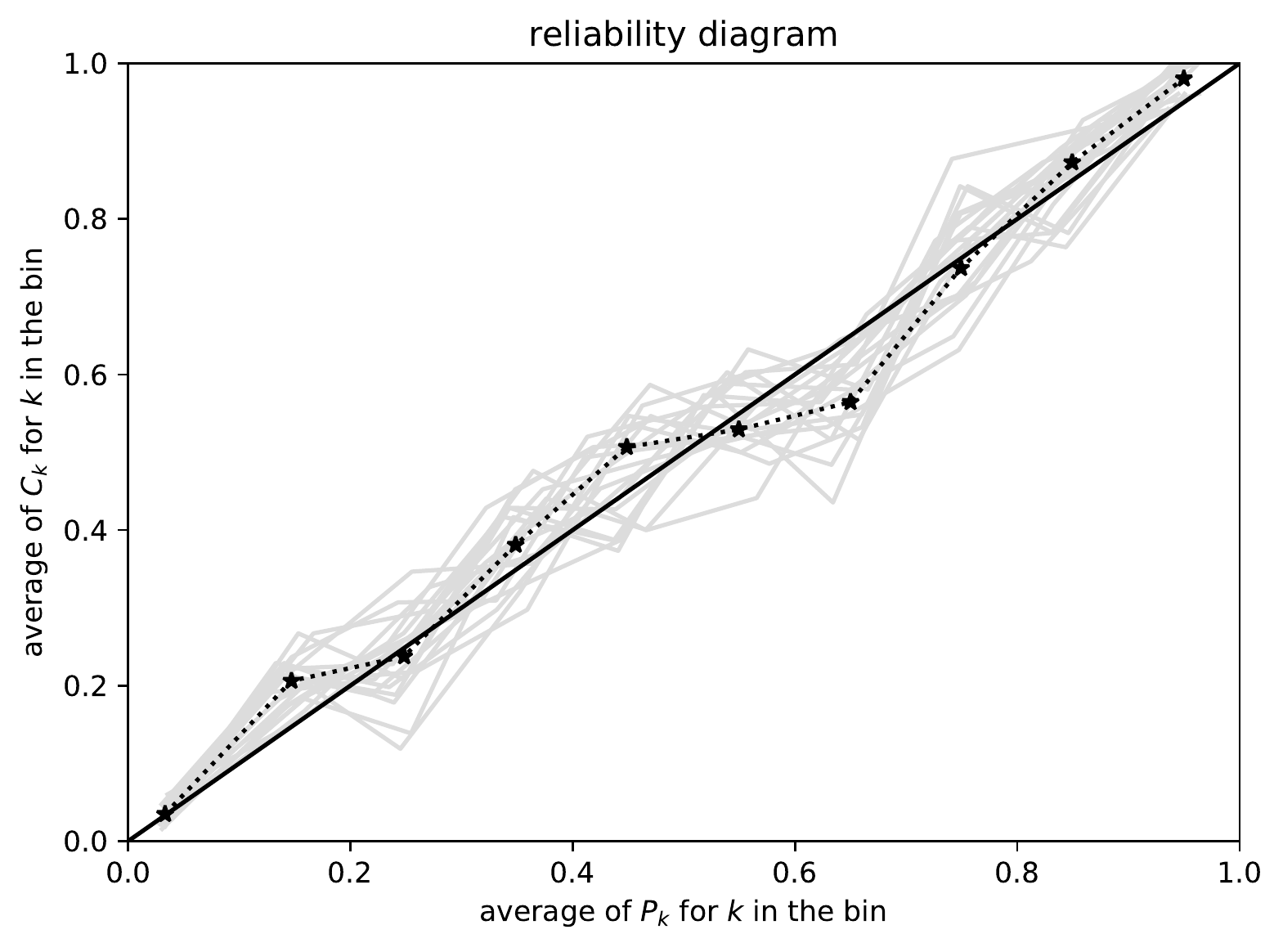}}
\quad\quad
\parbox{\imsize}{\includegraphics[width=\imsize]
                 {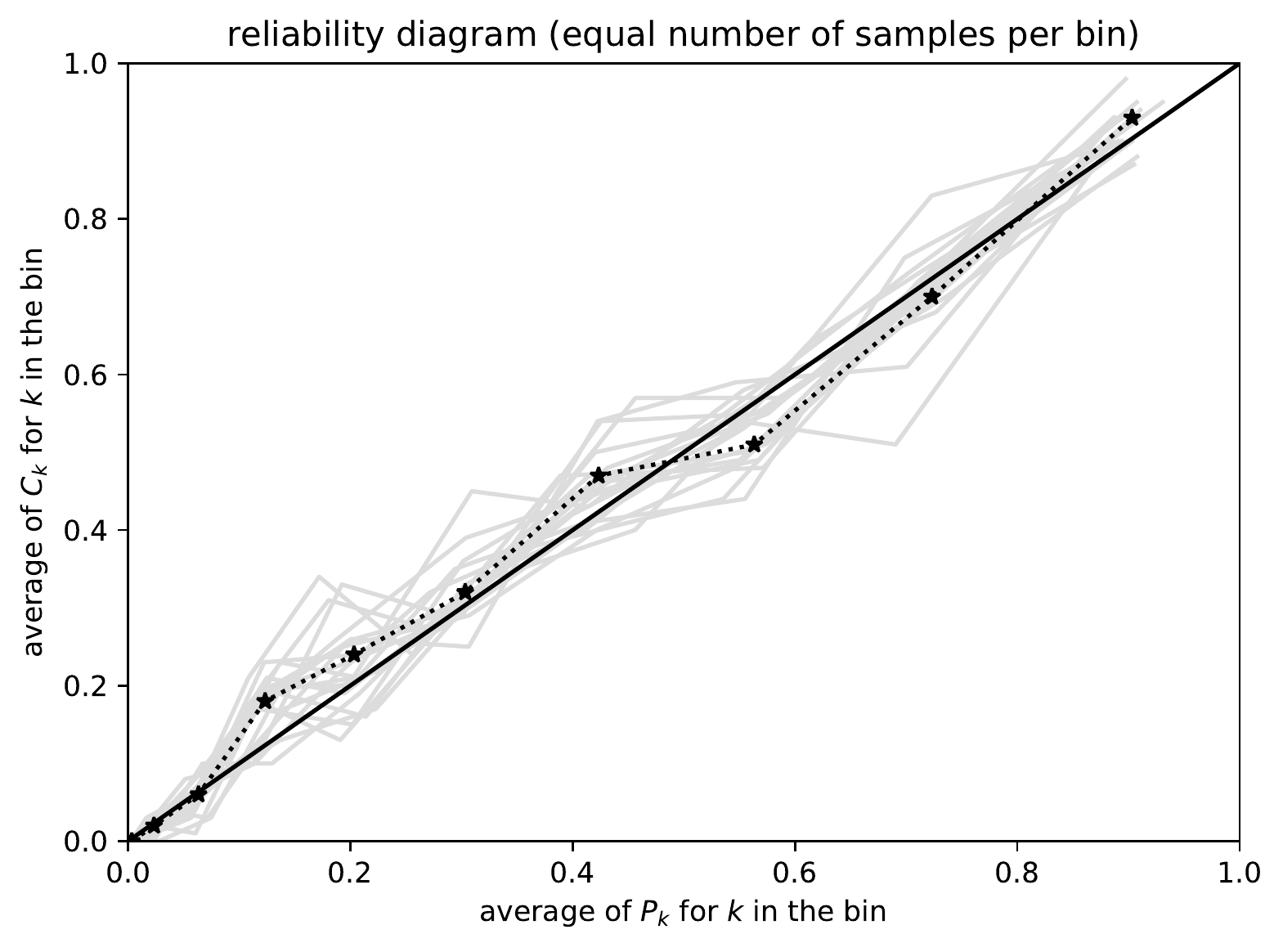}}

\vspace{\vertsep}

\parbox{\imsize}{\includegraphics[width=\imsize]
                 {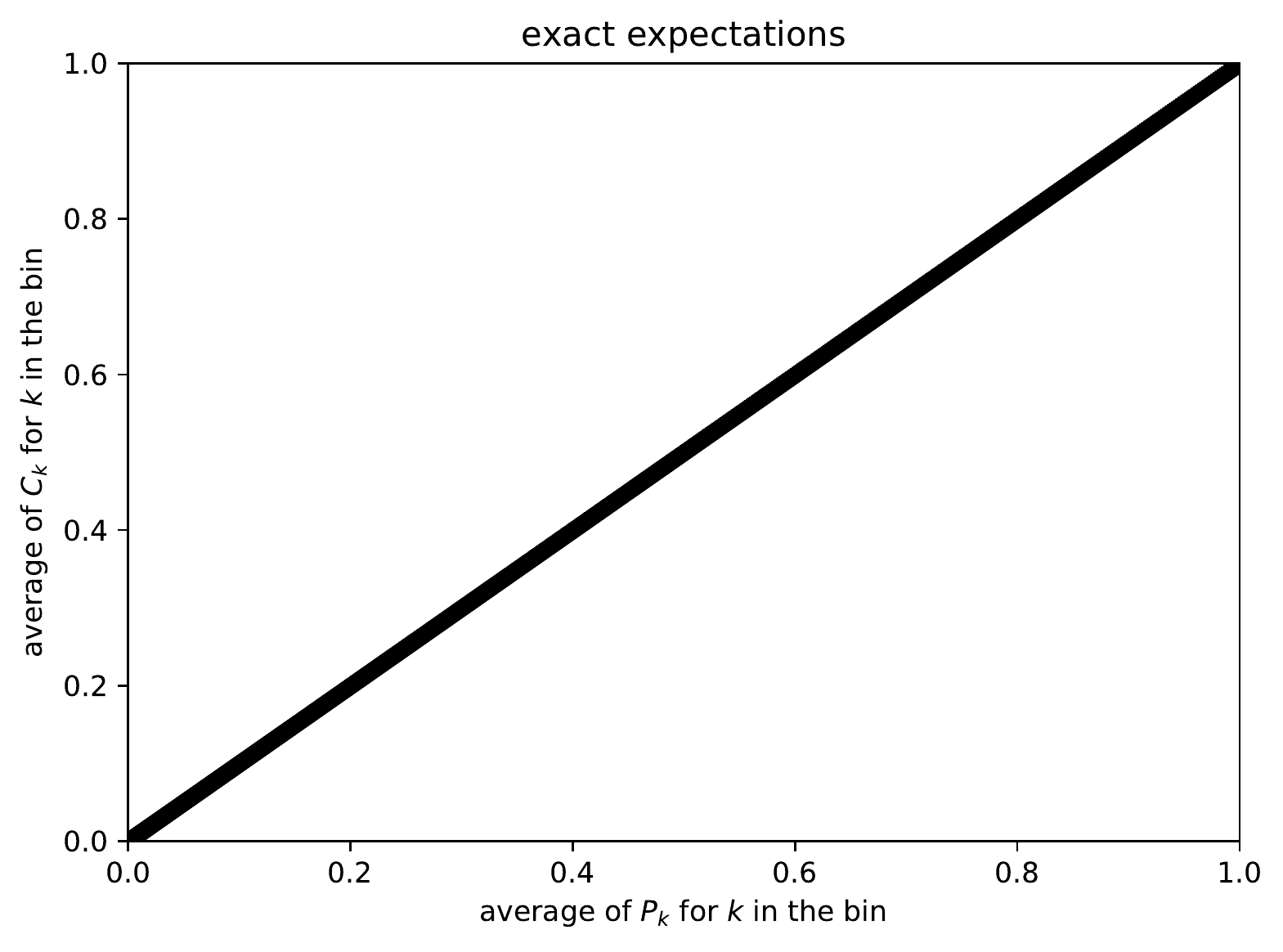}}

\end{centering}
\caption{$n =$ 1,000; $P_1$, $P_2$, \dots, $P_n$ are denser near 0}
\label{1000_00}
\end{figure}

\begin{figure}
\begin{centering}

\parbox{\imsize}{\includegraphics[width=\imsize]
                 {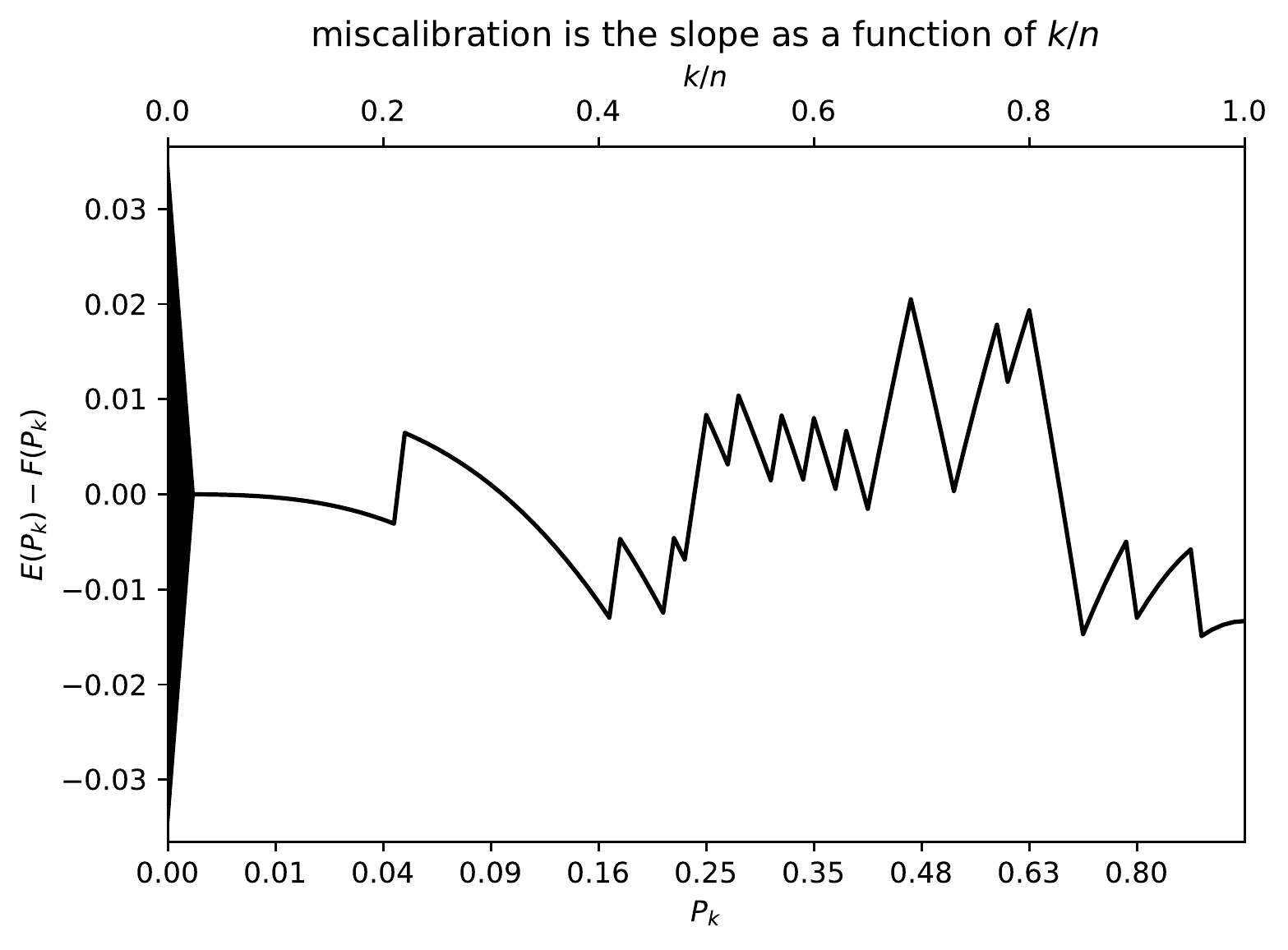}}
\quad\quad
\parbox{\imsize}{\includegraphics[width=\imsize]
                 {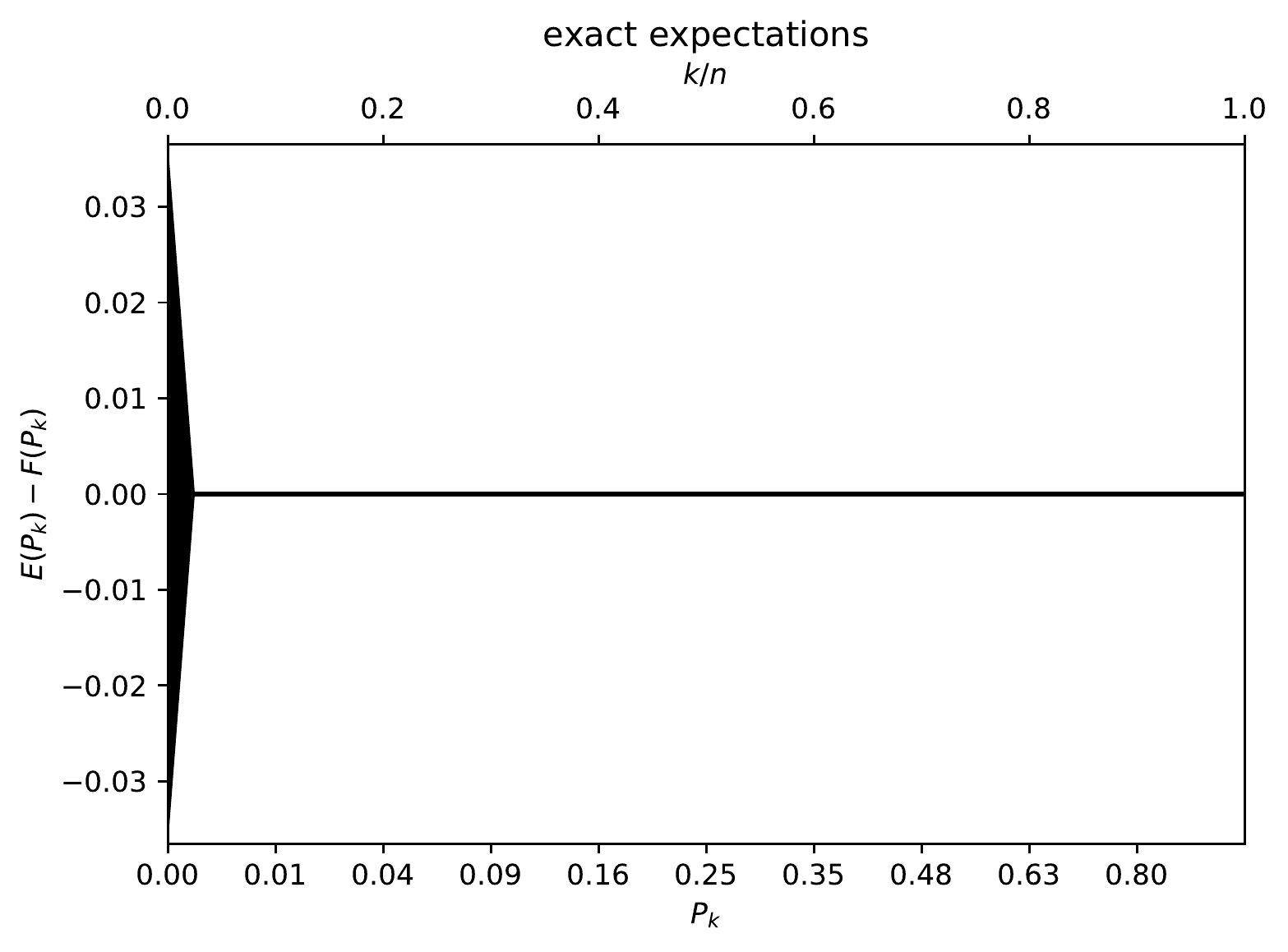}}

\vspace{\vertsep}

\parbox{\imsize}{\includegraphics[width=\imsize]
                 {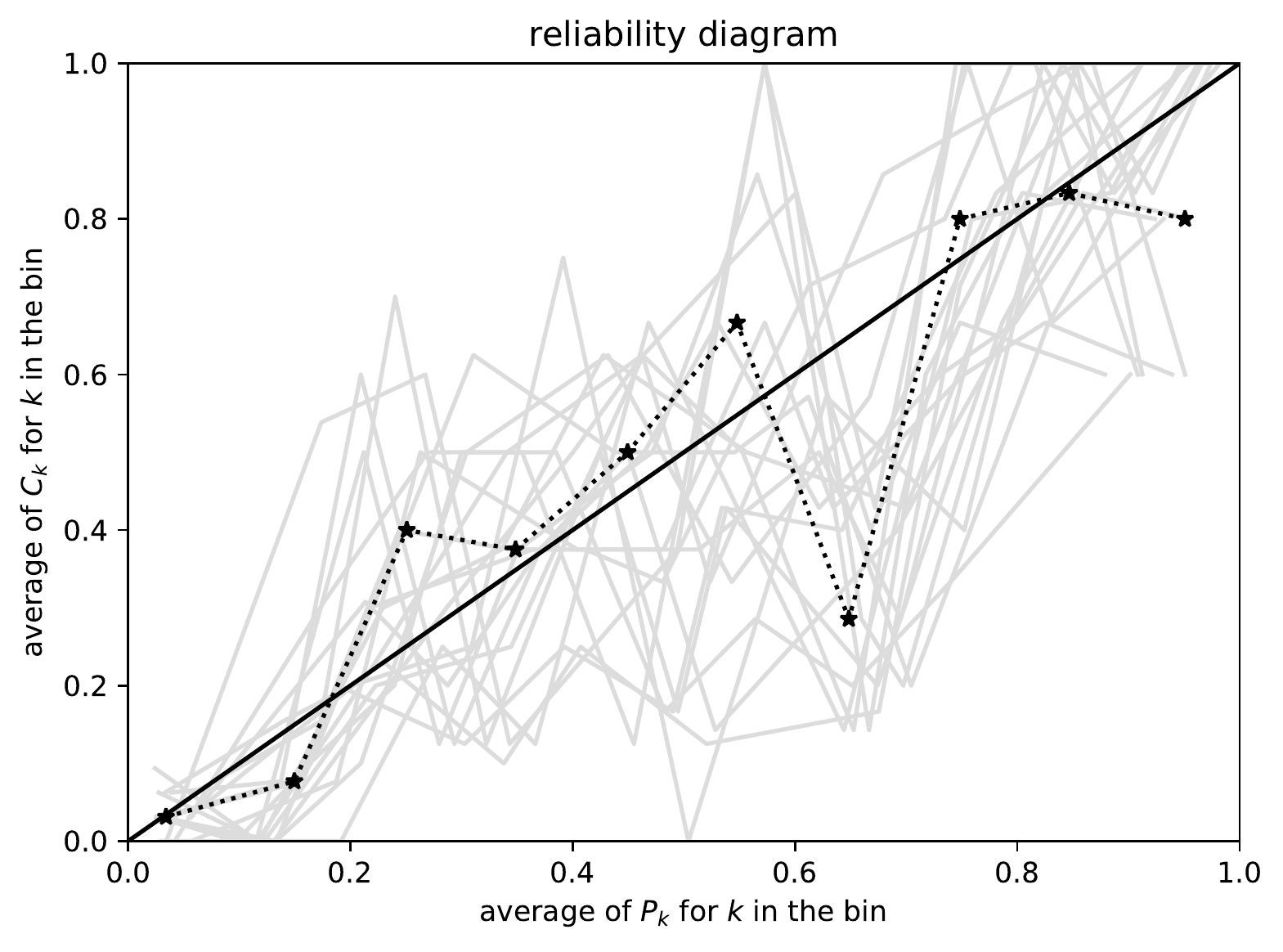}}
\quad\quad
\parbox{\imsize}{\includegraphics[width=\imsize]
                 {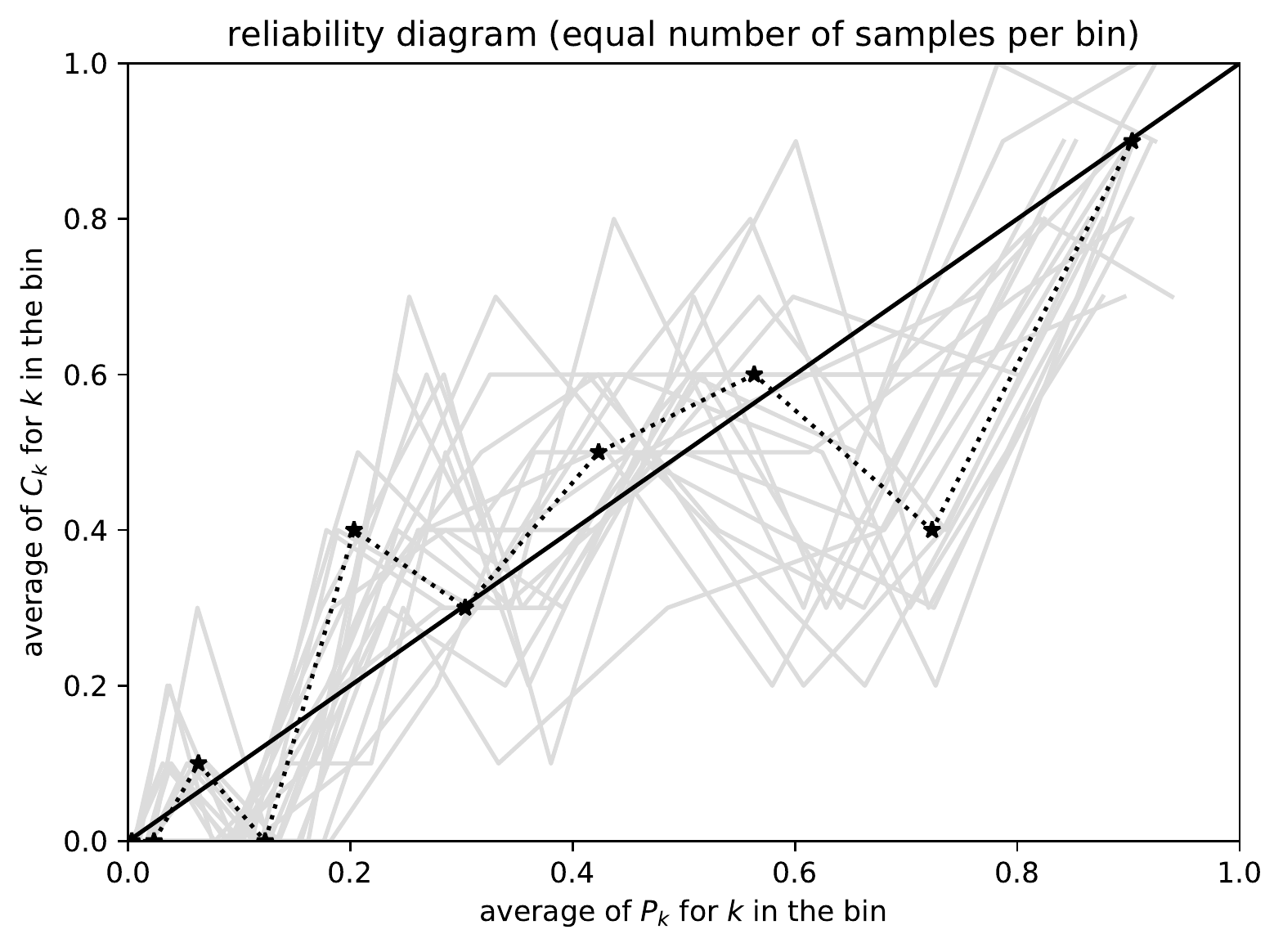}}

\vspace{\vertsep}

\parbox{\imsize}{\includegraphics[width=\imsize]
                 {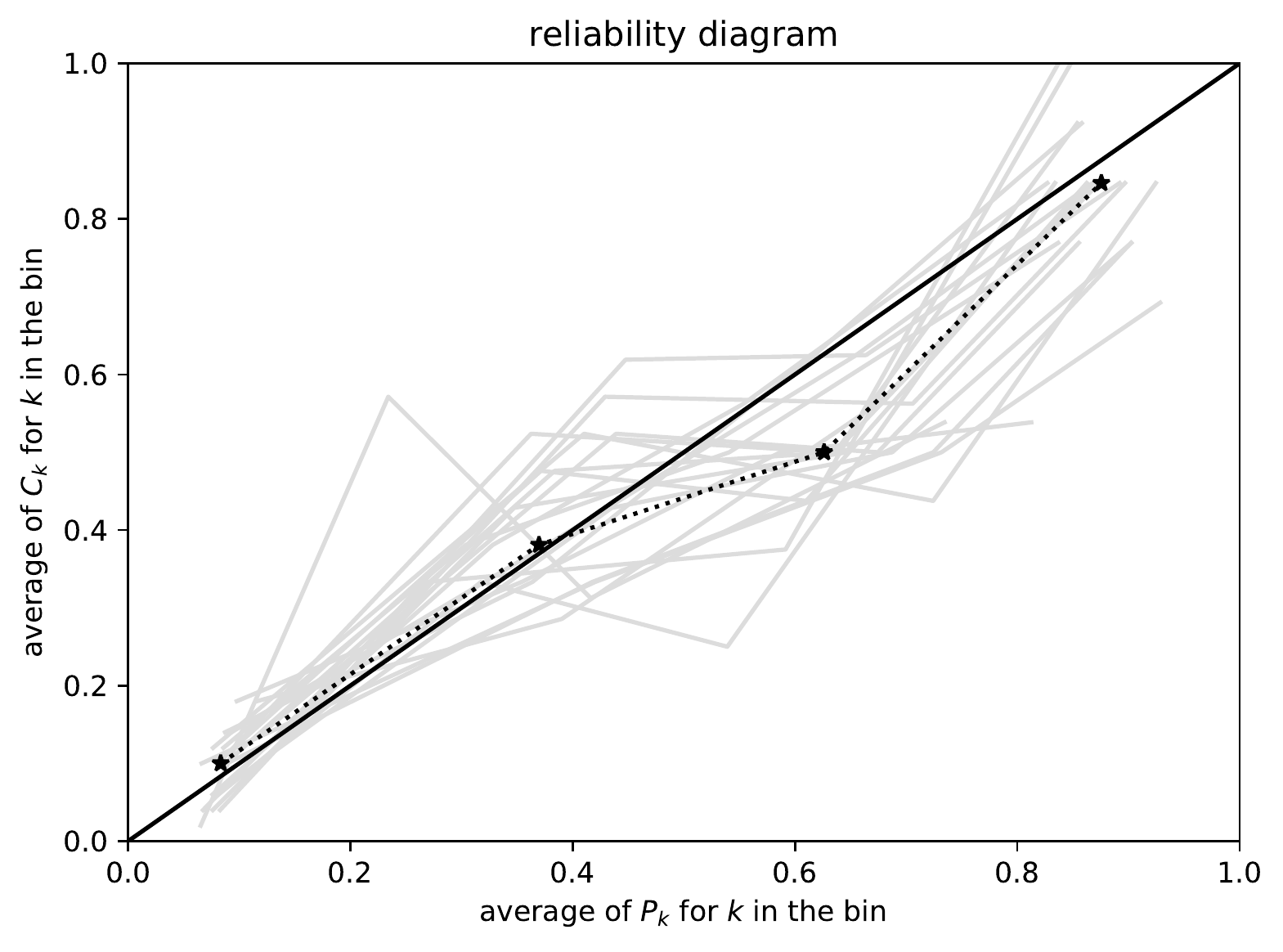}}
\quad\quad
\parbox{\imsize}{\includegraphics[width=\imsize]
                 {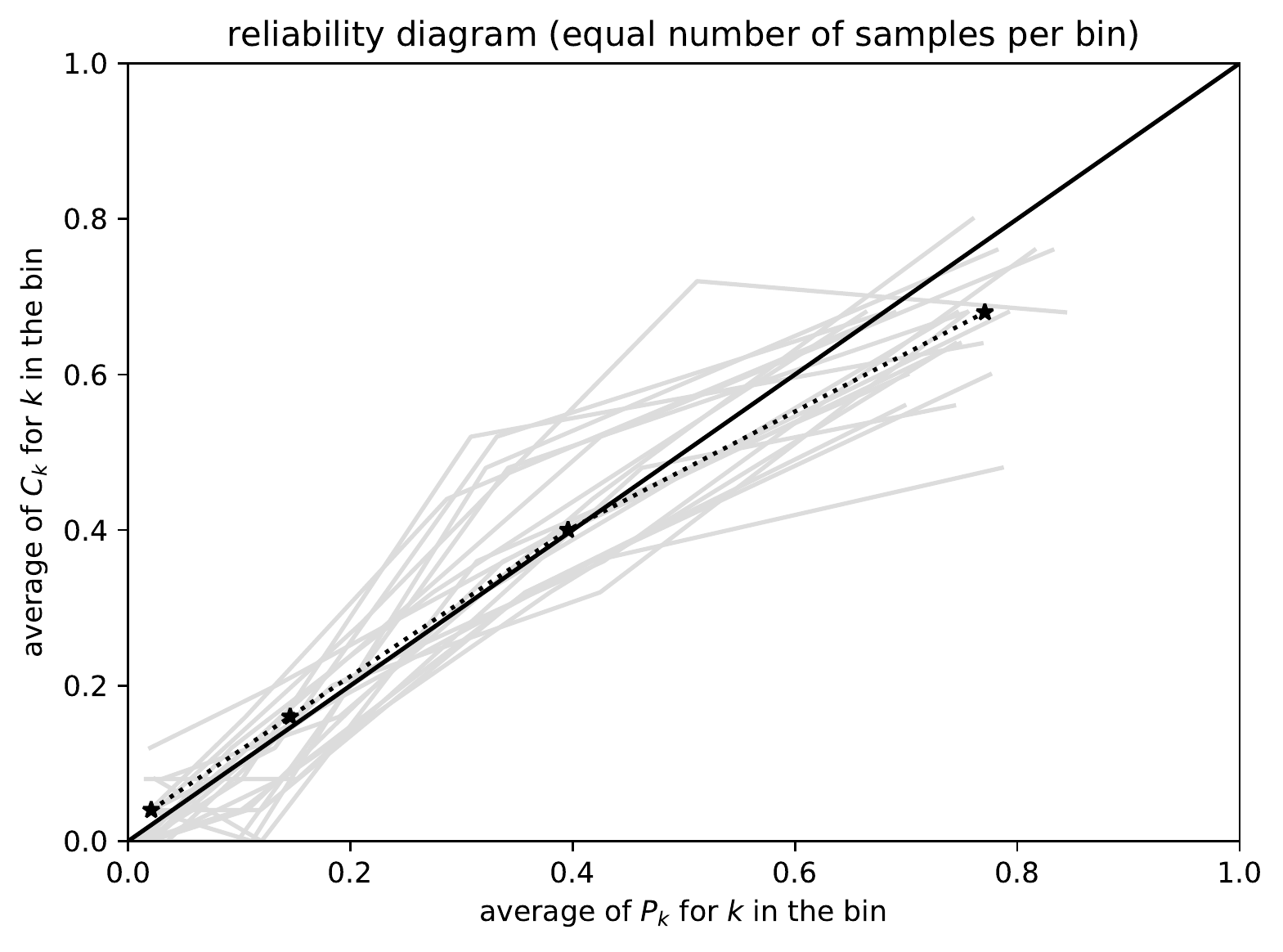}}

\vspace{\vertsep}

\parbox{\imsize}{\includegraphics[width=\imsize]
                 {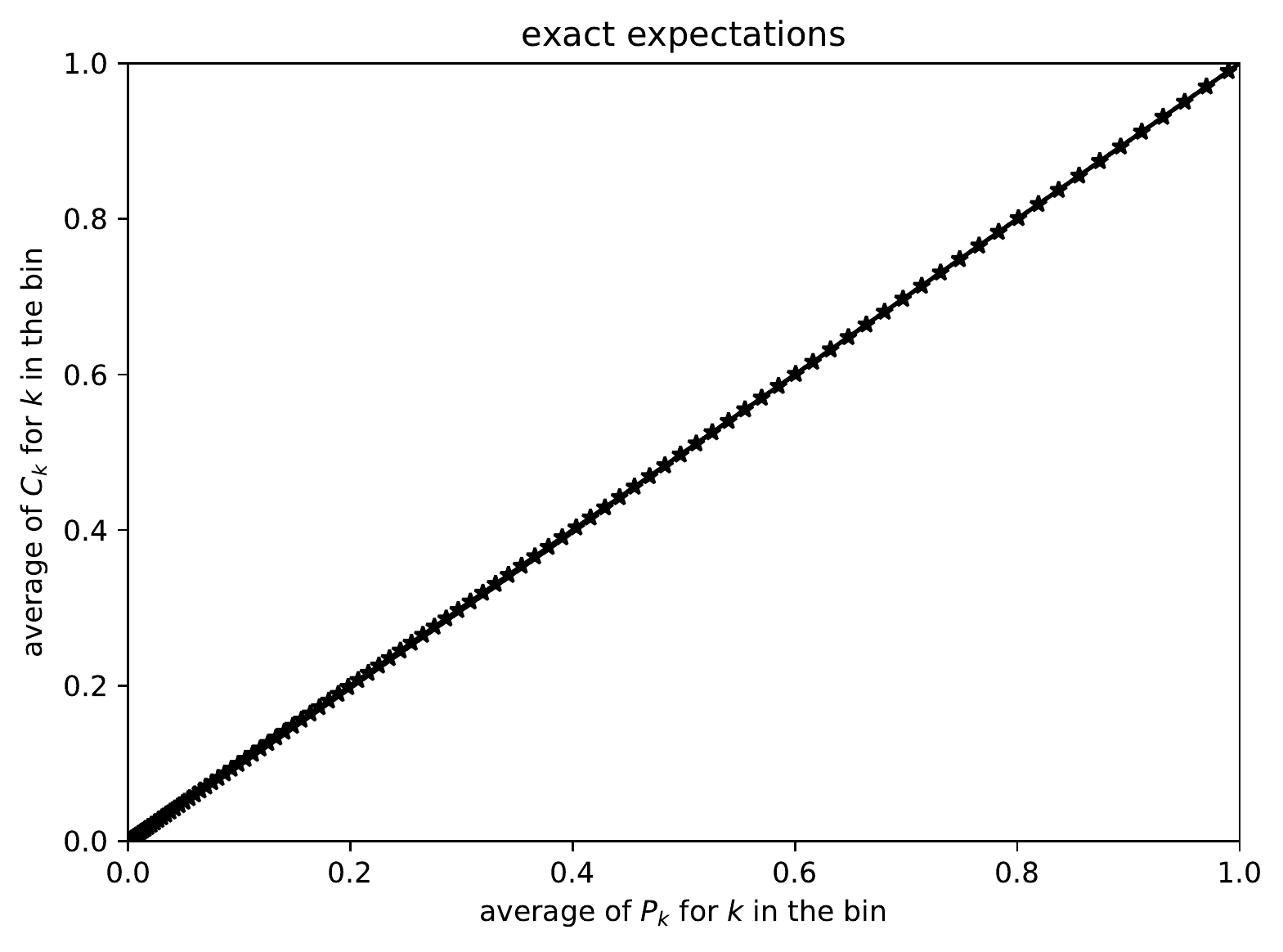}}

\end{centering}
\caption{$n =$ 100; $P_1$, $P_2$, \dots, $P_n$ are denser near 0}
\label{100_00}
\end{figure}

This appendix provides figures --- Figures~\ref{10000_00}--\ref{100_00}
--- analogous to those presented in Section~\ref{results},
but with the observations drawn from the same predicted probabilities
used to generate the graphs, so that the discrepancy from perfect calibration
should be statistically insignificant.
More precisely, Figures~\ref{10000_00}--\ref{100_00}
all set $P_k$ to be proportional to $(k-0.5)^2$
and draw $C_1$,~$C_2$, \dots, $C_n$ from independent Bernoulli distributions
with expected success probabilities $P_1$,~$P_2$, \dots, $P_n$, respectively;
this corresponds to setting $\tilde{P}_k = P_k$
for all $k = 1$,~$2$, \dots, $n$, in the numerical experiments
of Section~\ref{results}.
Figures~\ref{10000_00}, \ref{1000_00}, and~\ref{100_00}
consider $n =$ 10,000, $n =$ 1,000, and $n =$ 100, respectively.
Please note that the ranges of the vertical axes
for the top rows of plots are drastically smaller
in Figures~\ref{10000_00}--\ref{100_00}
than in Figures~\ref{10000}--\ref{100_1}.
The leftmost topmost plots in Figures~\ref{10000_00}--\ref{100_00}
look like driftless random walks;
in fact, they really are driftless random walks.
The variations of the graphs are comparable to the heights
of the triangles centered at the origins.
Comparing the second rows with the third rows
shows that the deviations from perfect calibration
are consistent with expected random fluctuations.
Indeed, all plots in this appendix depict only small deviations
from perfect calibration, as expected (and as desired).

\newpage

\bibliography{calibration}
\bibliographystyle{siam}

\end{document}